\documentclass[twocolumn, twocolappendix]{aastex631}

\usepackage{mathtools}
\usepackage{csvsimple}
\usepackage{booktabs}
\usepackage{longtable}

\usepackage{multirow}
\usepackage{tikz}

\newcommand*\circled[1]{\tikz[baseline=(char.base)]{
            \node[shape=circle,draw,inner sep=1pt] (char) {#1};}}

\submitjournal{PSJ}

\shorttitle{Mechanisms of atmospheric loss during giant impacts}
\shortauthors{Roche et al.}

\begin{document}

\title{Atmospheric loss during giant impacts: mechanisms and scaling of near- and far-field loss}




\author[0000-0002-2595-1393]{Matthew J. Roche}
\affiliation{School of Earth Sciences, Wills Memorial Building, University of Bristol, Bristol, BS8 1RJ, UK}

\author[0000-0001-5365-9616]{Simon J. Lock}
\affiliation{School of Earth Sciences, Wills Memorial Building, University of Bristol, Bristol, BS8 1RJ, UK}

\author[0000-0003-4518-8003]{Jingyao Dou}
\affiliation{School of Physics, HH Wills Physics Laboratory, University of Bristol, Bristol, BS8 1TL, UK}

\author[0000-0001-5065-4625]{Philip J. Carter}
\affiliation{School of Physics, HH Wills Physics Laboratory, University of Bristol, Bristol, BS8 1TL, UK}

\author[0000-0001-5383-236X]{Jacob A. Kegerreis}
\affiliation{NASA Ames Research Centre, MS 245-3, Moffett Field, CA 94035, USA}

\author[0000-0003-4813-7922]{Zo\"{e} M. Leinhardt}
\affiliation{School of Physics, HH Wills Physics Laboratory, University of Bristol, Bristol, BS8 1TL, UK}

\correspondingauthor{Matthew J. Roche}
\email{matthew.roche@bristol.ac.uk}

\begin{abstract} 

\noindent The primary epoch of planetary accretion concludes with giant impacts -- highly energetic collisions between proto-planets that can play a key role in shaping a planet's inventory of volatile elements. Previous work has shown that single giant impacts have the potential to eject a significant amount of a planet's atmosphere but that the efficiency of atmospheric loss depends strongly on the impact parameters and atmospheric properties. Fully quantifying the role of giant impacts in planetary volatile evolution requires a more complete understanding of the mechanisms driving loss during impacts. Here, we use a suite of 3D smoothed particle hydrodynamics simulations to show that loss in giant impacts is controlled primarily by ejecta plumes near the impact site and breakout of the impact shock in the far field, with the efficiency of the latter well approximated by 1D ground-kick calculations. The relative contributions of each mechanism to loss changes drastically with varying impact parameters. By considering the near and far field separately, we present a scaling law that precisely approximates (to within an average of $\sim$3\%) loss from 0.35 to 5.0 Earth mass planets with 5\% mass fraction H$_2$–He atmospheres for any combination of impactor mass, impact velocity, and angle. Finally, we apply our scaling law to the results of $N$-body simulations for different solar system formation scenarios. We find that while individual impacts rarely cause significant loss ($>$10\%) from roughly Earth-mass planets with such massive primary atmospheres, the cumulative effect of multiple impacts can be substantial (40–70\% loss).

\end{abstract}


\keywords{Impact phenomena, planet formation, atmospheric evolution, hydrodynamics}

\section{Introduction} \label{sec:intro}
Volatile elements (e.g., N, C, H, noble gases) play a fundamental role in the evolution and habitability of planets \citep[][]{crowley2011relative,kasting2003evolution}. The terrestrial planets of our solar system display a wide variety of volatile budgets and, consequently, atmospheric properties \citep[e.g.,][]{broadley2022origin}. Observations of exoplanetary systems have revealed an even larger range in planetary volatile inventories even within individual systems \citep[e.g.,][]{howard2012planet, lopez2014understanding, rogers2015most, fulton2017california, schlichting2018formation, bonomo2019giant, liu2015giant}. However, the fundamental question of how Earth and other terrestrial planets acquired and retained their volatile budgets, and ultimately what led to the diversity in planetary volatiles, remains poorly understood \citep[][and references therein]{broadley2022origin}. 

\begin{figure*}[t]
\begin{center}
    \includegraphics[scale=0.77]{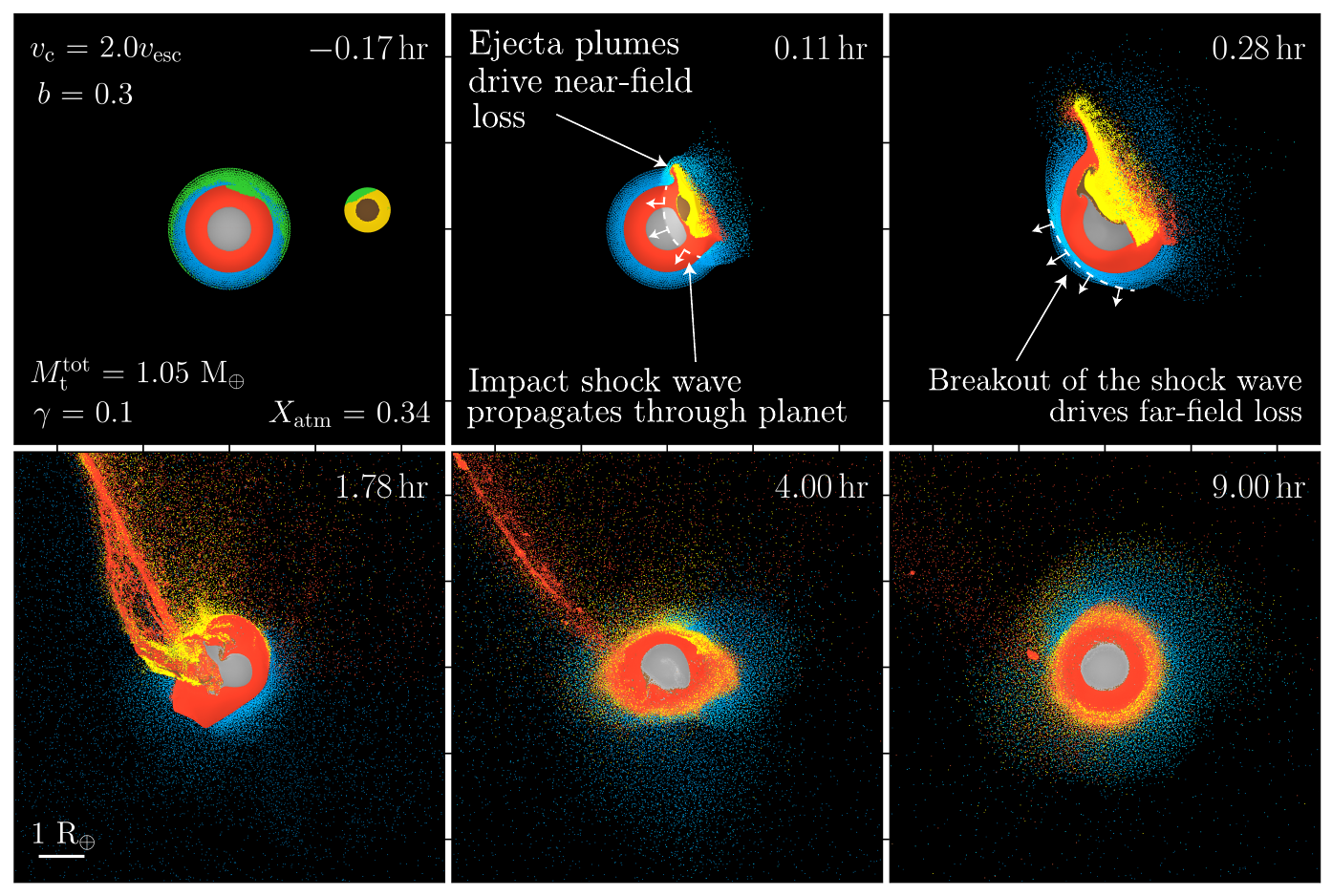}
    \caption{Giant impacts are highly complex events with different processes dominating the dynamics of different regions of the colliding bodies. Illustrative time snapshots from an example SPH giant impact simulation with a resolution on the order 10$^6$ particles. All particles in the lower hemispheres only of the colliding bodies are shown and coloured by material type as follows: target core (grey), mantle (orange), and atmosphere (blue); impactor core (brown) and mantle (yellow). Colour luminosity is scaled by the particles' internal energy. Impact parameters (defined in Section~\ref{subsec:impact_sim}) and atmospheric loss fraction ($X_{\rm atm}$) are detailed in the first panel, with lost particles (those that ultimately become unbound to the largest remnant) coloured in green. The simulation time corresponding to each snapshot is given in each panel, with t = 0.0~hr defined as the point of impact. White dashed lines in the second and third panels indicate the approximate location of the front of the impact shock wave. An animation of this figure is available in the Supplementary Material.}
    \label{fig:impact}
\end{center}
\end{figure*}

Large fractions of planetary volatiles are likely acquired during the main stages of accretion \citep[][]{halliday2013origins, marty2012origins, marty2022meteoritic, marty2016origins, hirschmann2016constraints, mukhopadhyay2019noble, grewal2024contribution} but, for the Earth at least, not all of those volatiles are retained by the final planet. We have observational constraints on volatile and isotope ratios of the modern bulk silicate Earth (BSE) and of the Earth's cosmochemical building blocks – meteorites and comets. Despite the isotopic similarities of Earth's N, C, and H to chondritic sources \citep[i.e., primitive material that most closely resembles the composition of the Sun;][]{marty2012origins, marty2022meteoritic}, the C/H, N/H, and C/N abundance ratios of the BSE are highly non-chondritic \citep{hirschmann2009h, marty2012origins, halliday2013origins, tucker2014evidence, hirschmann2016constraints, bergin2015tracing}. These signatures are inconsistent with terrestrial volatiles being acquired from a simple accumulation of volatile-rich chondritic material and other processes must have shaped the Earth's volatile budget.

Given the violent and highly stochastic nature of accretion, there are many mechanisms through which volatiles can be gained and lost during the early evolution of planets \citep[e.g.,][]{o2014water, schlichting2015atmospheric, massol2016formation, marty2016origins, schlichting2018atmosphere, odert2018escape}. The era of primary planetary accretion is concluded by giant impacts --- collisions between planet-sized bodies --- which likely play a crucial role in the chemical evolution of planets. Giant impacts are the most violent and energetic events that planets experience during their formation and can drastically alter planetary composition, often leaving large portions of the post-impact bodies melted or vaporised and rapidly rotating \citep[e.g.,][]{canup2004dynamics, carter2018collisional, carter2020energy, lock2018origin, lock2017structure,lock2019giant,caracas2023no,nakajima2015melting,nakajima2021scaling}. In our solar system, most growing terrestrial planets likely experience at least one giant impact during their formation \citep{quintana2016frequency}. The products of these impacts \citep[e.g., amorphous silica/melt glass, SiO vapour and CO gas;][]{lisse2009abundant, schneiderman2021carbon, su2019extreme} and the resulting post-impact bodies \citep{kenworthy2023planetary} have even potentially been observed in exoplanetary systems.
Critically, giant impacts have the potential to remove significant proportions of a planet's existing atmosphere and/or ocean \citep{genda2003survival, genda2005enhanced, schlichting2015atmospheric, inamdar2015formation, schlichting2018atmosphere, kegerreis2020atmosphericA, kegerreis2020atmosphericB, denman2020atmosphere, denman2022atmosphere, lock2024atmospheric} -- a process that could be recorded in the large density diversity observed amongst the exoplanet population \citep[][]{inamdar2016stealing, chance2022signatures}. Given the ubiquity of giant impacts in planet formation, quantifying the fraction of atmosphere that is lost in different styles of impact is imperative to understanding the origin and evolution of planetary volatiles.

It has previously been posited, based on a combination of early theoretical and 1D studies \citep[e.g.,][]{ahrens1990earth, melosh1989impact, ahrens1987impact, vickery1990atmospheric}, that atmospheric loss during giant impacts can be driven primarily by three key mechanisms: air shocks, ejecta plumes, and shock-kick. Air shocks are generated by the entry of the impactor into the target planet's atmosphere and by deposition of impact energy at the planet's surface, meaning atmospheric material near the impact site (in the near field) can reach particle velocities in excess of the escape velocity \citep{schlichting2015atmospheric, ahrens1989formation,ahrens1993impact,shuvalov2009atmospheric}. Vaporisation of the colliding planets also drives atmospheric gases (as well as crustal/mantle material and volatiles dissolved in it) from the near field to escape in high-temperature ejecta plumes \citep[][]{vickery1990atmospheric}. Away from the impact site (in the far field), atmospheric loss is driven by a process often referred to as `shock-kick' (see Section 2 of \citet{lock2024atmospheric} for a detailed description). The shock wave generated by the impact propagates through the interiors of the colliding bodies until it breaks out at the surface into the atmosphere. The atmosphere has a different impedance (resistance to shock compression) to that of the rocky surface and the surface isentropically decompresses (releases) and accelerates providing a `kick' to the bottom of the atmosphere. The shock accelerates up the hydrostatic atmosphere driving portions of the atmospheric column to reach velocities above the escape velocity and become ejected from the system. The strength of the impact shock varies significantly around the surface of the planet and depends on the initial strength of the shock at the contact point (i.e., the impact velocity), the length of time for which the shock is supported, and the geometry of the collision. The longer the shock is supported at the contact site, and for over a greater area, the longer it takes for release waves from the surfaces of the colliding bodies to catch up with the main impact shock and reduce its amplitude as it travels through the planet. Atmospheric loss due to shock-kick has previously been investigated using 1D hydrodynamic simulations \citep{chen1997erosion, genda2003survival, genda2005enhanced, lock2024atmospheric} and semi-analytical calculations \citep{schlichting2015atmospheric, inamdar2015formation}.

Advances in numerical techniques and the expansion in high-performance computing resources have recently allowed 3D smoothed particle hydrodynamics (SPH) simulations --- a Langrangian scheme where planets are modelled as a system of interacting particles that evolve under gravity and pressure forces --- to be employed at high enough resolutions to model atmospheric loss during giant impacts \citep[][]{liu2015giant,kegerreis2018consequences, kegerreis2020atmosphericA, kegerreis2020atmosphericB, denman2020atmosphere, denman2022atmosphere, lammer2020constraining, kurosaki2023giant, kurosaki2023evolution}, capturing loss via all possible mechanisms simultaneously (Figure~\ref{fig:impact}). Suites of 3D giant impact simulations have been used to produce scaling laws that predict the atmospheric loss fraction \citep{kegerreis2020atmosphericA, kegerreis2020atmosphericB, denman2020atmosphere, denman2022atmosphere} given the parameters of the impact (target and impactor masses, impact angle, and velocity). Such laws allow the effect of atmospheric loss due to giant impacts to be included in larger scale simulations of planet formation \citep[e.g.,][]{gu2024composition}. However, previous studies did not explore in detail the relative contributions of the different mechanisms driving loss and how they varied with impact parameters. Without understanding the mechanisms driving loss it is not possible to extend the results of current studies to explore parameter spaces not directly modelled in SPH simulations, such as the thin atmospheres of the terrestrial planets in our solar system. Also, existing scaling laws do not explicitly account for the complex interplay between the changing loss in the near and far field and their different sensitivity to impact parameters, leading to errors in predicted loss as large as 40\% (see Section \ref{subsec:limitations}). A more complete understanding of the mechanisms of atmospheric loss in giants impacts, as well as more precise scaling relations which model each mechanism of loss separately, are required to determine the role of giant impacts in setting the volatile budgets of terrestrial planets.

In this study, we focus on building an understanding of the mechanisms driving atmospheric loss in different giant impacts. We define giant impacts as collisions in which the target mass is at least 0.1 Earth masses and the impactor-to-target mass ratio is greater than 0.01, as proposed by \citet{carter2020energy}. We present the results of a suite of 479 3D SPH simulations of giant impacts onto planets between 0.35–5.0 Earth masses (M$_{\oplus}$) with 5\% mass fraction primordial H$_2$–He atmospheres at varying combinations of impact parameters, impact velocities, and impactor-to-total system mass ratios. Simulations are run with (and a subset without) pre-impact tidal deformation to constrain the effect of pre-impact deformation on loss (Section~\ref{subsec:tot_loss}). We examine in detail the atmospheric loss experienced and the mechanisms by which atmosphere is lost in different styles of impact. In particular, we de-convolve the total (hereafter `combined') loss into a sum of near-field loss, driven by air shocks and ejecta plumes, and far-field loss, driven by shock-kick (Section~\ref{subsec:NF_loss}). We find that the relative contributions to loss from these mechanisms change significantly as a function of the impact parameters. We derive a high-precision scaling law for loss that models total loss as a sum of near- and far-field loss separately, which can readily be incorporated into models of volatile accretion during planet formation (Section~\ref{subsec:scaling}). Future work will expand our scaling law to account for the effect of varying atmospheric properties (composition, temperature, and mass fraction) known to be key controlling factors on loss \citep{lock2024atmospheric}. We also present a framework for convolving loss from 3D SPH simulations with 1D impedance match and loss calculations (Section~\ref{subsec:1D} and Appendix~\ref{sec:NF_sep}). We then discuss the limitations, applicability and implications of our results (Section~\ref{sec:discussion}). In particular, to understand the potential role of giant impacts in volatile evolution, we apply our scaling law for loss to the results of existing $N$-body simulations of four solar system formation scenarios to provide insight on the significance of atmospheric loss due to giant impacts during accretion of the solar system terrestrial planets (Section \ref{subsec:implications_volatiles}). Finally, we conclude in Section~\ref{sec:conclusions}. Appendix~\ref{sec:appendix} contains tables of the results of the SPH impact simulations and the fitted scaling law parameters. Supplementary material containing all SPH input parameter files and analysis scripts is available on Zenodo.

\section{Methods} \label{sec:methods}
In this section, we first outline the setup, initial conditions, and input parameters for our model SPH planets and impact simulations. All SPH impact simulations were run with the \texttt{SWIFT} (version 0.9.0) hydrodynamics and gravity code \citep{schaller2016swift, kegerreis2019planetary, schaller2023swift} using the \texttt{subtask\_speedup} branch \citep{matthieu_schaller_2024_14230347}. The version of the code used is archived on Zenodo. We then describe our methods for calculating atmospheric loss from the output of these simulations (Section~\ref{subsec:atmos_loss}), define the near and far field for different impacts (Section~\ref{subsec:NF_def}), and finally outline how the results of the 3D SPH simulations can be convolved with the 1D impedance match calculations to predict loss due to shock-kick in the far field (Sections~\ref{subsubsec:ground_surf} \& \ref{subsec:impedance_match}).


\subsection{Initial Conditions} \label{subsec:init_cond}
We considered impacts with four target masses and five impactor-to-total system mass ratios (Section~\ref{subsec:impact_sim}), thus requiring initialization of four target planets and 20 impactors. The targets had pre-impact atmospheres but the impactors did not. The mass of the target planets neglecting atmosphere ($M_{\rm t}^{\rm r}$, i.e., the refractory target mass) were 0.35, 1.0, 2.0, and 5.0~M$_{\oplus}$. The impactor-to-total system mass ratio, $\gamma$, is given by 

\begin{equation}
    \gamma = \frac{M_{\rm i}^{\rm r}}{M_{\rm i}^{\rm r} + M_{\rm t}^{\rm r}} = \frac{M_{\rm i}^{\rm r}}{M_{\rm tot}^{\rm r}} \; ,
\end{equation}

\noindent where $M_{\rm i}^{\rm r}$ and $M_{\rm tot}^{\rm r}$ are the refractory impactor and total refractory masses, respectively. Impactor masses for each refractory target mass were calculated according to

\begin{equation}
    M_{\rm i}^{\rm r} = \frac{\gamma}{1 - \gamma}M_{\rm t}^{\rm r} \; .
\end{equation}

\noindent Interior thermal profiles for both target and impactor planets were generated using the \texttt{WoMa} (version 1.2.0) planetary structure code \citep{kegerreis2019planetary, ruiz2021effect} by iteratively solving the equation of hydrostatic equilibrium. Both target and impactor planets are differentiated into an iron core and forsterite mantle with a core mass fraction (CMF) of 0.3 and have no pre-impact rotation. The ANEOS pure iron \citep{sarah_t_stewart_2020_3866507} and forsterite \citep{stewart_2019_3478631,10.1063/12.0000946} equations of state (EoSs) were used to model the two materials. The mantle was modelled as isentropic with a surface temperature of 1,700~K (somewhat below the melting temperature of the mantle), giving a corresponding specific entropy of $\sim$2.6~kJ~kg$^{-1}$K$^{-1}$. The core was isentropic with a specific entropy of $\sim$1.8~kJ~kg$^{-1}$K$^{-1}$ corresponding to core–mantle boundary temperatures in the range of 3,000–4,500~K. 

The target planets were initialised with a 5\% mass fraction adiabatic H$_2$–He atmosphere using the EoS from \citet{hubbard1980structure} (hereafter HM80) with a fixed basal temperature of 500~K. An early implementation of the HM80 EoS into \texttt{SWIFT} that was used in previous studies \citep{kegerreis2018consequences, kegerreis2020atmosphericA, kegerreis2020atmosphericB} contained an error in the calculation of internal energy. This error was rectified in \citet{lock2024atmospheric} and we use the correct formulation here to generate the EoS table used. The EoS table, available in the Supplementary Materials, was also re-gridded at a finer resolution (200 $\times$ 200 grid) and extended down to a lower minimum density ($\rho_{\rm min} = 10^{-5}$~kg~m$^{-3}$) and a lower minimum specific internal energy ($u_{\rm min}$ = 1.0~J~kg$^{-1}$). Atmospheric profiles were then generated with \texttt{WoMa} by integrating outwards from the solid surface to a minimum density of 30~kg~m$^{-3}$ and the surface pressure iteratively updated until the desired mass of atmosphere was reached. This procedure resulted in surface pressures of 0.72, 1.90, 4.26, and 11.59~GPa for the targets with refractory masses of 0.35, 1.0, 2.0, and 5.0~M$_{\oplus}$, respectively, and refractory radii (defined as the radius to the mean of the outermost 100 mantle particles) of 0.75, 1.0, 1.22, and 1.54 R$_{\oplus}$, respectively. The total target planet masses and radii, $M_{\rm t}^{\rm tot}$ and $R_{\rm t}^{\rm tot}$ (defined as the radius to the mean of the outermost 100 atmosphere particles), were 0.37, 1.05, 2.1, and 5.25~M$_{\oplus}$, and 1.15, 1.34, 1.57, 2.0~R$_{\oplus}$, respectively. See Table \ref{table:initial_conditions} for a full breakdown of target and impactor properties.

We used roughly equal-mass SPH particles (in most cases the difference between particle masses was $<5$\%) that were then arranged according to the calculated initial thermal profiles using the stretched equal-area (SEA) method \citep[][\texttt{SEAGen} version 1.4.2]{kegerreis2019planetary}. Planet initialisation was then concluded by running a 20~hr settling simulation for each body in isolation. The particles' specific entropies were fixed for the first 10 simulation hours to ensure adiabatic relaxation and to produce planets with isentropic layers. For the atmosphere particles, specific internal energies were fixed for the first 10 simulation hours since the \citet{hubbard1980structure} EoS for H$_2$–He mixtures does not natively include entropy. Then, the planets were evolved for a further 10 simulation hours without fixing any parameters using the same SPH parameters as in the full impact simulations, to ensure a hydrostatic profile. This settling procedure resulted in final mean core and mantle specific entropies over all planets of 1.80~kJ~kg$^{-1}$K$^{-1}$ and 2.57~kJ~kg$^{-1}$K$^{-1}$, respectively -- very close to the desired values (see above). 

\subsection{Impact Simulations} \label{subsec:impact_sim}
\begin{figure}
\begin{center}
    \includegraphics[width=\columnwidth]{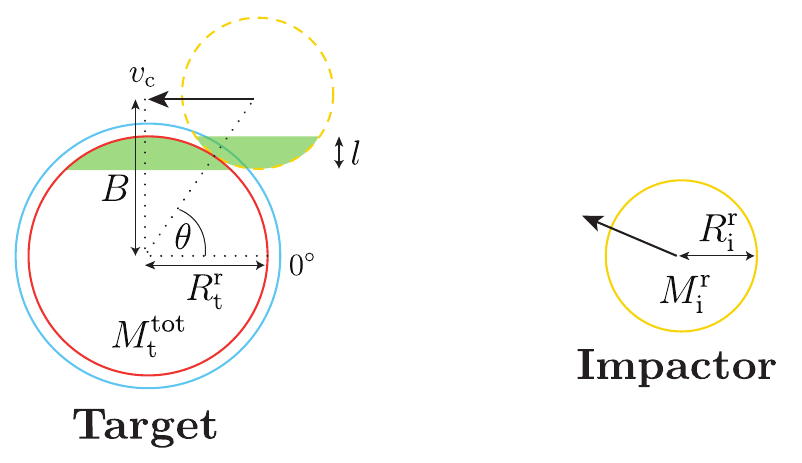}
    \caption{Schematic of the initial conditions and geometry for an example impact. The yellow circles show the top of the impactor's mantle at a time during approach (solid) and upon collision (dashed) at an impact angle, $\theta$ (with corresponding impact parameter $b = \sin{\theta}$), and velocity, $v_{\rm c}$. The orange and blue circles show the top of the target's mantle and atmosphere, respectively. The green shaded regions highlight the total interacting refractory mass -- the portion of the total planetary mass whose kinetic energy actually participates in the initial collision  \citep{leinhardt2012collisions}. The interacting mass is described by the parameters $B = \left(R_{\rm t}^{\rm r} + R_{\rm i}^{\rm r}\right)b$ and $l$, the projected length of the impactor which overlaps the target planet (Equation~\ref{eqn:l}). 0$^{\circ}$ longitude on the surface of the target is defined as the point on the surface whose radial vector is parallel to the impactor's velocity vector at time of first contact (neglecting tidal deformation).}
    \label{fig:initial_conditions}
\end{center}
\end{figure}

We considered impact scenarios for every combination of: (i) impact parameter, $b$ $\in$ \{0.0, 0.3, 0.5, 0.7, 0.9\} ($b = \sin{\theta}$, where $\theta$ is the impact angle); (ii) velocity, $v_{\rm c}$ $\in$ \{1.0, 1.5, 2.0, 3.0\}$v_{\rm esc}$; (iii) impactor-to-total system mass ratio, $\gamma$ $\in$ \{0.1, 0.2, 0.3, 0.4, 0.5\}; and (iv) total target mass, $M_{\rm t}^{\rm tot}$ $\in$ \{0.37, 1.05, 2.1, 5.25\}~M$_{\oplus}$. The impact parameter and velocity at the time of impact were defined as the point at which the mantles of the colliding bodies touch, ignoring the initial entry of the impactor into the target's atmosphere and neglecting any tidal deformation (illustrated in Figure \ref{fig:initial_conditions}). The impact velocity is set as a multiple of the mutual escape velocity between the colliding bodies, which is defined in this case as

\begin{equation}
    v_{\rm esc} = \sqrt{2G\left (\frac{M_{\rm t}^{\rm tot} + M_{\rm i}^{\rm r}}{R_{\rm t}^{\rm r} + R_{\rm i}^{\rm r}}\right )} \; ,
\end{equation}

\noindent where $G$ is the gravitational constant and $R_{\rm t}^{\rm r}$ and $R_{\rm i}^{\rm r}$ are the refractory target and refractory impactor radii, respectively (Figure \ref{fig:initial_conditions}). As determining the extent of pre-impact tidal deformation a priori is non-trivial, the particle placements, velocities, etc., required to provide the desired impact properties and timings were set based on the collision of spherical bodies. The initial position of the impactor was set to either allow or restrict tidal deformation. In most simulations, the initial position was set such that first contact occurs 1~hr after the start of the simulation to allow for tidal deformation of the interacting bodies before collision. To examine any effects of tidal deformation, a subset of impacts for every combination of $b$, $v_{\rm c}$, and $\gamma$ onto the 1.05~M$_{\oplus}$ target were run such that first contact occurred 5 minutes after the start of the simulation. 

\begin{figure}
\begin{center}
    \includegraphics[width=\columnwidth]{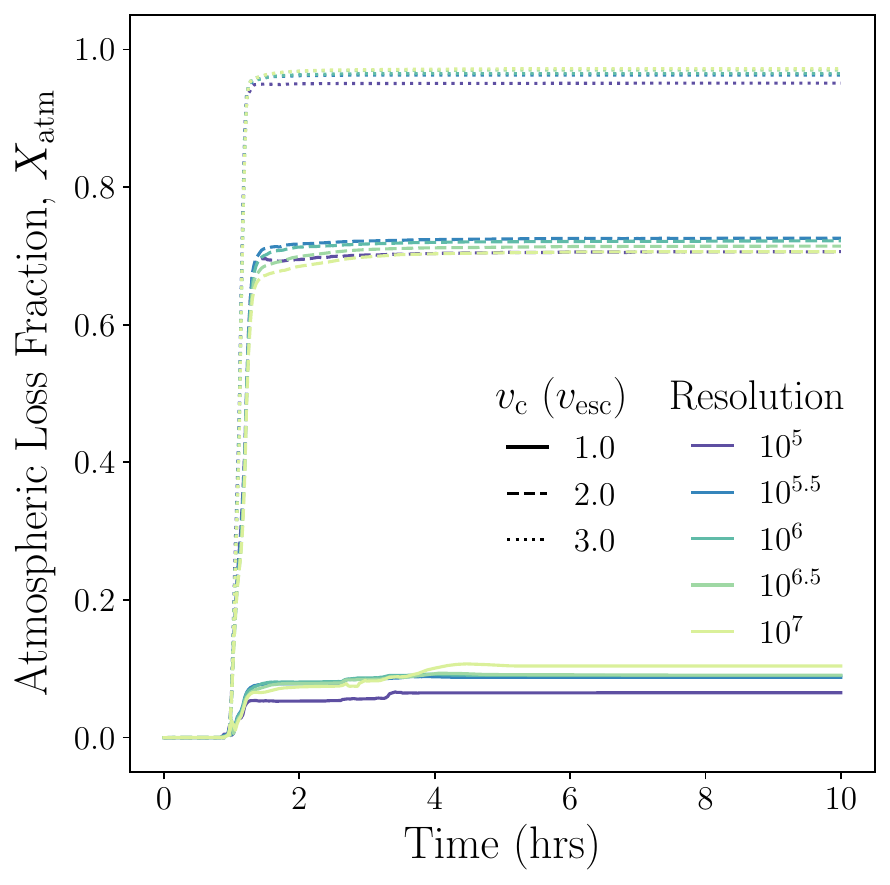}
    \caption{Final atmospheric loss converges for simulations at a resolution for the target of $\geq$10$^{5.5}$. Atmospheric loss fraction is plotted as a function of time for an example impact onto the 1.05 M$_{\oplus}$ target planet ($b$ = 0.3, $\gamma$ = 0.3) performed at five different particle resolutions (colours) and three velocities (line styles).}
    \label{fig:resolution}
\end{center}
\end{figure}

We ran a total of 479 impact simulations with $\sim$10$^6$ SPH particles per total target mass. This resolution ensures that there are at least 7 particle shells across the atmosphere for the planets considered here. Initial test simulations showed that loss converged for model atmospheres that contained at least 5 particle shells. Further test atmospheric loss simulations, which considered impacts for a single $M_{\rm t}^{\rm tot}$ and $\gamma$ at $b \in$ \{0.0, 0.3, 0.5, 0.7\} and $v_{\rm c}$ $\in$ \{1.0, 1.5, 2.0, 3.0\}$v_{\rm esc}$, were run with $\sim$10$^{5}$, 10$^{5.5}$, 10$^6$, 10$^{6.5}$, and 10$^7$ SPH particles per total target mass. The final atmospheric loss fraction converges at and beyond a resolution of 10$^{5.5}$ (Figure~\ref{fig:resolution}). We note that \citet{kegerreis2020atmosphericA} found that loss varied in their simulations by a few percent depending on resolution for a given impact scenario up to a resolution of $10^8$. It is not absolutely clear the reason for the discrepancy in convergence between our simulations, but it may result from the fact that they used a smaller target atmosphere mass fraction and less-advanced EoSs for the mantle and core which are known to affect convergence \citep{meier2021eos,10.1063/12.0000946}. 

We use a mass resolution of 10$^6$ particles per total target mass which translates into the number of particles in the impactor being scaled according to 

\begin{equation}
    N_{\rm i} = \frac{M_{\rm i}^{\rm r}}{M_{\rm t}^{\rm tot}}N_{\rm t} \; ,
\end{equation}

\noindent where $N_{\rm i}$ and $N_{\rm t}$ are the number of particles in the impactor and target planets, respectively. The total number of particles in each simulation therefore varied between $\sim$1.1 and 2$\times 10^6$ particles. 

All simulations were run with a cubic spline kernel plus the \citet{balsara1995neumann} switch for artificial viscosity. \texttt{SWIFT} uses a parameter $h_{\rm max}$ which sets the maximum allowed smoothing length of SPH particles, which in turn sets a density floor, $\rho_{\rm min}$ (see \texttt{SWIFT} documentation\footnote{\url{https://swift.strw.leidenuniv.nl/docs/index.html}}). This density limit prevents particles from interacting with one another through pressure forces at large separations (gravitational forces are still considered), optimizing code performance. It is, however, known that $\rho_{\rm min}$ can greatly influence the properties of an under-resolved post-collision disk \citep{hull2024effect}. All simulations were run with a $h_{\rm max}$ of 0.1~R$_{\oplus}$, resulting in a density floor of 3.1–48.6~kg~m$^{-3}$ for the planet masses considered. Since we only consider bulk ejection of material in this study, our results are not sensitive to $\rho_{\rm min}$, shown by the convergence of loss at resolutions of $\geq$10$^{5.5}$ (Figure~\ref{fig:resolution}). 
Simulations were run with a cubic boxsize of 1000~R$_{\oplus}$ and for a total of 10~hrs since test simulations showed that most atmospheric erosion was completed by $\sim$6~hrs into the simulation for all styles of impact (Figure~\ref{fig:resolution}). Any particles which leave the box are deleted from the simulation and recorded as lost. Full snapshots, which record all the dynamic and thermodynamic information about the particles, were recorded at 100~s intervals throughout the entire simulation, with additional `snipshots' containing only information about the particle IDs and pressures output every 1~s between 0.5–3~hrs to allow resolution of the impact shock (see Section~\ref{subsubsec:ground_surf}). See Table~\ref{table:initial_conditions} for a full breakdown of the parameters and results from each impact.

\subsection{Atmospheric Loss} \label{subsec:atmos_loss}
At each time step, atmospheric loss was determined by tracking the change in atmospheric mass bound to the largest remnant and normalising this change by the initial atmospheric mass (Figure~\ref{fig:resolution}). The atmosphere particles bound to the largest remnant were found following the method also detailed in \citet{marcus2009collisional} and \citet{carter2018collisional}, available from \citet{dou_2023_10260841}. The particle with the most negative gravitational potential was located and then the potential and kinetic energies of the other particles were computed to determine which other particles were gravitationally bound to this `seed' particle. The position and velocity of the centre of mass for these bound particles was computed and used as the new seed for subsequent iterations. This process was repeated for any remaining unbound particles until convergence (set by when the change in the fraction of the system mass that is bound at the current iteration is smaller than a given tolerance $\ll1/N_{\rm t}$, defaulting to 10$^{-10}$). For most impacts the loss fraction plateaued to a value within a few percent of its final value within 1--2~hrs after first contact, with some minor variation caused by re-impacting material or readjustment of the mass distribution in the largest remnant (e.g., at $\sim$4.5~hrs in Figure~\ref{fig:resolution}), and so we took the final loss as that at 10~hrs. This approach may somewhat misestimate the loss from long-period ($\gtrsim$9~hrs) graze-and-merge style impacts, but as the atmosphere is greatly disturbed by the initial impact, the effect on final loss of the second impact in graze-and-merge impacts is generally minor. We discuss this further in Section~\ref{subsec:mechanisms}.

To examine the dependence of loss efficiency on the initial position of material on the surface of the target (i.e., the spatial dependence), we divided the atmospheric particles into a grid in latitude–longitude space across the target's surface at the start of the simulation, with a grid cell size of 10$^{\circ} \times 10^{\circ}$. The change in the fraction of the atmosphere mass that started in each grid cell was then calculated independently.

\subsection{Definition of near and far field} \label{subsec:NF_def}

\begin{figure*}[t]
\begin{center}\includegraphics[width=\textwidth]{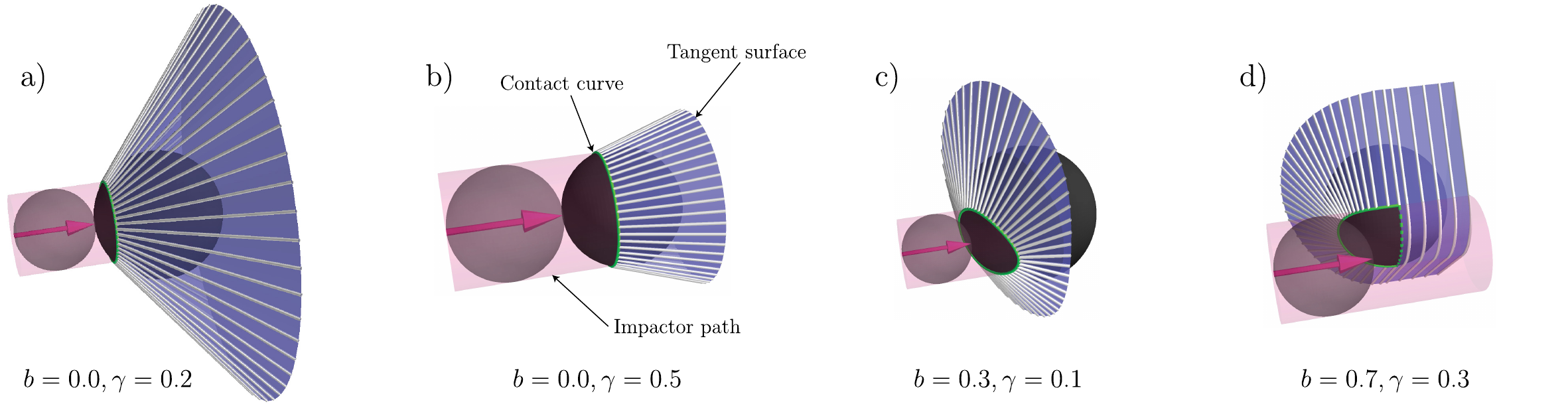}
    \caption{The near- and far-field regimes are separated using a tangent surface (blue) to the target–impactor contact curve (green curves) considering just the refractory parts of the colliding bodies and neglecting pre-impact deformation. The target and impactor planets are shown in dark and light grey, respectively, at the time of first contact between the two bodies. The pink cylinders and arrows illustrate the path of the impactor and the volume impactor material would sweep through if its motion was unaltered by the collision. The white lines illustrate a discrete set of tangents used in constructing the tangent surface itself (see Appendix~\ref{sec:NF_sep}). The geometric parameters of each impact are given below each panel, and the target is 1.05~M$_{\oplus}$ in each case.}
    \label{fig:NF_schematic}
\end{center}
\end{figure*}

\begin{figure}
\begin{center}
    \includegraphics[width=\columnwidth]{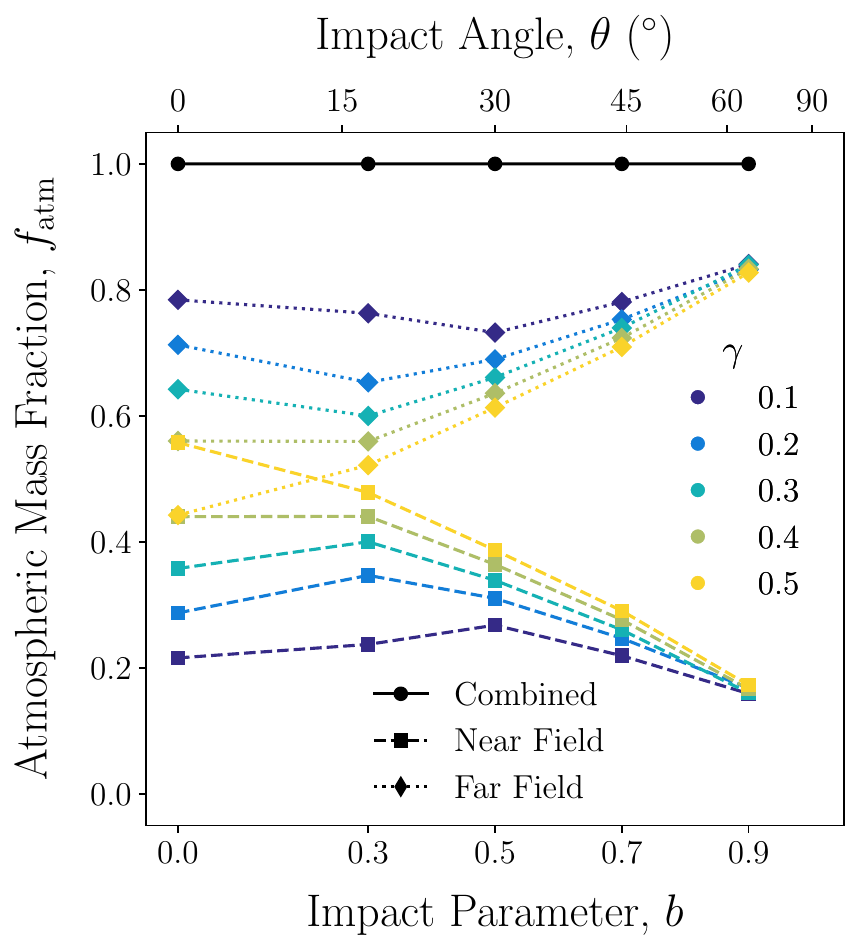}
    \caption{Mass fraction of atmosphere in the near and far field (lines/shapes) varies significantly between impacts. Shown are the masses of atmosphere in the two regions for all of our collisions on to a 1.05~M$_{\oplus}$ target as a function of $b$ and $\gamma$ (colours).}
    \label{fig:NF_mass}
\end{center}
\end{figure}

In this section, we define a routine for demarcating the near- and far-field atmosphere to approximately separate and explore the different mechanisms of loss. A full mathematical derivation of the near-/far-field separation can be found in Appendix~\ref{sec:NF_sep} but we provide a conceptual summary here. In giant impacts, the concept of a simple tangent plane at the point of contact between the colliding bodies, often used when describing smaller impacts \citep{schlichting2015atmospheric,shuvalov2009atmospheric}, breaks down as the contact area between the two bodies becomes a significant fraction of the size of the target. Here we extend the definition of the tangent plane to a tangent surface that is defined by the set of tangents to the contact curve of the target with the projected path of the impactor (Figure~\ref{fig:NF_schematic}). In the case of grazing impacts, the contact curve is completed by a great circle path between the contact extrema. Points initially above the tangent surface (in the direction from which the impactor approached) are then considered to be in the near field, and points initially below the surface are considered to be in the far field. The amount of atmosphere mass in the near and far field changes as a function of impact parameter and impactor-to-total system mass ratio (Figure~\ref{fig:NF_mass}), and we use this formulation to consider loss in the near and far field independently. It is important to note that our definition of near and far field is formulated assuming collisions between spherical planets and so ignores any tidal deformation or other redistribution of the atmosphere pre-impact. In addition, the refractory target radius used in defining the planet's near- and far-field atmosphere is slightly different than that used when initialising the impacts (see Section~\ref{subsec:impact_sim}). The refractory radius here is instead defined as the base of the atmosphere, taken as the smallest radius of any atmosphere particle, since using a radius defined by the mantle particles leads to an artificially extended contact site.

\subsection{1D–3D loss coupling}\label{subsec:1D_3D_method}

We compared the efficiency of loss in the far field from our simulations to previous 1D studies of loss from shock-kick \citep{lock2024atmospheric} by determining the atmospheric loss predicted by 1D calculations in response to the impact shock waves reaching the surface in our 3D simulations. By doing so, we gain insight into the loss mechanisms acting in the far field as well as the applicability of results and conclusions from lower-dimensional studies. 

\begin{figure}
\begin{center}
    \includegraphics[width=\columnwidth]{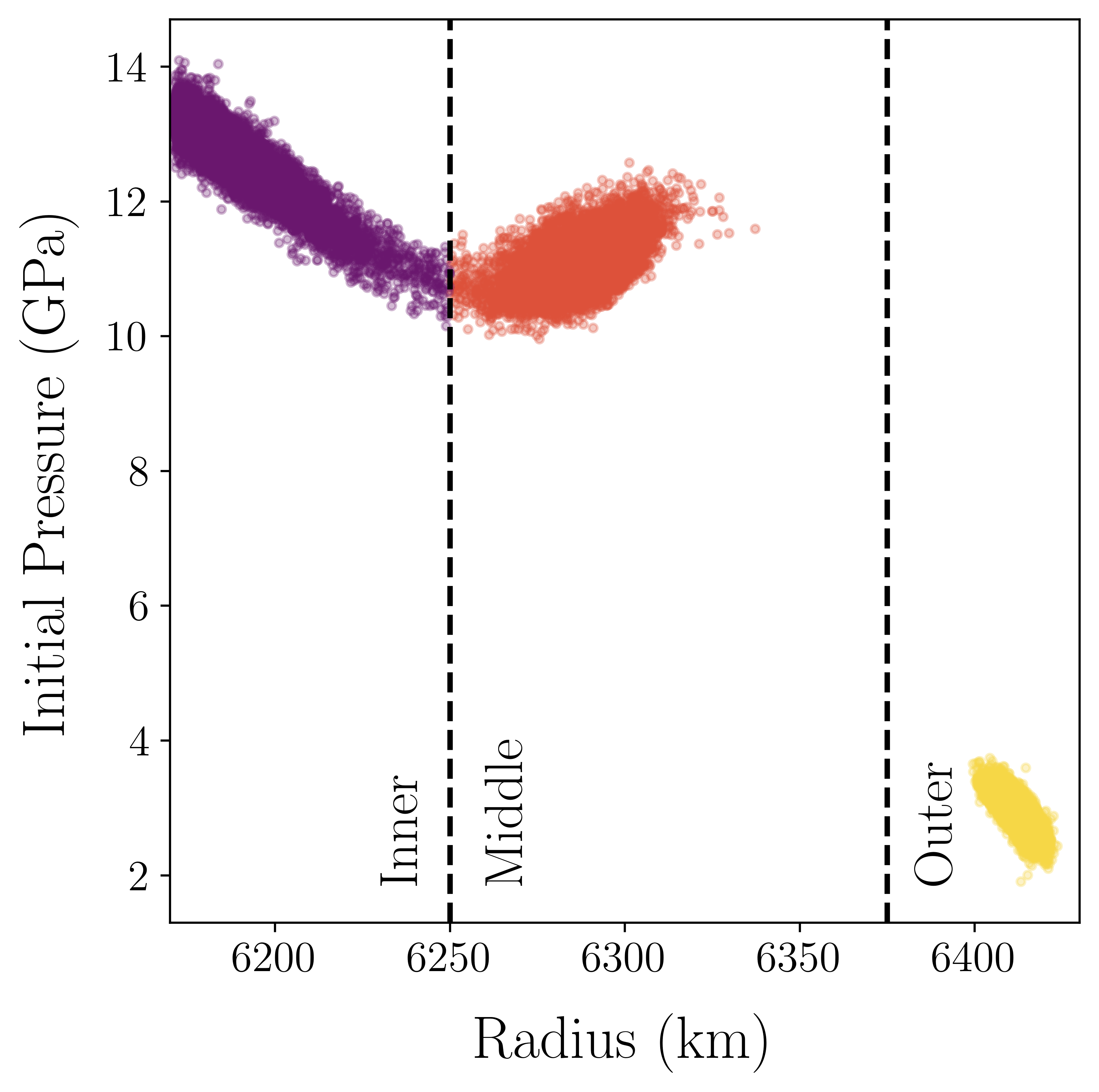}
    \caption{Smoothing across the mantle–atmosphere boundary and the influence of nearby atmosphere particles causes the outermost layers of initialised SPH planets to show anomalous pressures. Initial pressure of upper mantle particles vs. radius for the initialised 1.05~M$_{\oplus}$ planet, grouped and coloured by means of a $k$-means clustering algorithm with $N$ = 3, using the \texttt{KMeans} function from the \texttt{scikit.learn} (version 1.5.2) package.}
    \label{fig:init_p_r}
\end{center}
\end{figure}

Previous work \citep{genda2003survival, genda2005enhanced,lock2024atmospheric} modelled a situation that is considered to be analogous to loss via shock-kick in the far field. They determined the efficiency of atmospheric loss as a function of the ground velocity imparted by breakout of the impact shock wave at the surface of the planet, assuming radial shock propagation in 1D. It is possible to extract the ground velocity from 3D impact simulations and use these 1D loss scalings to make a prediction for loss \citep[as done by][]{kegerreis2020atmosphericA}.  However, accurately determining the particle velocity of the ground upon breakout of the shock wave from SPH simulations is extremely challenging. First, for reasons of numerical stability, SPH uses an artificial viscosity to smooth rapid temporal changes in material properties (such as the leading edge of shock waves) over some number of particles and hence some time. The peak ground velocity is only reached when the impact shock wave reaches the surface and the mantle releases. The effect on the ground velocity of the mantle shock, mantle release wave, and the continued evolution of the atmospheric pressure, are smeared together and it can be difficult to identify the velocity that corresponds to the initial velocity of the ground upon release for comparison to 1D simulations. The situation is further complicated due to numerical artifacts arising from the sharp density discontinuity between the mantle and atmosphere. In the standard density-smoothed formulation of SPH used by \texttt{SWIFT}, such discontinuities cannot be accurately modelled due to the inherent `smoothing' of the density field defined in the fundamental equation of SPH. This smoothing gives rise to a number of known problems in SPH, including suppressed mixing between materials (see \citet{ruiz2022dealing} and \citet{sandnes_remix_2025} for a more thorough review). Of primary importance in our case, is that the smoothing leads to an underestimate of the pressures of mantle particles and an overestimate of the pressures of atmosphere particles directly either side of the mantle–atmosphere boundary (Figure~\ref{fig:init_p_r}). The shock and release paths followed by those particles are not those that would be followed by material at the interface in a `real' planet.

\begin{figure}
\begin{center}
    \includegraphics[width=\columnwidth]{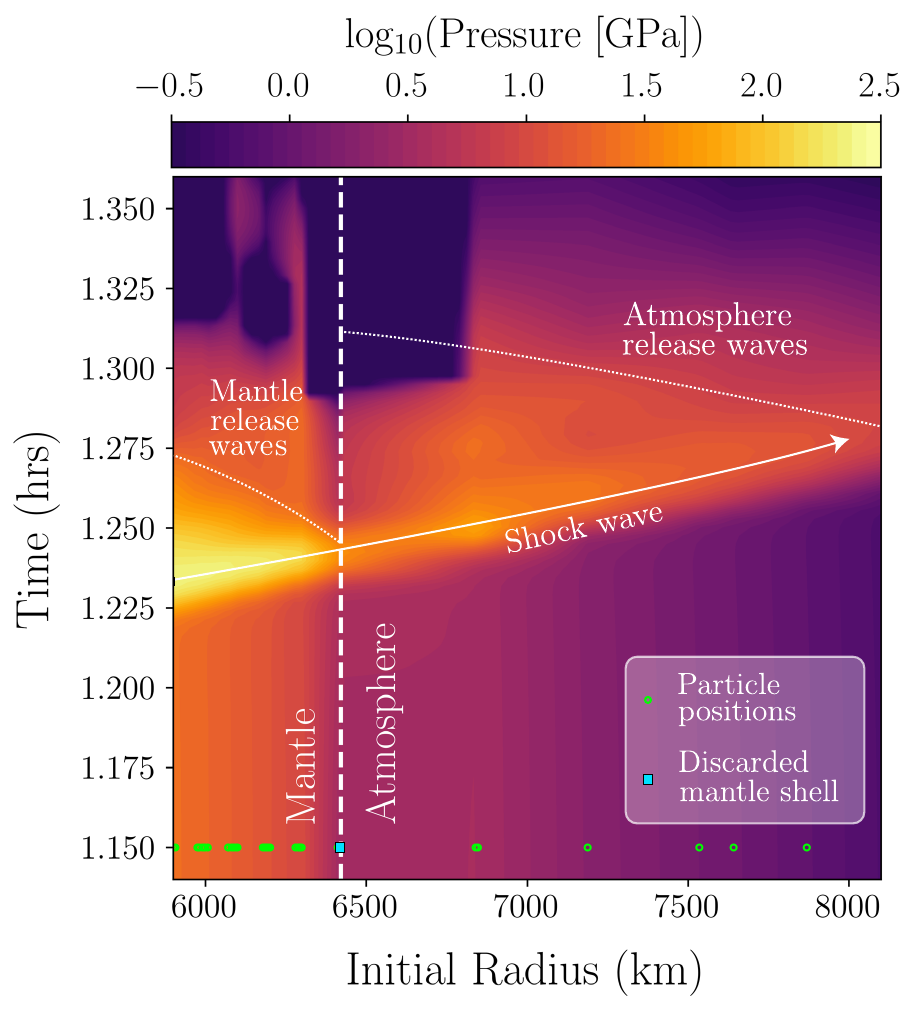}
    \caption{The propagation of the impact shock wave across the mantle–atmosphere boundary is shown in an $x$–$t$ diagram of upper mantle and atmosphere particle shells for the 1.05~M$_{\oplus}$ planet, coloured by pressure. The initial radius of the `outer' shell mantle particles illustrated in Figure \ref{fig:init_p_r} is shown as a filled cyan square on the white dashed line. Other initial particle positions are shown as green open circles. The white dashed line marks the mantle–atmosphere boundary whilst the white arrow illustrates the propagation of the shock wave front through the mantle and up into the atmosphere.}
    \label{fig:x-t}
\end{center}
\end{figure}

The issues with simulating the thermodynamic paths of material at the mantle-atmosphere boundary can be seen in the $x$–$t$ diagram in Figure~\ref{fig:x-t}. The passage of the shock wave across the interface can be clearly seen (white arrow), but the smoothed pressure profile (colours) and the pressure in the shock first falling and then rising across the outer mantle and lower atmosphere particles is unphysical and arises from the numerical issues described above. Despite the numerical complications at the boundary, standard SPH does an excellent job of correctly reproducing the transmission of the shock across the boundary at a longer spatial scale. 

Here we take an alternative approach that avoids many of the numerical issues associated with temporal and spatial smoothing in SPH to accurately determine the ground velocity exactly at the mantle--atmosphere interface, equivalent to the sudden impulse imposed in 1D simulations. We identify the strength of the shock in the mantle below the surface and before breakout of the shock wave at the surface, and then semi-analytically propagate the shock to the surface of the planet and perform an impedance match with the atmosphere. In this section we describe how we determine the strength of the shock wave near the planet's surface (Section~\ref{subsubsec:ground_surf}), and how the strength of the shock wave is then converted to a predicted loss based on impedance-match and 1D loss calculations (Section~\ref{subsec:impedance_match}). 

\subsubsection{Finding the impact shock pressure}
\label{subsubsec:ground_surf}
We quantify the strength of the impact shock wave using pressure as the key variable. The arrival and passage of the shock wave in our SPH simulations is much clearer in pressure space rather than, for example, particle velocity. It is also straightforward to correct peak shock pressure for the initial pressure of particles in the mantle and the effects of atmospheric properties. As a result, it is possible to use the shock pressures from one impact to predict loss for another impact with different atmospheric properties -- a process that we will explore in future work (see Section~\ref{subsec:1D_3D_coupling}). 

To avoid the near-surface numerical artifacts, we neglect a fraction of the outermost mantle SPH particles when determining the peak shock pressure. Figure~\ref{fig:init_p_r} shows the particles in the upper 10\% of the mantle of an initialised SPH planet split into three discreet groups (hereafter referred to as inner, middle, and outer particles), identified using a \textit{k}-means clustering algorithm. Each group is influenced to varying degrees by density smoothing across the mantle–atmosphere boundary. The yellow `outer' particles are most severely affected, recording unrealistically low pressures and a large separation in radius from the other mantle particles (Figure \ref{fig:init_p_r}). Furthermore, tracking the propagation of the shock wave across the mantle–atmosphere boundary reveals that the last mantle shell of particles predict anomalous and unphysically low pressures (Figure \ref{fig:x-t}). As a result, we discard this shell of particles (cyan square in Figure \ref{fig:x-t}) for the purposes of our 1D–3D comparison. The upper 3\% of the remaining mantle particles, most of which form the orange `middle' group of particles, are then used to define the strength of the impact shock near the planet's surface (we henceforth will refer to this group of particles as being `near-surface' particles).  Taking the near-surface cut-off as being anywhere between $\sim$1–5\% of the mantle after removing the outermost anomalous particles makes little difference to our results. However, for percentages $>$5\%, higher peak pressures are recorded since the shock itself is recorded earlier at deeper depths in the mantle and so shows less influence from release waves or geometric expansion -- effects that we wish to avoid.

\begin{figure}
\begin{center}
    \includegraphics[width=\columnwidth]{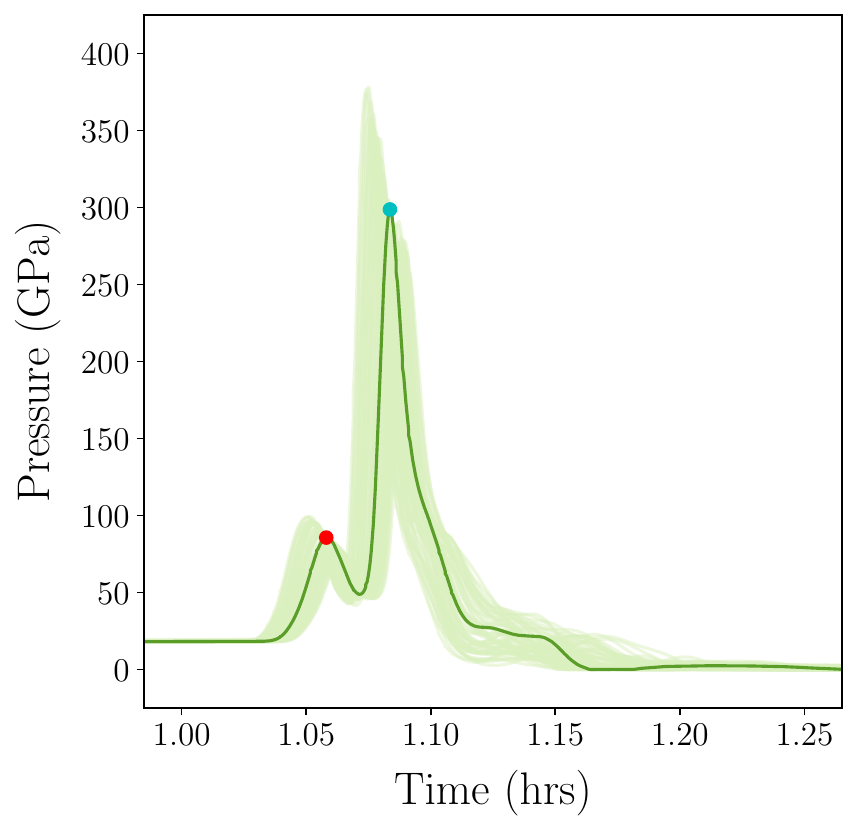}
    \caption{Pressure vs. time traces for all near-surface particles within one latitude–longitude grid cell (light green, with the trace for one example particle highlighted in dark green). Corrections are made in the 1D loss calculations to ensure that the peak shock pressure of the ground surface (blue circle) rather than the air shock in the atmosphere (red circle) is recorded as the true first peak pressure.}
    \label{fig:p_vs_time}
\end{center}
\end{figure}

To find the peak shock pressure of each near-surface particle  --- and not simply the highest pressure experienced during the impact --- we used the \texttt{find\_peaks} function from the \texttt{scipy.optimize} package (version 1.9.1). The function finds the first peak in the pressure–time curve for each particle, an example of which is shown in Figure~\ref{fig:p_vs_time}.  In grid cells closer to the impact site, such as that shown in Figure~\ref{fig:p_vs_time}, we observe a precursor shock wave that arises from the air shock driven by the initial entry of the impactor into the target's atmosphere (red point in Figure~\ref{fig:p_vs_time}) in addition to the shock wave driven by the collision of the mantles of the colliding bodies that has travelled through the mantle (blue point in Figure~\ref{fig:p_vs_time}). To isolate the shock propagating through the mantle, the time of the first pressure peak found is compared to the maximum pressure  within a time window bounded by the initial peak and an estimate for the crossing time of the impactor from the top of the atmosphere to the mantle surface, $\delta t$, given by

\begin{equation}
    \delta t = \frac{d}{v_{\rm c}},
\end{equation}

\noindent where $v_{\rm c}$ is the impact velocity in km~s$^{-1}$ and $d$ is the distance travelled by the impactor through the atmosphere prior to contact with the target's surface, itself given by

\begin{equation}
    d = \begin{cases}
        \mathcal{H} & \text{if } b = 0.0 \\
        -R_{\rm t}^{\rm r}\cos{\left(\theta\right)}
        \\ \;\; + \sqrt{{R_{\rm t}^{\rm r}}^2\cos{\left(\theta\right)}^2 + \mathcal{H}^2 + 2\mathcal{H}R_{\rm t}^{\rm r}} & \text{otherwise},
    \end{cases}
\end{equation}

\noindent where $\mathcal{H}$ = $R_{\rm t}^{\rm tot} - R_{\rm t}^{\rm r}$ (i.e., the total height of the atmosphere) and $\theta$ is the impact angle. If the maximum pressure within this time window is more than 5\% higher than the pressure of the first peak, a second peak is found using the same algorithm starting from the time of the first peak, else the first peak is taken as the peak pressure corresponding to the mantle shock.

To resolve the temporal pressure evolution and pressure peaks accurately requires a very high output cadence from the SPH simulations, with pressure data at least every 10~s. Here we have used the `snipshot' capability of \texttt{SWIFT} to provide pressure data for every particle at every second during the first 2~hrs after first contact to ensure that the peak shock pressure is captured whilst reducing the total volume of output.

To calculate the loss using the results of 1D studies requires knowing the local escape velocity. The escape velocity at each near-surface particle is given by

\begin{equation}
    v_{\rm esc}^i = \sqrt{2\left|\phi\right|} \; ,
\end{equation}

\noindent where $\phi$ is the gravitational potential of each SPH particle at the full output snapshot, and $i$ refers to a single particle. We linearly interpolate between snapshots to find the escape velocity for each particle at the time of each `snipshot'. Snapshots are output every 100~s, which is comparable to the width of the shock pressure peak, and there is only a modest change in the escape velocity between snapshots.

\subsubsection{Impedance match and coupling to 1D}
\label{subsec:impedance_match}
\begin{figure}
\begin{center}
    \includegraphics[width=\columnwidth]{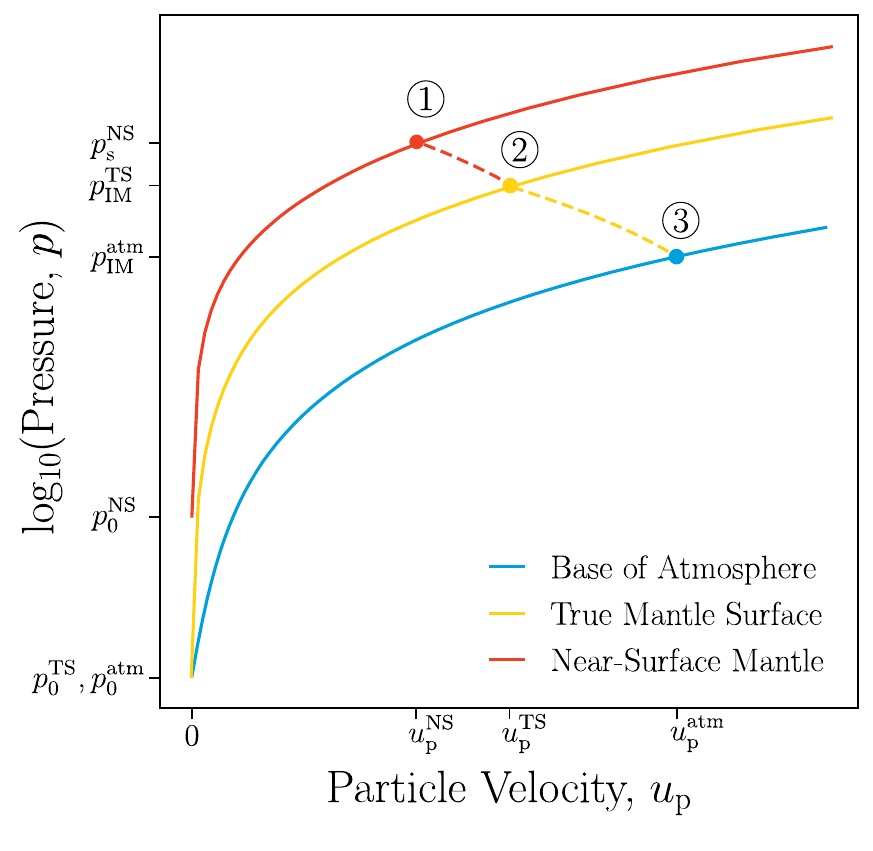}
    \caption{The 1D loss code converts near-surface mantle pressures to velocities by calculating a two-stage impedance match between the mantle and atmosphere. Upon impact, the near-surface mantle (NS) in the far field is shocked to a pressure, $p_{\rm s}^{\rm NS}$, on the mantle (forsterite) Hugoniot (orange, \circled{1}). The shocked NS then releases (orange dashed line) to a point on the lower pressure `true' mantle surface (TS) Hugoniot  at $p_{\rm IM}^{\rm TS}$ (yellow, \circled{2}), before releasing to a point (yellow dashed line) on the atmosphere Hugoniot at $p_{\rm IM}^{\rm atm}$ with a particle velocity $u_{\rm p}^{\rm atm}$ (blue, \circled{3}).}
    \label{fig:hugoniot}
\end{center}
\end{figure}

To translate the peak shock pressure found for the near-surface SPH particles to a ground velocity and then to a 1D predicted loss, we use impedance match calculations and the loss functions from \citet{lock2024atmospheric}. For a thorough discussion of the application of impedance match calculations to atmospheric loss by shock-kick, the reader is referred to \citet{lock2024atmospheric}. The procedure followed in the current study (illustrated schematically in Figure~\ref{fig:hugoniot}) is to take the shock recorded in the near-surface (NS) mantle (Section~\ref{subsubsec:ground_surf}) and semi-analytically propagate it, first to the `true' surface (TS) of the mantle and then into the atmosphere. For each near-surface SPH particle, we extract the peak shock pressure at a point ($p_{\rm s}^{\rm NS}$, \circled{1} in Figure~\ref{fig:hugoniot}) on the mantle (forsterite) Hugoniot --- the locus of shocked states --- with an initial shock pressure ($p_{\rm 0}^{\rm NS}$) taken as that for the same particle in the pre-impact initialised target body (Section~\ref{subsubsec:ground_surf}). The peak shock pressure before release at the `true' surface of the mantle ($p_{\rm IM}^{\rm TS}$) is then determined by an impedance match (IM) between the release of the near-surface mantle to a point on a mantle Hugoniot with an initial pressure of the base atmospheric pressure (\circled{2}, $p_{\rm 0}^{\rm TS}$ = $p_{\rm 0}^{\rm atm}$). This determination also gives the particle velocity in the ground before release of the shock ($u_{\rm p}^{\rm TS}$). It is important to note that requirement for this step comes from the fact that the `true' surface of the mantle is not directly captured in SPH. Given the small change in radius between the near-surface particles and the `true' surface of the body, the impedance match calculations were all performed in a planar geometry. Finally, we calculate the impedance match between this `true' surface-mantle Hugoniot and the base of the atmosphere to find the release pressure ($p_{\rm IM}^{\rm atm}$) and velocity ($u_{\rm p}^{\rm atm}$) of the mantle (\circled{3}) upon release of the shock into the atmosphere. This impedance match velocity is the equivalent of the ground velocity used in 1D shock-kick calculations. We can then use a relationship between ground velocity and atmospheric loss \citep[Equations C1–C3 in][]{lock2024atmospheric} to determine the loss that would be predicted from 1D calculations based on the impact shock simulated in 3D SPH simulations. It is important to note that here we assume that the total ground velocity, not just the radial component \citep[e.g.,][]{kegerreis2020atmosphericA}, contributes to driving loss and we will return to consider the validity of this assumption in Section~\ref{subsec:limitations}.

We followed this procedure to calculate the predicted loss for each near-surface SPH particle. We then divided the SPH particles into a latitude--longitude grid across the planet's surface as described in Section \ref{subsec:atmos_loss}, and found the median fractional loss for particles in each grid cell. To determine a total far-field loss we then multiplied this median loss fraction by the mass of atmosphere in the far field in each grid cell.

\section{Results}

We now describe our results, first for the combined (i.e., total) loss (Section~\ref{subsec:tot_loss}) and then for loss in the near and far field separately (Section~\ref{subsec:NF_loss}). We also present a scaling law that can predict loss as a function of impact parameters (Section~\ref{subsec:scaling}), and compare our full 3D results with coupled 1D–3D estimates of loss (Section~\ref{subsec:1D}). 

\subsection{Combined loss results} \label{subsec:tot_loss}

\begin{figure}
\begin{center}
    \hspace{-0.7cm}\includegraphics[scale=0.39]{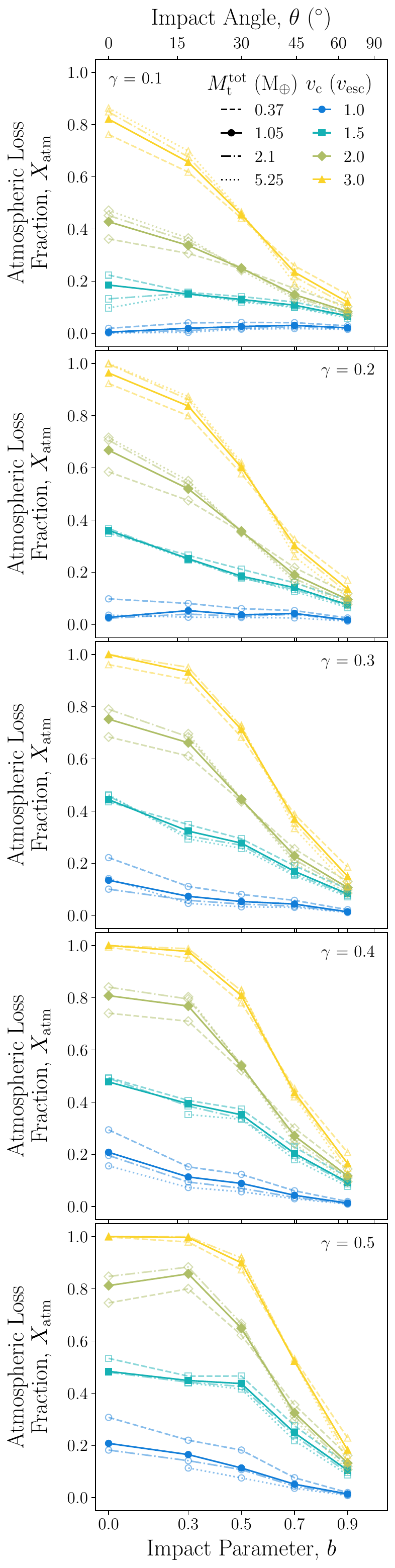}
    \caption{Atmospheric loss fraction decreases with increasing impact angle and decreasing impact velocity, and is largely independent of target mass. Combined atmospheric loss fraction as a function of impact parameter for impacts onto all target planets (line styles) with each panel showing one impactor-to-total system mass ratio. The results for impacts onto the 1.05 M$_{\oplus}$ target are highlighted with a solid line and closed symbols. Colours/shapes indicate the results for different impact velocities.}
    \label{fig:total_loss}
\end{center}
\end{figure}

\begin{figure*}[t]
\begin{center}
    \includegraphics[width=\textwidth]{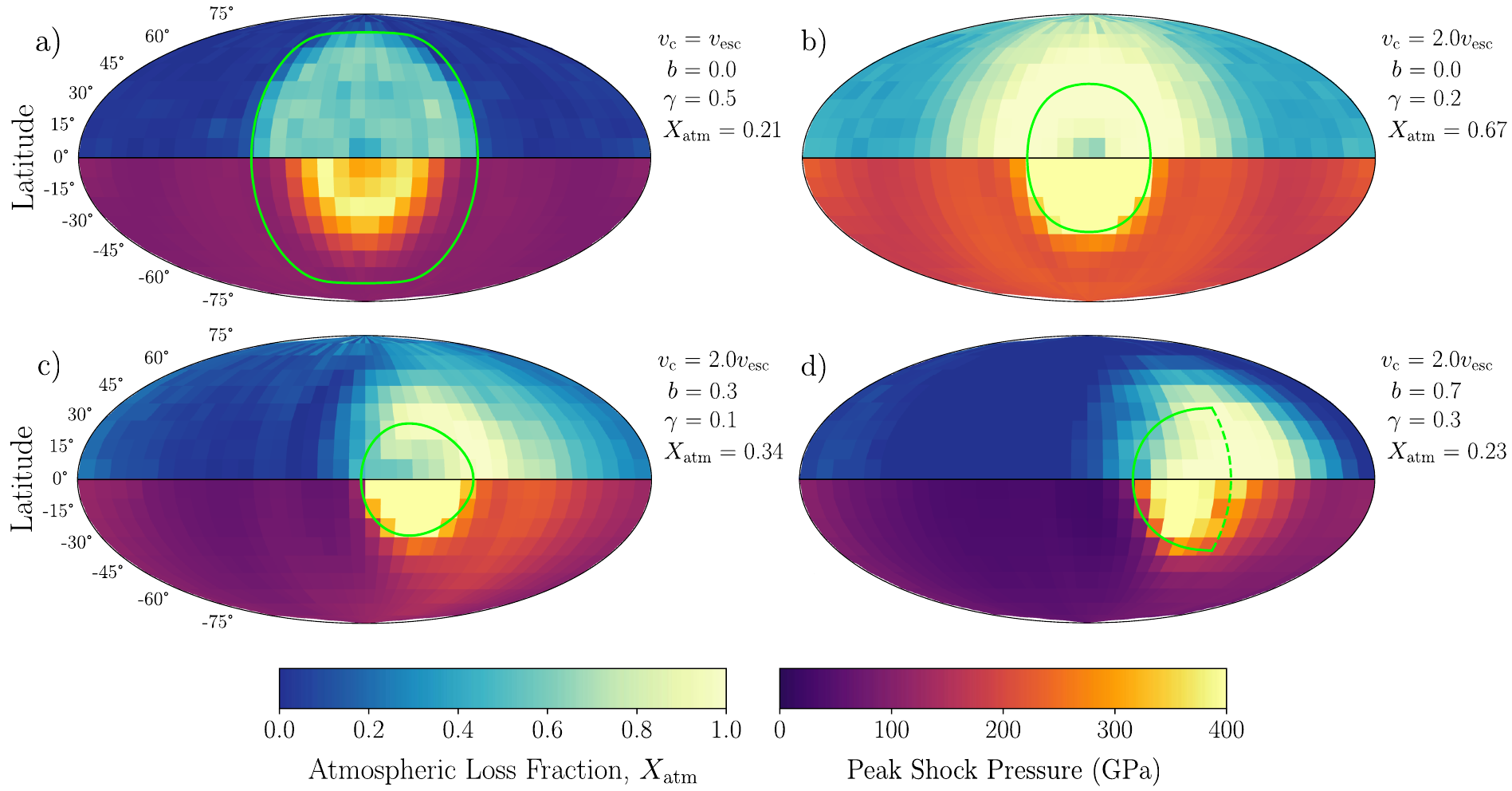}
    \caption{Atmospheric loss fraction correlates well with peak shock pressure. Mollweide projections centred on 0$^{\circ}$ longitude (see Figure \ref{fig:initial_conditions}) of combined atmospheric loss fraction (upper hemispheres) vs. peak shock pressure (lower hemispheres) across the 1.05 M$_{\oplus}$ planet's surface for four different combinations of $v_{
    \rm c}$, $b$, and $\gamma$ (given in each panel). The green curves are the intersection curves of the target with the path of the impactor (see Figure~\ref{fig:NF_schematic}). The dashed lines give the great-circle path of the edge of the target when the impactor is grazing (see Appendix \ref{sec:NF_sep}). Note: one might expect that in the case of panel a) where $\gamma$ = 0.5, the intersection curve of the target with the path of the impactor should stretch entirely to the poles since nominally $R_{\rm t}^{\rm r} \equiv R_{\rm i}^{\rm r}$ in this case. However, when determining the contact curve we define $R_{\rm t}^{\rm r}$ as the smallest radius of any atmosphere particle, since using a radius defined by the mantle particles leads to an artificially extended contact site, but $R_{\rm i}^{\rm r}$ is defined based on the mantle particles, as elsewhere in the paper, as all impactors have no atmosphere (see Appendix \ref{sec:NF_sep}). As a result, the radius of the impactor is slightly smaller than that of the target when defining the contact curve.}
    \label{fig:loss_vs_p_peak}
\end{center}
\end{figure*}

We have calculated atmospheric loss fractions during giant impacts for a range of values of total target mass ($M_{\rm t}^{\rm tot}$), impact parameter ($b$), impact velocity ($v_{\rm c}$), and impactor–to-total system mass ratio ($\gamma$). The results for impacts onto all target planet masses are shown in Figure~\ref{fig:total_loss}. Overall, the trends in loss we observe are in good agreement with those from previous work that also employed 3D SPH simulations \citep[e.g.,][]{kegerreis2020atmosphericA, kegerreis2020atmosphericB, denman2020atmosphere, denman2022atmosphere}: atmospheric loss fraction decreases with increasing $b$, but increases with increasing $v_{\rm c}$ (see also Figure~\ref{fig:loss_vs_p_peak}). In addition, we find that atmospheric loss fraction is largely independent of $M_{\rm t}^{\rm tot}$ and is more strongly dependent on $\gamma$. 

We find that in most collisions, the atmospheric loss fraction approaches its final value within $\sim$6~hrs of first contact (e.g., Figure~\ref{fig:resolution}). A small, and generally negligible, amount of later atmospheric loss is driven by subsequent interactions between impact debris and the largest remnant or oscillations in the distribution of bound mass \citep[see][]{carter2020energy}. As first noted by \citet{kegerreis2020atmosphericA}, these interactions can include repeated collision of material during head-on impacts, and secondary impacts during higher angle and/or velocity, graze-and-merge collisions. These secondary impacts cause minimal additional atmospheric loss since the secondary impactors are typically smaller than the main impactor, the impacts occur at a much lower velocity, and the collisions are onto a largest remnant with a significantly altered dynamic and thermal structure. The mantle is substantially extended and vaporised, meaning that energy from secondary collisions is deposited more gradually in the outer portions of the body, rather than dominantly at a clear mantle--atmosphere interface, and ejecta plumes that are formed, if any, are much less significant. There is no longer an atmosphere with a simple hydrostatic structure and atmosphere and mantle material is mixed together, reducing the ability of shocks to accelerate material to escape \citep{genda2003survival,lock2024atmospheric}. As a result, we take the loss at $\sim$9~hrs after first contact to be the final loss fraction and ignore any small additional loses from long-period debris or secondary collisions.

We now describe the results for head-on (Section~\ref{subsubsec:headon_tot}), intermediate (Section~\ref{subsubsec:accrete_tot}), and grazing (Section~\ref{subsubsec:grazing_tot}) impacts separately, and also describe the results of the subset of simulations run to examine the effects of tidal deformation (Section~\ref{subsubsec:tidal_def}). We explore the factors driving these trends in Section~\ref{subsec:NF_loss} when we break down the combined loss into near- and far-field components.

\subsubsection{Head-on impacts, $b$ = 0.0}\label{subsubsec:headon_tot}
Head-on impacts are an extreme end member that rarely occur in planet formation, but provide a useful limiting case to explore impact processes \citep[e.g.,][]{dou2024exploring,reinhardt2022forming}. Head-on impacts that take place at the mutual escape velocity of the colliding bodies ($v_{\rm esc}$) remove almost no atmosphere when $\gamma$ = 0.1 (i.e., the impactor mass is small), but remove an increasing amount of atmosphere with increasing $\gamma$. When $\gamma$ $\geq$ 0.3 up to 20\% of the atmosphere can be removed for $M_{\rm t}^{\rm tot}$ $\in$ \{1.05, 2.1\} M$_{\oplus}$ and up to 30\% for $M_{\rm t}^{\rm tot}$ = 0.37 M$_{\oplus}$. Figure~\ref{fig:loss_vs_p_peak}a (upper hemisphere) shows the spatial distribution of loss across the surface of the planet in an example low-velocity, head-on collision. The atmosphere is primarily lost from around the impact site and within the intersection path between the target and impactor (green outlines) for these lowest velocity impacts. Furthermore, the loss is well-correlated with the geometry of the peak pressure field (lower hemisphere), since the highest pressures are also experienced at the impact site.

Atmospheric loss increases with increasing $v_{\rm c}$, with near-complete loss reached in impacts at 3.0$v_{\rm esc}$ when $\gamma$ $\geq$ 0.3 for $M_{\rm t}^{\rm tot}$ $\in$ \{1.05, 2.1\} M$_{\oplus}$ and when $\gamma$  $\geq$ 0.4 for $M_{\rm t}^{\rm tot}$ = 0.37 M$_{\oplus}$. Atmospheric loss is much more globally widespread at higher impact velocities (Figure \ref{fig:loss_vs_p_peak}b, upper hemisphere) rather than being restricted to near the impact site, but loss is still well-correlated with the peak shock pressures experienced around the surface of the planet (lower hemisphere).

\subsubsection{Intermediate impact angles, $b$ = 0.3, 0.5}\label{subsubsec:accrete_tot}
For a given $M_{\rm t}^{\rm tot}$–$v_{\rm c}$–$\gamma$ combination, increasing the impact parameter (and thus impact angle) generally results in decreased atmospheric loss. When $v_{\rm c}$ = $v_{\rm esc}$ for $\gamma$ $\in$ \{0.1, 0.2\}, loss can however first increase with impact parameter by at most a few percent compared to the head-on case, then decreases by up to 10\% beyond $\gamma \approx 0.3$. Increasing $v_{\rm c}$ results in higher loss, but never exceeds the loss for a head-on impact with the same $M_{\rm t}^{\rm tot}$–$v_{\rm c}$–$\gamma$ combination. Near-complete loss for intermediate impact angles is only reached when $v_{\rm c}$ = 3.0$v_{\rm esc}$ and $\gamma$ $\geq$ 0.4. Loss again correlates well with the geometry of the pressure field and remains globally widespread for high-velocity impacts even for these higher-angle impacts (Figure \ref{fig:loss_vs_p_peak}c). The highest loss is experienced down-range of the contact site, and there can be a `shadow' of low loss towards the antipode of the impact site -- an effect well-illustrated in Figure \ref{fig:loss_vs_p_peak}c (upper hemisphere).

\subsubsection{Grazing impacts, $b$ = 0.7, 0.9}\label{subsubsec:grazing_tot}
For more grazing impacts, the fraction of atmosphere lost continues to decrease with increasing impact angle for a given $M_{\rm t}^{\rm tot}$–$v_{\rm c}$–$\gamma$ combination. When $\gamma$ = 0.1, impacts at $v_{\rm c}$ = $v_{\rm esc}$ remove almost no atmosphere for $b$ $\in$ \{0.7, 0.9\} and even for impacts with $v_{\rm c}$ = 3.0$v_{\rm esc}$ the maximal loss fractions are only $\sim$25\% and $\sim$15\% for $b$ = 0.7 and $b$ = 0.9, respectively. When $\gamma$ $\geq$ 0.3, impacts at $v_{\rm c}$ = $v_{\rm esc}$ remove between 0–10\% atmosphere, increasing to between $\sim$20–55\% when $v_{\rm c}$ = 3.0$v_{\rm esc}$. At these higher impact angles, much lower peak shock pressures are found across much of the planet and more atmosphere is lost down-range of the impact site than directly from the impact site, when compared to lower angle impacts -- an effect well-illustrated in Figure \ref{fig:loss_vs_p_peak}d.

\subsubsection{Effect of tidal deformation}
\label{subsubsec:tidal_def}

As described in Section~\ref{subsec:impact_sim}, a subset of impacts were run such that each simulation starts 5 minutes before the collision between the target and impactor planets occurs, instead of the default 1~hr, to explore the effects of tidal deformation (or a lack thereof, in this case) on loss. The final loss fractions and their trends closely mirror the data for the the impacts that did allow the bodies to tidally deform on approach, and in general show no significant deviation in final loss (typically less than $\sim$1\%). From this we conclude that tidal deformation has no substantial impact on atmospheric loss, at least for entirely fluid planets of this size range, and no further analysis was performed on this subset of low-deformation simulations.

\subsection{Near- and far-field loss results} \label{subsec:NF_loss}

\begin{figure}
\begin{center}
    \includegraphics[width=\columnwidth]{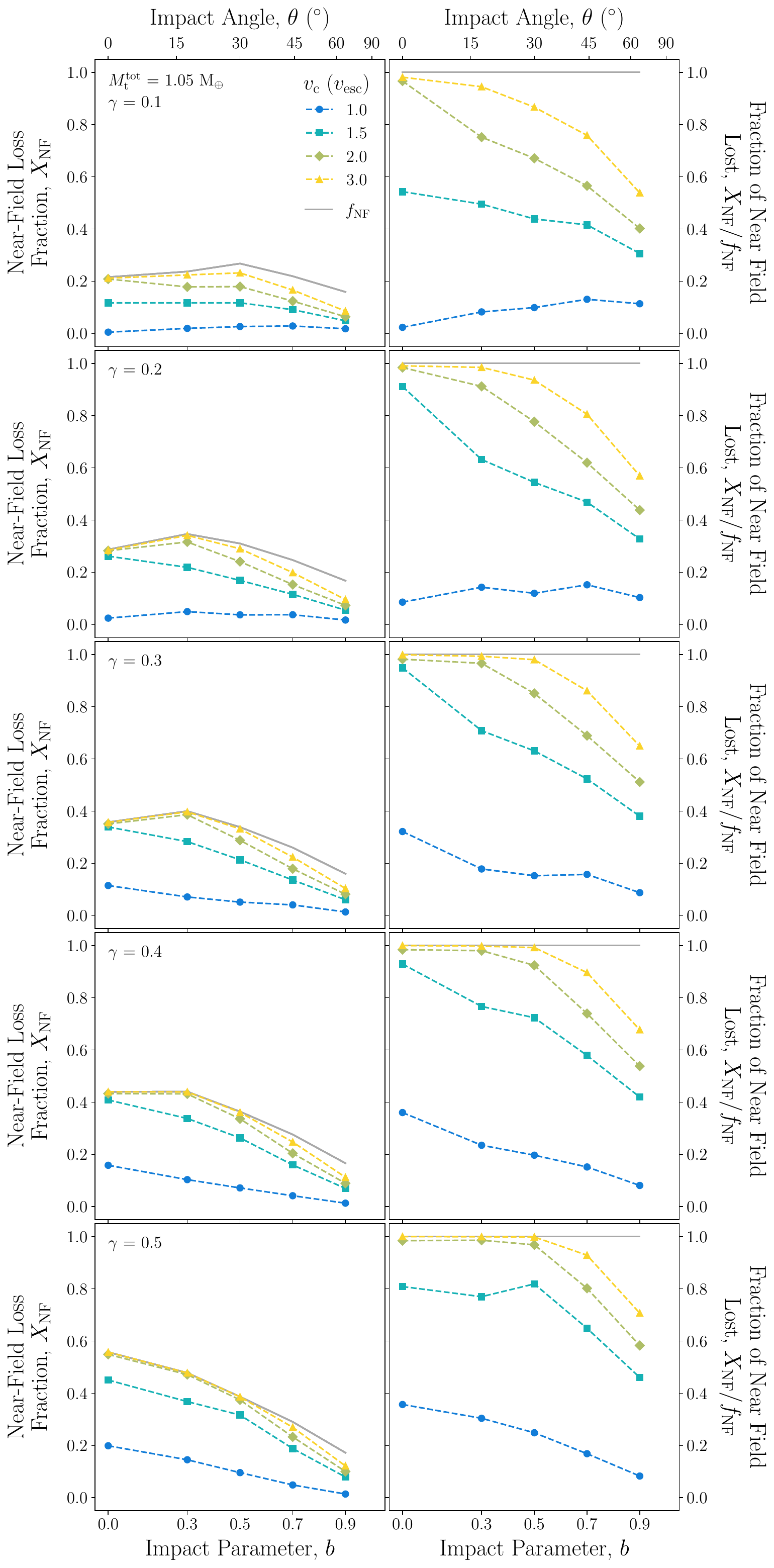}
    \caption{The near-field loss fraction is controlled primarily by the impact geometry and impact velocity. The near-field loss fraction (first column) and the near-field loss normalised to the near-field mass (second column) as a function of impact parameter for the 1.05 M$_{\oplus}$ target planet, all impactor-to-total system mass ratios (rows), and all impact velocities (colours/shapes). The grey solid line shows how the mass in the near field changes with $b$, and marks the maximum possible atmosphere than can be lost from the near field.}
    \label{fig:NF_loss}
\end{center}
\end{figure}

\begin{figure}
\begin{center}
    \hspace{-1.5cm}\includegraphics[width=\columnwidth]{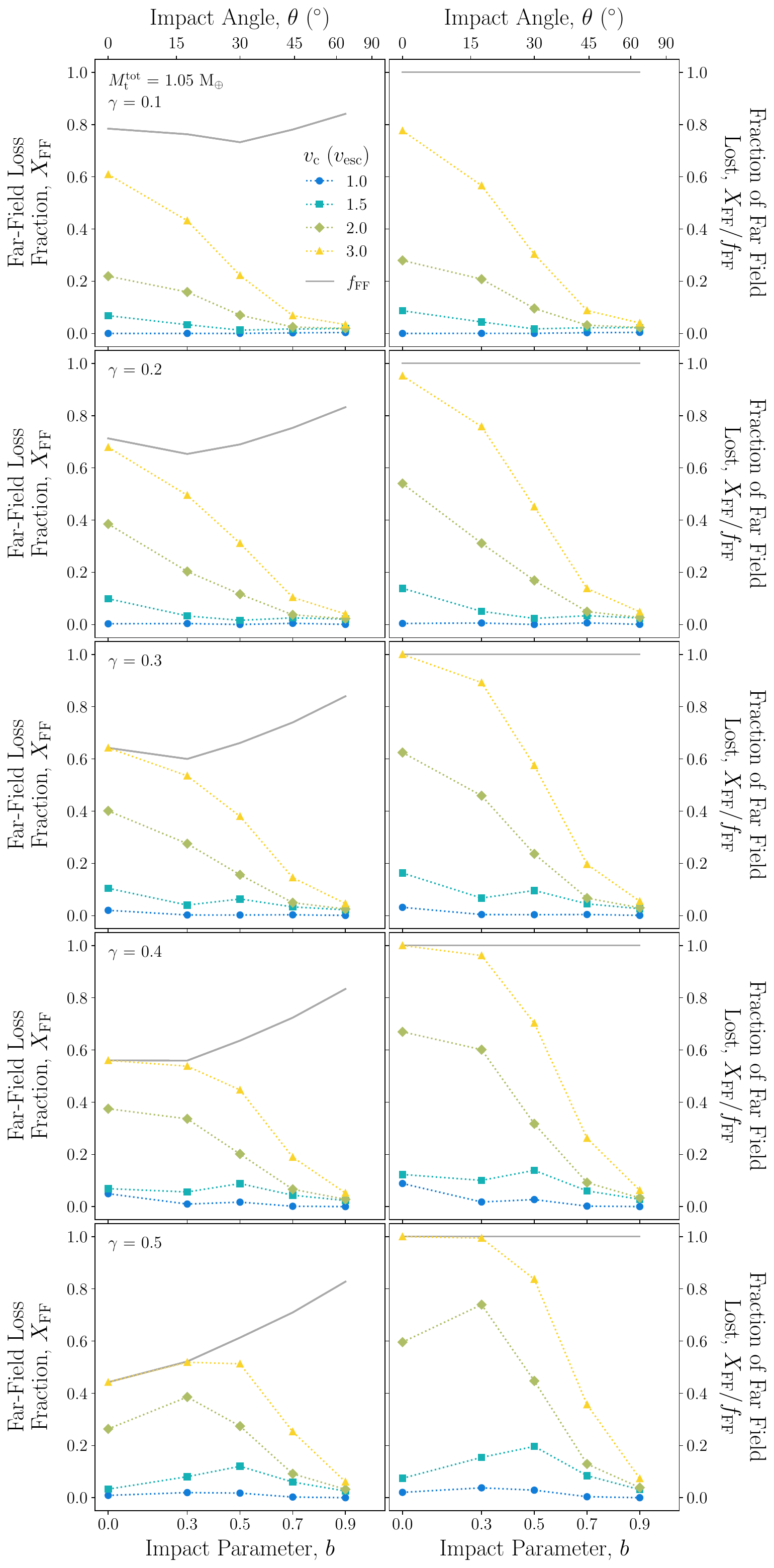}
    \caption{The far-field loss fraction decreases with increasing impact angle but increases with increasing impactor–to-total system mass ratio. The far-field loss fraction (first column) and the far-field loss normalised to the far-field mass (second column) as a function of impact parameter for the 1.05 M$_{\oplus}$ target planet, all impactor-to-total system mass ratios (rows), and all impact velocities (colours/shapes). The grey solid line shows how the mass in the far field changes with $b$, and marks the maximum possible atmosphere than can be lost from the far field.}
    \label{fig:FF_loss}
\end{center}
\end{figure}

\begin{figure*}[t]
\begin{center}
    \hspace{-0.7cm}\includegraphics[scale=0.465]{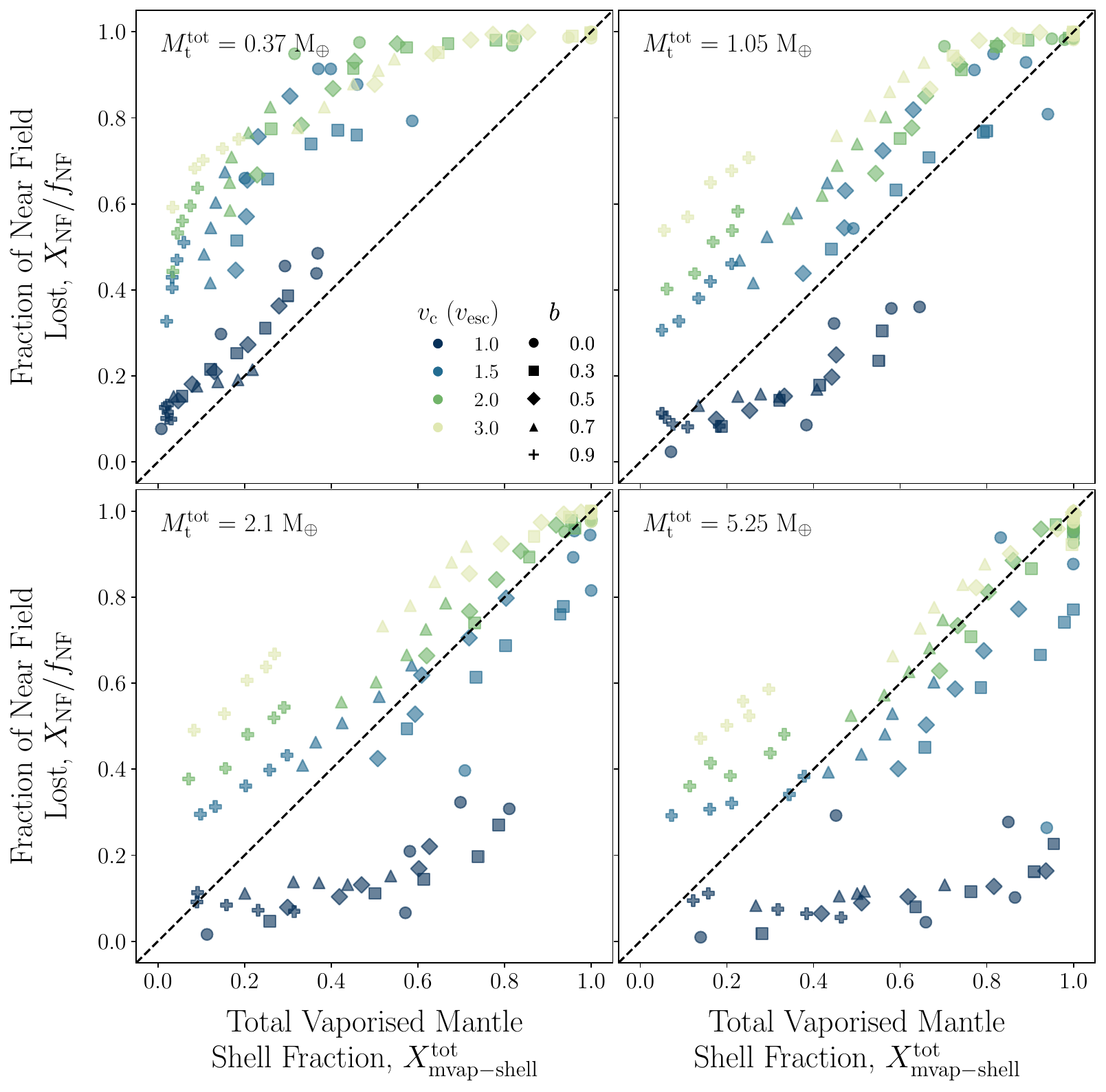}
    \caption{The near-field loss fraction increases with increasing vaporisation of the near-field surface mantle. Near-field loss fraction normalised to the near-field mass against the vapour fraction of bound and unbound mantle material from the 3\% surface shell (see Section~\ref{subsubsec:ground_surf}) for cells that underlay the near-field atmosphere, for each value of $v_{\rm c}$ (colours) and $b$ (shapes). Note that marker colours and shapes are used differently in this figure than in others in the paper to better highlight the $v_{\rm c}$ dependence. Each panel shows the results for a different target mass.}
    \label{fig:Xmvap_tot_vs_NF_frac}
\end{center}
\end{figure*}

\begin{figure*}[t]
\begin{center}
    \includegraphics[width=\textwidth]{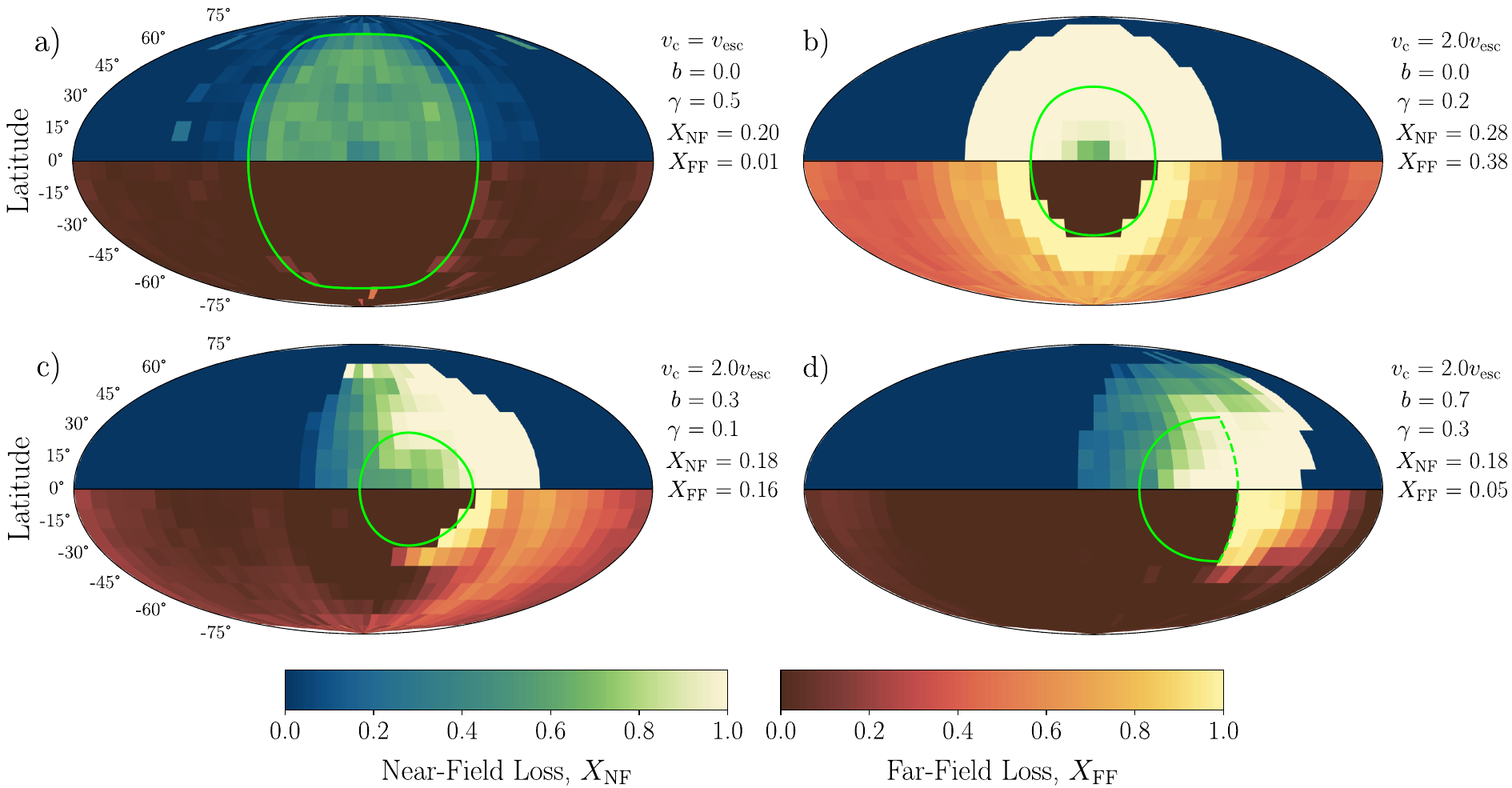}
    \caption{Using the tangent surface constructed around the target–impactor contact curve (see Section~\ref{subsec:NF_def}), the combined atmospheric loss can be decomposed into its near- and far-field components. Mollweide projections centred on 0$^{\circ}$ longitude (see Figure \ref{fig:initial_conditions}) of near-field loss fraction (upper hemispheres) and far-field loss fraction (lower hemispheres) across the 1.05 M$_{\oplus}$ planet's surface for four different combinations of $v_{
    \rm c}$, $b$, and $\gamma$ (given in each panel). The green ellipses are the intersection curves of the target with the path of the impactor (see Figure~\ref{fig:NF_schematic} and Appendix \ref{sec:NF_sep}). The tangent surface and normal to the planet's surface are generally not aligned, meaning each grid cell can contain some amount of both near- and far-field atmosphere.}
    \label{fig:NF_vs_FF}
\end{center}
\end{figure*}

\begin{figure*}[t]
\begin{center}
    \includegraphics[width=\textwidth]{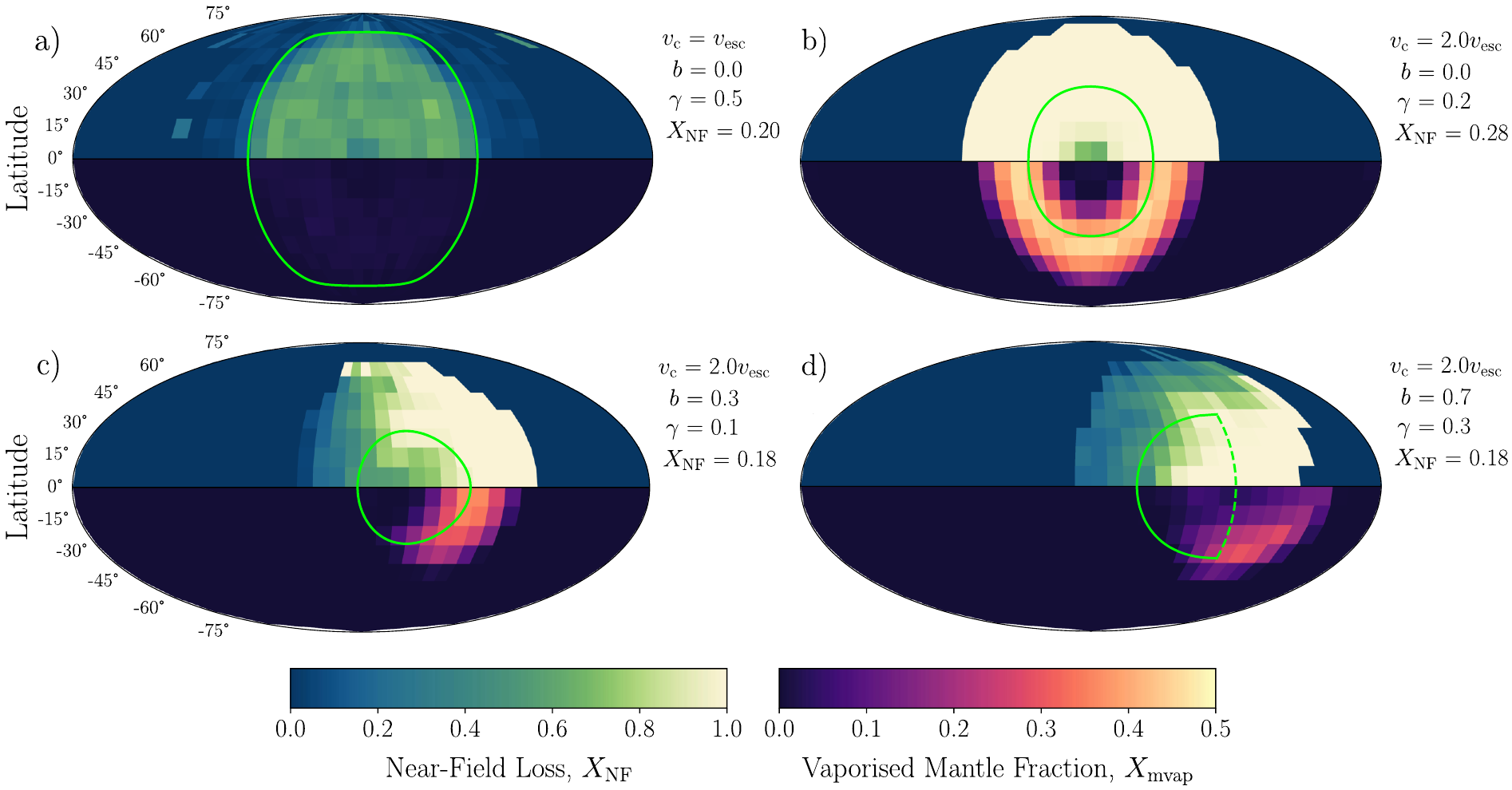}
    \caption{The geometry of near-field loss correlates well with the geometry of ejected and vaporised mantle material. Mollweide projections centred on 0$^{\circ}$ longitude (see Figure \ref{fig:initial_conditions}) of near-field loss fraction (upper hemispheres) and mantle fraction that is both vaporised and lost (lower hemispheres) across the 1.05 M$_{\oplus}$ planet's surface for four different combinations of $v_{
    \rm c}$, $b$, and $\gamma$ (given in each panel). The green ellipses are the intersection curves of the target with the path of the impactor (see Figure~\ref{fig:NF_schematic} and Appendix \ref{sec:NF_sep}).}
    \label{fig:NF_vs_Xmvap}
\end{center}
\end{figure*}

Using the method detailed in Section \ref{subsec:NF_def}, we now consider the loss in the near and far field separately to elucidate the mechanisms driving atmospheric loss. As discussed in Section \ref{sec:intro}, it has been theorised that loss is controlled primarily by three mechanisms: air shocks and ejecta plumes, which predominantly drive loss in the near field close to the impact site; and shock-kick, which predominantly drives loss elsewhere \citep{ahrens1990earth,vickery1990atmospheric}. By examining the loss from the near and far field separately, we can explore how the contribution to loss from these different mechanisms varies between impact scenarios. 

The near-field and far-field loss for a subset of $\gamma$ and $M_{\rm t}^{\rm tot}=1.05$~M$_{\oplus}$ are shown in Figures \ref{fig:NF_loss} and \ref{fig:FF_loss}, respectively. In general, although the overall sense of trends with respect to different impact parameters can be similar, the near-field and far-field loss show quite different functional forms. As for the combined loss (Figure~\ref{fig:total_loss}), both the near- and far-field loss fractions appear to be only weakly dependent on target mass and instead are much more strongly dependent on the impact velocity and impactor-to-total system mass ratio. For both the near and far field, loss generally decreases with increasing $b$ but increases with increasing $v_{\rm c}$. In addition, the near-field loss can saturate for almost all angles when $v_{\rm c} \geq 2.0v_{\rm esc}$ but the far-field loss only begins to saturate when $\gamma \geq 0.4$ for $b \leq 0.3$ and $v_{\rm c} = 3.0v_{\rm esc}$ 
(Figures \ref{fig:NF_loss} \& \ref{fig:FF_loss}). 

The near-field loss also strongly correlates with vaporisation. Figure~\ref{fig:Xmvap_tot_vs_NF_frac} shows that the degree of near-field loss increases with increasing vaporisation of the mantle that underlays the near-field atmosphere, which itself increases with $v_{\rm c}$. We note that there is a significant offset in the trend of loss with near-field vaporisation between the cases where $v_{\rm c} = v_{\rm esc}$ and at higher velocities. When $M_{\rm t}^{\rm tot}$ = 0.37 M$_{\oplus}$ however, the much lower absolute (i.e., not normalised to $v_{\rm esc}$) collision velocities in general lead to less vaporisation and a weaker trend between vaporisation and near-field loss (Figure~\ref{fig:Xmvap_tot_vs_NF_frac}).

We now describe the near- and far-field loss results for head-on (Section \ref{subsubsec:headon_NF}), intermediate angle (Section \ref{subsubsec:accrete_NF}), and grazing (Section \ref{subsubsec:grazing_NF}) impacts in more detail.

\subsubsection{Head-on impacts, $b$ = 0.0}\label{subsubsec:headon_NF}
In the head-on case, the near-field loss increases with increasing $\gamma$ and $v_{\rm c}$, and saturates at total loss of the near-field atmosphere when $v_{\rm c}$ $\geq$ 2.0$v_{\rm esc}$ (Figure \ref{fig:NF_loss}). Far-field loss also increases with increasing $\gamma$ but only up to a point (e.g., $\gamma \geq$ 0.4 for $v_{\rm c} \geq$ 2.0$v_{\rm esc}$) when loss starts to decrease with increasing $\gamma$. Loss at 3.0$v_{\rm esc}$ begins to saturate at the total far-field atmosphere beyond $\gamma \geq$ 0.3 (Figure \ref{fig:FF_loss}). Very minimal far-field loss occurs for impacts at $v_{\rm c}$ = $v_{\rm esc}$ for any value of $\gamma$. 

As was evident from the spatial distribution of combined loss (Figure \ref{fig:loss_vs_p_peak}a), Figure \ref{fig:NF_vs_FF}a shows that the majority of the loss when $v_{\rm c}$ = $v_{\rm esc}$ is confined to the near field with little to no loss recorded in the far field.  This effect is consistent with the fact that the highest peak shock pressures are also confined to within the impact site (Figure \ref{fig:loss_vs_p_peak}a) and those in the far-field are not strong enough to drive much loss by shock-kick. Comparatively lower loss is also seen in the near field close to the central point of contact as atmospheric gases can become inertially trapped between the impactor and target and shocked mantle cannot readily release to low enough pressures to vaporise, at least until later in the impact when the atmosphere is already significantly disturbed. Nearer the edge of the impact site, shocked atmosphere and mantle material can more easily release to low pressure early in the impact, and plumes of vaporised surface mantle drive loss of atmosphere. 

Figure \ref{fig:NF_vs_FF}b illustrates a higher-velocity case where there is near total loss in the near field --- as suggested by the grey curves in Figure \ref{fig:NF_loss} --- and also significant far-field loss. Since the tangent surface and normal to the planet's surface are generally not aligned, each grid cell can contain some amount of both near- and far-field atmosphere. The near-field loss is spatially well-correlated with significant vaporisation of mantle material close to the contact site (much of which is accelerated to escape, Figure~\ref{fig:NF_vs_Xmvap}b) that generates strong ejecta plumes capable of driving significant loss of atmosphere. Note that even in this high-velocity case, there is still an amount of atmosphere that is inertially trapped between the impactor and target. The far-field loss itself is well-correlated with the globally widespread peak shock pressures (Figure \ref{fig:loss_vs_p_peak}b). In particular, the slightly increased far-field loss at the antipode is mirrored by the higher antipodal peak shock pressure due to geometric focusing of the impact shock. 

Having separated the near- and far- field loss, it is relatively easy to understand the trends in loss with different impact parameters. Firstly, changes in impact parameters change the fraction of atmosphere in the near and far field (Figure~\ref{fig:NF_mass} and grey lines in Figures~\ref{fig:NF_loss} and \ref{fig:FF_loss}), changing the amount of atmosphere that is susceptible to loss by different processes. For example, the fraction of atmosphere in the near field increases with increasing $\gamma$ which explains the trend described above for head-on impacts of increasing absolute (not fractional) near-field loss with increasing $\gamma$. The impact parameters also affect the shock physics of the impact. A higher $v_{\rm c}$, and thus a higher specific impact energy, generates a stronger shock wave into the atmosphere and refractory planet. In the near field, a stronger shock leads to increased vaporisation, and so stronger ejecta plumes as well as a stronger air shocks (Figures~\ref{fig:NF_vs_Xmvap} \& \ref{fig:Xmvap_tot_vs_NF_frac}). In the far-field, a stronger shock wave accelerates the ground to higher velocity and so produces more loss by shock-kick -- a conclusion supported by the correlation of far-field loss with peak shock pressure (Figure~\ref{fig:loss_vs_p_peak}). At higher values of $\gamma$, the shock wave is also sustained for longer as the release wave from the surface has further to travel before it can catch up with the impact shock, explaining the trend of increasing fractional near-field atmospheric loss with increasing $\gamma$ for a given $v_{\rm c}$. We return to consider the relative roles of each of these mechanisms of loss in Section~\ref{subsec:mechanisms}. 

\subsubsection{Intermediate impact angles, $b$ = 0.3, 0.5}\label{subsubsec:accrete_NF}
For impacts with intermediate impact angles and $v_{\rm c}$ $\leq$ 1.5$v_{\rm esc}$, the near-field loss typically decreases from the head-on value as $b$ increases. However, when $v_{\rm c}$ $\geq$ 2.0$v_{\rm esc}$, the near-field loss first increases for $b$ = 0.3 and then decreases again for $b$ = 0.5 when $\gamma$ $\leq$ 0.4. The non-monotonic trends in near-field loss with increasing $b$ and $\gamma$ result from a combination of two effects. Firstly, the mass in the near field is changing as a function of $b$ and $\gamma$ (Figure \ref{fig:NF_mass} and grey lines in Figure~\ref{fig:NF_loss}). The mass of atmosphere in the near field typically increases to a maximum at $b$ = 0.3 and then begins to decrease. The outer edge of the impactor travels through more of the atmosphere for slightly off-centre impacts, and there is simply more near-field atmosphere available to be lost at $b=0.3$. This effect is most clearly seen when the near-field loss reaches near-saturation for $v_{\rm c}$ $\geq$ 2.0$v_{\rm esc}$ (Figure~\ref{fig:NF_loss}). Secondly, at higher impact parameters, the shock wave is less efficiently coupled into the mantle. Less mantle material is vaporised and less of the vaporised material is ejected, indicating that weaker ejecta plumes are generated compared to the head-on case (Figure~\ref{fig:NF_vs_Xmvap}c and Figure~\ref{fig:Xmvap_tot_vs_NF_frac}). The ejecta plumes, and the motion of any unvaporised impactor debris, are highly focused in the direction of travel of the impactor, leading to a shadow of loss in the near field up-range of the impact site where air shocks are likely the dominant loss mechanism (Figure~\ref{fig:NF_vs_FF}c, Section~\ref{subsec:mechanisms}).  

The trends in far-field loss are generally less complex, such that loss generally decreases with increasing $b$. The geometry of the far-field loss when $b$ = 0.3 (Figure \ref{fig:NF_vs_FF}c) is well-captured in the geometry of the peak shock pressure field (Figure \ref{fig:loss_vs_p_peak}c) with the gradual decay in shock peak pressure away from the impact site mirroring the gradual decay in degree of atmospheric loss. The simpler trends are due to the fact that the effect of the decreasing shock strength dominates over changes in the absolute far-field mass. This is partly the case because the fractional far-field loss does not saturate for non-head-on impacts except at very high velocities ($v_{\rm c}=3.0v_{\rm esc}$) and $\gamma\geq 0.4$ (right column of Figure~\ref{fig:FF_loss}). 

\subsubsection{Grazing impacts, $b$ = 0.7, 0.9}\label{subsubsec:grazing_NF}
Further increasing $b$ to 0.7 and 0.9 results in a decrease in loss from both the near and far field for a given $v_{\rm c}$–$\gamma$ combination (Figures~\ref{fig:NF_loss} and \ref{fig:FF_loss}). Whilst up to $\sim$90\% and $\sim$35\% of the near- and far-field atmosphere, respectively, can be lost in specific cases, this translates to only a maximum of $\sim$55\% combined loss. Even for $v_{\rm c}$ = 3.0$v_{\rm esc}$, the near-field loss does not completely saturate (Figure \ref{fig:NF_loss}) as it does for lower-angle impacts. The peak in combined loss for an impact at $b$ $\geq$ 0.7 is slightly geometrically offset down-range from the areas of highest peak pressure (Figure \ref{fig:loss_vs_p_peak}d). 

For grazing impacts, the total interacting mass participating in the impact is much smaller than for lower angle impacts, and so the specific impact energy, and the strength and duration of the shock, is less (Figure~\ref{fig:loss_vs_p_peak}d). The weaker shock drives weaker ejecta plumes with less surface vaporisation (Figure~\ref{fig:Xmvap_tot_vs_NF_frac}) and less of the vaporised material reaches escape (Figure~\ref{fig:NF_vs_Xmvap}d). The ejecta plumes are concentrated in the down-range direction and the non-interacting impactor mass (i.e., the side of the impactor whose trajectory would `miss' the refractory core of the target) can also accelerate atmosphere to escape. As a result, the near-field loss is concentrated down-range with, as for the intermediate cases, a shadow of less loss up-range of the impact site. The weaker impact shock also leads to little loss in the far field from shock-kick (Figure~\ref{fig:NF_vs_FF}d).

\subsection{Scaling law for loss} \label{subsec:scaling}
Scaling laws provide a means for efficiently predicting a given outcome of an impact based on different impact parameters. For example, the widely-used scaling law of \citet{leinhardt2012collisions} is used to predict the outcome of planetesimal impacts at any angle or mass ratio. Such laws are invaluable as they allow the effects of giant impacts to be incorporated into larger-scale models of accretion \citep[e.g.,][]{quintana2016frequency,carter2015compositional,carter2018collisional,esteves2022breaking,clement2019early,gu2024composition}. Here, we present a new scaling law for predicting combined atmospheric loss, $X_{\rm atm}$, from giant impacts using a linear least squares fit (implemented using the \texttt{curve\_fit} function from the \texttt{scipy.optimize} package, version 1.9.1). A \texttt{python} function to calculate the fitting parameters and subsequent atmospheric loss for any given impact scenario is included in the Supplementary Material. Our model is constructed as a sum of the near-field loss, $X_{\rm NF}$, and the far-field loss, $X_{\rm FF}$:

\begin{equation}
    X_{\rm atm} = X_{\rm NF} + X_{\rm FF} \; .
\end{equation}

\noindent This allows the different mechanisms of loss and their changing relative contributions for different impact parameters to be more accurately parametrised than in previous work \citep[e.g.,][]{kegerreis2020atmosphericB,denman2020atmosphere,denman2022atmosphere}. 

\begin{figure*}[t]
\begin{center}
    \includegraphics[width=\textwidth]{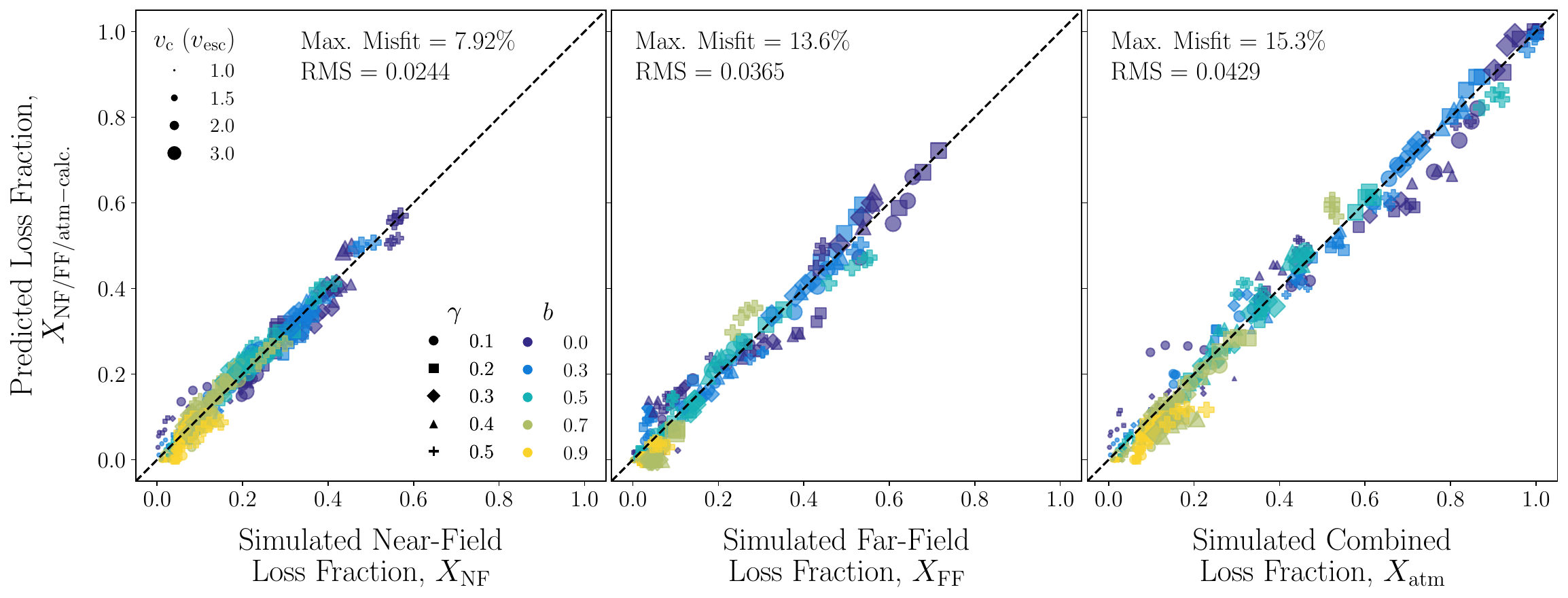}
    \caption{Our new scaling law can predict the combined atmospheric loss with an average misfit in absolute atmospheric loss of 3.2\%. Simulated (from our SPH simulations) near-field, far-field, and combined loss fraction data are plotted against the corresponding values predicted using our scaling law for each value of impact parameter (colours), impactor-to-total system mass ratio (shapes), and impact velocity (marker sizes) considered in this study, with the maximum misfits and root-mean squared (RMS) misfit reported in each panel.}
    \label{fig:scaling_law}
\end{center}
\end{figure*}

Given that the definition of the near-field atmosphere itself is controlled by impact geometry, we fit the near-field loss using an empirical function. We find that the data are well-modelled by a quadratic function of the impact parameter, b, given by

\begin{equation}\label{eq:NF}
    X_{\rm NF} = \xi_1 - \xi_2\left(b + \xi_3\right)^2 \;,
\end{equation}

\noindent where $\xi_{i}$ are functions of $\gamma$, $v_{\rm c}$, and $M_{\rm t}^{\rm tot}$ with each $\xi_{i}$ a linear combination of a 2nd-order polynomial in $\gamma$, a power law function of $v_{\rm c}$ normalised to $v_{\rm esc}$, and a linear function of $M_{\rm t}^{\rm tot}$:

\begin{equation}\label{eq:NF_params}
    \begin{aligned}
        \xi_{i \in \{1, 2, 3\}} &= k_{i1} + k_{i2}\,\gamma + k_{i3}\,\gamma^2 + k_{i4}\left(\frac{v_{\rm c}}{v_{\rm esc}}\right)^{k_{i5}} \\
        &\quad + k_{i6}\left(\frac{M_{\rm t}^{\rm tot}}{M_{\oplus}}\right) \;,
    \end{aligned}
\end{equation}

\noindent where $k_{ij}$ are fitted parameters. $i$ refers to the relevant $\xi_{i}$ whose dependence is being described and $j$ to the $j^{\rm th}$ fitting parameter (Table \ref{table:NF_params}). Note, as near-field loss is only weakly dependent on target mass we set $k_{26}=k_{36}=0$.

The far-field loss is dependent on the strength of the impact shock, which itself is related to the energy of the impact, and we find that the far-field loss is well-modelled by an exponential function of the centre of mass-modified specific impact energy, $Q_{\rm R}^{\prime}$ (in MJ~kg$^{-1}$), given by

\begin{equation}\label{eq:FF}
    X_{\rm FF} = \psi_1 \exp\left[-\psi_2\,Q_{\rm R}^{\prime}\left(1 + \frac{M_{\rm i}^{\rm r}}{M_{\rm t}^{\rm tot}}\right)\left(1 - b\right)^{\psi_4}\right] + \psi_3 \;,
\end{equation}

\noindent where $\psi_{i}$ are power-law functions of $\gamma$ and $M_{\rm t}^{\rm tot}$:

\begin{equation}\label{eq:FF_params}
    \begin{aligned}
        \psi_{i \in \{1, 2, 3, 4\}} = s_{i1} + s_{i2}\,\gamma^{s_{i3}} + s_{i4}\left(\frac{M_{\rm t}^{\rm tot}}{M_{\oplus}}\right)^{s_{i5}} \;,
    \end{aligned}
\end{equation}

\noindent and $s_{ij}$ are additional fitting parameters. $i$ refers to the $\psi_{i}$ whose dependencies are being described and $j$ to the $j^{\rm th}$ fitting parameter (Table \ref{table:FF_params}). $Q_{\rm R}^{\prime}$, first defined in \citet{leinhardt2012collisions}, is given by

\begin{equation}
    Q_{\rm R}^{\prime} = \frac{\mu_{\alpha}}{\mu}Q_{\rm R} \;.
\end{equation}

\noindent $\mu$, $\mu_{\alpha}$, and $Q_{\rm R}$ are the reduced mass, the modified reduced mass, and the unmodified centre of mass specific impact energy, which allow us to consider only the interacting mass of the impactor planet. These are given by

\begin{equation}
    \mu = \frac{M_{\rm i}^{\rm r}M_{\rm t}^{\rm tot}}{M_{\rm i}^{\rm r} + M_{\rm t}^{\rm tot}} \;,
\end{equation}

\begin{equation}
    \mu_{\alpha} = \frac{\alpha M_{\rm i}^{\rm r} M_{\rm t}^{\rm tot}}{\alpha M_{\rm i}^{\rm r} + M_{\rm t}^{\rm tot}} \;,
\end{equation}

\noindent and

\begin{equation}
    Q_{\rm R} = \frac{\mu v_{\rm c}^2}{2\left(M_{\rm i}^{\rm r} + M_{\rm t}^{\rm tot}\right)} \;,
\end{equation}

\noindent where $v_{\rm c}$ is the impact velocity. $\alpha$ is the mass fraction of the impactor involved in the collision:

\begin{equation}
    \alpha = \frac{m_{\rm interact}}{M_{\rm i}^{\rm r}} = \frac{3R_{\rm i}^{\rm r}l^2-l^3}{4R_{\rm i}^{\rm r ^ 3}} \; \;,
\end{equation}

\noindent where $m_{\rm interact}$ is the interacting mass of the impactor and $l$ is the projected length of the projectile which overlaps the target:

\begin{equation}
    l = \begin{cases}
        R_{\rm t}^{\rm r} + R_{\rm i}^{\rm r} - B &\text{when } B + R_{\rm i}^{\rm r} \;\textgreater\; R_{\rm t}^{\rm r} \\
        2R_{\rm i}^{\rm r} &\text{when } B + R_{\rm i}^{\rm r} \leq R_{\rm t}^{\rm r} \; \;,
    \end{cases}
    \label{eqn:l}
\end{equation}

\noindent where $B = \left(R_{\rm t}^{\rm r} + R_{\rm i}^{\rm r}\right)b$. Note that the modified energy used in our scaling law (Eqn.~\ref{eq:FF}) is reminiscent of the modified energy, $Q_{\rm S}$, used by \citet{quintana2016frequency} and \citet{lock2017structure}, given by

\begin{equation}\label{eq:QS}
    Q_{\rm S} = Q_{\rm R}^{\prime}\left(1 + \frac{M_{\rm i}^{\rm r}}{M_{\rm t}^{\rm tot}}\right)\left(1-b\right) \;.
\end{equation}

The loss fractions from our SPH simulations are plotted against the corresponding values calculated using our scaling law in Figure \ref{fig:scaling_law}. Our new scaling law does an excellent job at modelling both near- and far-field losses, and can predict the combined loss from a giant impact to within an absolute average of 3.2\% with a maximum misfit of 15.3\% -- a significant improvement in accuracy over existing scaling laws. It is important to note, however, that like all existing scaling laws, our scaling law is only applicable to the exact atmospheric parameters from our SPH simulations (see Section~\ref{subsec:limitations}). The scaling law does not yet include terms that model atmospheric properties (i.e., mass, composition, temperature), presence of an ocean, planetary rotation, or allow for an impactor to have its own atmosphere and/or ocean, for example. Future work will seek to extend our scaling law to incorporate such parameters. 

\subsection{Comparison to 1D simulations} \label{subsec:1D}

\begin{figure*}[t]
\begin{center}
    \hspace{-0.7cm}\includegraphics[scale=0.465]{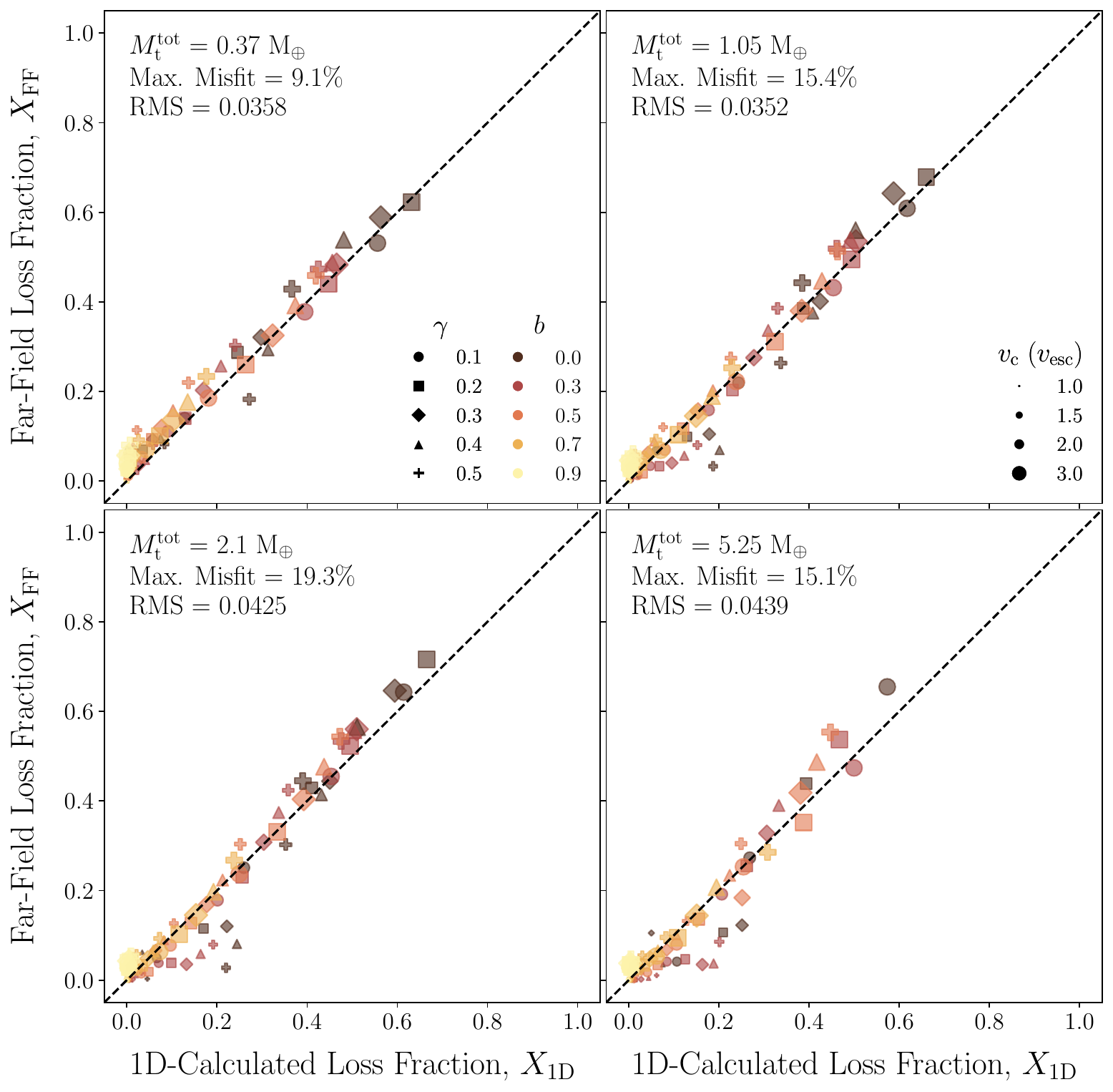}
    \caption{The far-field loss fraction can be well-approximated by combining 1D impedance match and loss calculations with the 3D pressure field derived from SPH simulations. Far-field loss fraction calculated in SPH against their corresponding values calculated using coupled 1D–3D calculations for each value of $b$ (colours), $\gamma$ (shapes), and $v_{\rm c}$ (marker sizes), with the maximum misfits and root-mean squared (RMS) misfit reported in each panel. Each panel shows the results for a different target mass.}
    \label{fig:1D_loss_fractions}
\end{center}
\end{figure*}

We calculated the loss that would be predicted from coupling the shock field from our 3D SPH simulations with 1D simulations of loss from shock-kick \citep{lock2024atmospheric,genda2003survival} to quantify the extent to which far-field loss is driven by shock-kick, and to test the applicability of conclusions from simpler 1D models to giant impacts. The loss calculated from SPH simulations has previously been compared to that from 1D simulations performed by \citet{genda2003survival} by using the ground motion imparted by the breakout of the shock wave generated during the impact directly from the SPH simulations \citep{kegerreis2020atmosphericA}. In this study, we build on previous work to develop a more accurate method for relating the strength of the shock in SPH simulations to the loss that would be predicted in 1D (see Section~\ref{subsec:1D_3D_method}). Using this new method, combined with our large suite of high output cadence simulations, allows us to systematically explore the role of shock-kick in giant impacts. Since 1D simulations consider only the upward passage of a shock wave from the surface of a planet into its atmosphere (i.e., modelling the process of shock-kick which drives loss in the far field), we only make comparisons between the calculated and predicted far-field loss.

\begin{figure*}[t]
\begin{center}
    \includegraphics[width=\textwidth]{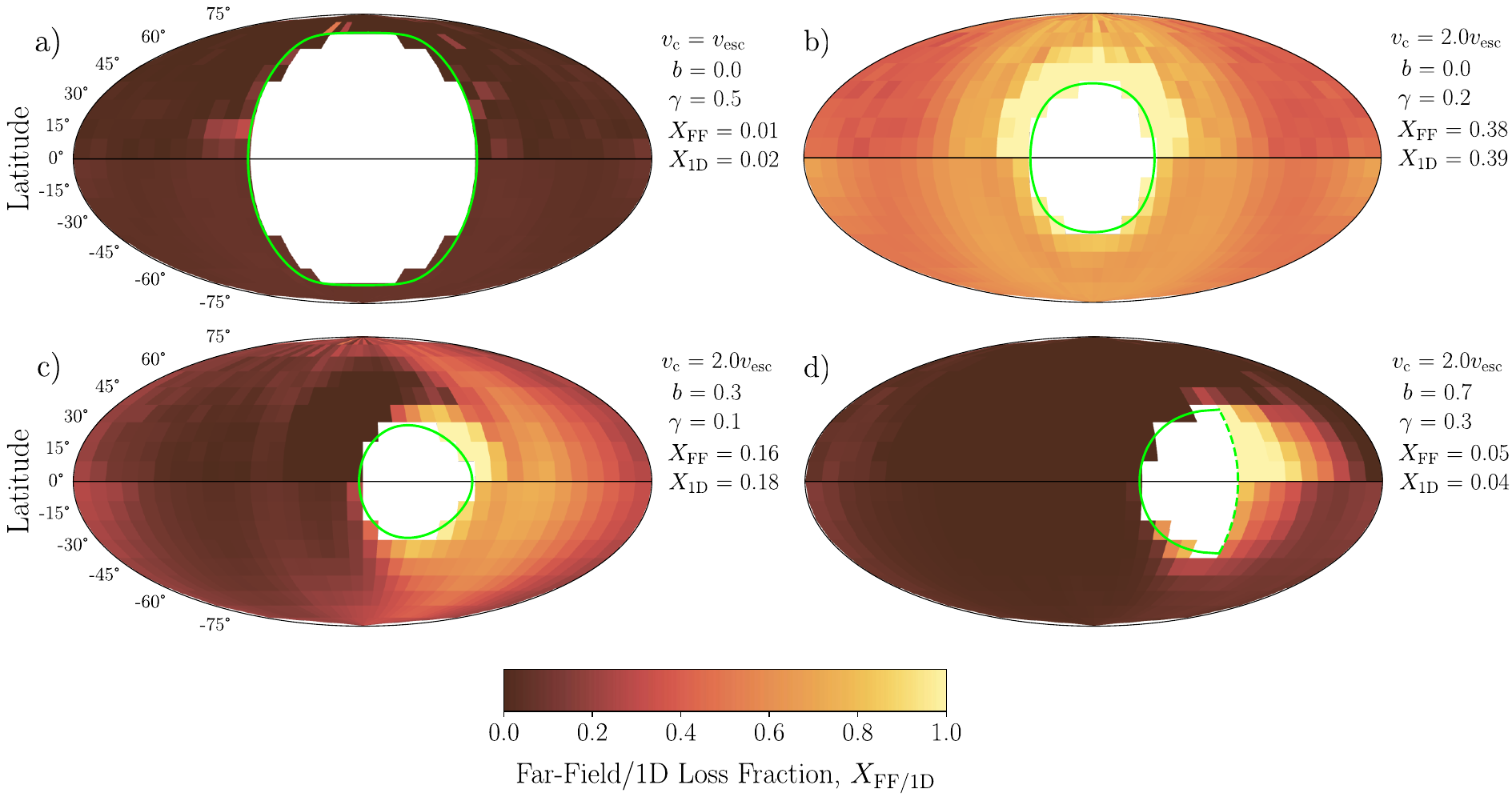}
    \caption{Coupled 1D–3D loss calculations well-approximate the magnitude and geometry of far-field loss. Mollweide projections centred on 0$^{\circ}$ longitude (see Figure \ref{fig:initial_conditions}) of far-field loss fraction (upper hemispheres) from 3D SPH simulations and the far-field loss fraction predicted by coupled 1D–3D calculations (lower hemispheres) across the 1.05 M$_{\oplus}$ planet's surface for four different combinations of $v_{
    \rm c}$, $b$, and $\gamma$. Note the colour bar is the same in each case. The green ellipses are the intersection curves of the target with the path of the impactor (see Figure~\ref{fig:NF_schematic} and Appendix \ref{sec:NF_sep}).}
    \label{fig:FF_vs_1D}
\end{center}
\end{figure*}

The far-field loss from each of our SPH simulations is plotted against the corresponding values predicted from our coupled 1D–3D calculations in Figure~\ref{fig:1D_loss_fractions}. We find very good agreement between the 3D and coupled 1D–3D calculations with a maximum absolute difference of $\sim$20\%. Furthermore, we show in Figure \ref{fig:FF_vs_1D} that the far-field loss geometry is also generally well-captured by the coupled 1D–3D calculation. However, it is important to note that the coupled approach generally underestimates loss close to the impact site and overestimates loss further away, with the two effects approximately balancing each other out. We will discuss the limitations of the coupled approach further in Section~\ref{subsec:limitations}.

\section{Discussion}
\label{sec:discussion}
We now discuss the implications of our findings for improving our understanding of the processes driving atmospheric loss during giant impacts (Section \ref{subsec:mechanisms}), the assumptions made in this work and over what parameter space the results of our study are applicable (Section \ref{subsec:limitations}), and how our results could impact future atmospheric loss studies (Section \ref{subsec:1D_3D_coupling}). We then place our findings in the wider context of planet formation and and discuss the role of giant impacts in controlling the volatile budgets of terrestrial planets (Section \ref{subsec:implications_volatiles}).

\subsection{The mechanisms of atmospheric loss during giant impacts}
\label{subsec:mechanisms}
The complex dynamics of giant impacts can make it challenging to draw direct mechanistic links between SPH simulations and the physical drivers of loss. However, we now present multiple lines of evidence, using SPH simulations and coupled 1D–3D calculations, to make these connections and show that atmospheric loss is controlled principally by two mechanisms --- ejecta plumes and shock-kick --- that operate primarily in the near and far field, respectively.

In the near field, there is a strong correlation between the magnitude and spatial distribution of loss and vaporisation of the surface mantle around the impact site (Figures~\ref{fig:NF_vs_Xmvap} \& \ref{fig:Xmvap_tot_vs_NF_frac}). Material around the impact site is highly shocked and vaporises when it is released to low pressure during the early stages of the impact, with substantial fractions of the vaporised material ejected from the system entirely (Figure~\ref{fig:NF_vs_Xmvap}). A majority of near-field atmosphere is typically only lost in higher-velocity impacts, when a significant fraction of the near-surface mantle is vaporised generating strong vapour plumes (Figure~\ref{fig:Xmvap_tot_vs_NF_frac}). Although a substantial fraction of the surface mantle is vaporised in low-velocity ($v_{\rm c}$ = $v_{\rm esc}$), head-on collisions (up to $\sim$70\% for 1.05 M$_{\oplus}$ and nearly 90\% for 5.25 M$_{\oplus}$, Figure~\ref{fig:Xmvap_tot_vs_NF_frac}), only a small fraction of the vaporised mantle is ejected from the system (Figure~\ref{fig:NF_vs_Xmvap}a). This suggests that surface mantle material does not necessarily need to be lost itself to drive loss of the overlying near-field atmosphere. The effect of vapour plumes on loss can also be seen in the focusing of vapour plumes down-range of the impact site in off-centre impacts, resulting in a notable `shadow' of loss in the up-range direction (Figure~\ref{fig:NF_vs_Xmvap}c, d).  The strong dependence of near-field loss on the degree of vaporisation, and the spatial correlation of vaporisation and loss, lead us to conclude that vapour plumes are the key driver for atmospheric loss in many giant impacts.

However, there can be significant contribution from other mechanisms --- such as air shocks and acceleration of the atmosphere by unvaporised impactor material --- in certain types of giant impacts. For impacts at velocities above the mutual escape velocity, it is possible to have a substantial fraction of near-field loss without significant vaporisation, as can be seen in the relatively large offset in the degree of near-field loss with vaporisation of the surface mantle between impacts at $v_{\rm c} = v_{\rm esc}$ and those at higher velocities when $M_{\rm t}^{\rm tot}$ $\in$ \{1.05, 2.1, 5.25\} M$_{\oplus}$ (Figure~\ref{fig:NF_vs_Xmvap}). The non-vapour-plume-related loss is likely driven by air shocks produced by the entry of the impactor into the target's atmosphere, which directly accelerate the upper atmosphere to velocities above the escape velocity \citep{ahrens1987impact,vickery1990atmospheric,ahrens1990earth,schlichting2015atmospheric,shuvalov2009atmospheric}. In some cases, this contribution from air shocks means that the loss from many impacts is greater than would be possible from just vapour plumes alone. The reason that this additional loss is not seen for impacts at escape velocity is that the impedance-match velocity between the impactor's mantle and the target's atmosphere is always somewhat lower than the impact velocity. The shocks generated during entry of the impactor into the target's atmosphere do not have the opportunity to accelerate up the hydrostatic atmospheric profile, a process which plays a significant role in enhancing loss by the shock-kick mechanism \citep{genda2003survival,genda2005enhanced,lock2024atmospheric}. At impact velocities near the mutual escape velocity the impedance match velocity is therefore simply not high enough for atmosphere accelerated by the entry air shock to escape the system. Air shocks may play a more significant role for impacts onto smaller planets, where the lower absolute impact velocity (rather than as a fraction of the escape velocity) lead to less vaporisation and weaker vapour plumes, as suggested by the weaker trend between vaporisation and near-field loss when $M_{\rm t}^{\rm tot}$ = 0.37 M$_{\oplus}$ in Figure~\ref{fig:NF_vs_Xmvap}. 

In other situations, different loss mechanisms can play a more significant role in loss from certain parts of the target. For example, in grazing collisions, the focusing of loss down-range of the contact site may be extenuated by the inertia of the non-interacting, and usually unvaporised, mass of the impactor accelerating atmosphere to escape. Conversely the focusing of vapour plumes (and unvaporised impactor material) in the down-range direction of significantly off-centre collisions, means that in the up-range direction air shocks are likely the dominant mechanism of loss. Therefore, although ejecta plumes are the dominant mechanism for substantial loss of near-field atmosphere, the importance of other mechanisms, particularly air shocks, in removing near-field atmosphere cannot be ignored. 

In the far field, the geometry of the pressure field mirrors that of loss (e.g., Figures \ref{fig:loss_vs_p_peak} \& \ref{fig:NF_vs_FF}), suggesting a strong relationship between the breakout of the impact shock and loss. This is quantitively supported by the fact that the far-field loss found in SPH simulations matches, both spatially and in total, predictions made by combining 1D impedance match and loss calculations of shock-kick with the 3D SPH pressure field (Figure \ref{fig:FF_vs_1D}). The predictive ability of combined 1D–3D shock-kick calculations is discussed further in Section~\ref{subsec:1D_3D_coupling}. The correlation and mechanistic connection made between loss and the impact shock provides strong evidence that shock-kick is the main driver of loss in the far-field \citep[see also][]{kegerreis2020atmosphericA}.

We find that additional loss in both the near and far field driven by secondary impacts from debris or the original impactor in graze-and-merge collisions, is generally a minor contribution to the combined loss. Secondary impacts are of much lower energy than the initial collision and the atmospheric and mantle structure of the largest remnant is significantly perturbed when the impact occurs. The surface of the body is often already vaporised and so strong ejecta plumes are not generated. In addition, the process of shock-kick is substantially less efficient due to the reduced impedance contrast at the surface of the body \citep[if in fact the body still has a clearly defined surface; see][]{lock2017structure,caracas2023no} and the disruption of the steep density gradient in the atmosphere. The major process generally acting to remove atmosphere in secondary impacts is akin to atmospheric shocks and, similarly for primary impacts, is not very effective at removing volatiles in sub-escape velocity collisions.

We have shown that the relative contributions of the near- and far-field loss, driven by these different mechanisms, to the combined loss change as a function of the parameters of the impact (i.e., target and impactor masses, angle, and velocity). Both near- and far- field loss generally decrease with impact angle and are highly sensitive to impact velocity, but the functional form of these dependences is quite different. By separately parametrising the near- and far-field loss as a function of the impact parameters, we have shown that considering loss from both regions of the target (and hence that driven by different mechanisms) separately can allow the construction of highly precise scaling laws to predict atmospheric loss from giant impacts.

Here, we have geometrically defined the near and far field based on the tangents to the surface at the point of contact between the impactor and target (see Section~\ref{subsec:NF_def}). However, it is possible that, at least in certain scenarios, this definition does not provide a good demarcation between the mechanisms driving loss, such that atmosphere material could be being lost via ejecta plumes or shock-kick outside of the regions initially designated as near and far field, respectively. For example, the air shock in the atmosphere from the impactor's initial entry can move atmospheric material across the surface, and material that was initially in the near field can be moved into the far field affecting its susceptibility for loss \citep[a related phenomena was observed for smaller impacts by][]{shuvalov2009atmospheric}. Future improvements to the definition of near and far field would further aid in isolating the role of different loss mechanisms and development of predictive scaling laws for loss.

It is important to note that the final loss fractions, and thus the final atmospheric masses, reported in Figure \ref{fig:total_loss}, are not the final states of planets. Following the cessation of primary and secondary impacts (i.e., the main collision between the target and impactor planets and the re-collision of impact debris of the following several hours, respectively), further atmospheric loss may be driven by processes such as re-accretion of debris, or the runner from hit-and-run collisions, over solar-centric orbital timescales \citep[e.g.,][]{emsenhuber2021collision}, and core-powered mass loss (e.g., via Parker winds or photoevaporation) -- rapid mass loss through heating of the atmosphere by the hot post-impact remnant's interior \citep[e.g.,][]{biersteker2019atmospheric, biersteker2021losing, ginzburg2018core, gupta2019sculpting, gupta2020signatures}. However, since hydrogen and other volatiles are likely mostly dissolved in the post-impact body's silicate envelope, their susceptibility to post-impact loss by such mechanisms is difficult to determine \citep{lock2018origin, caracas2023no,fegley2023chemical}. Quantifying the potential for additional losses following the first few hours of the impact is left for future work.

\subsection{Limitations and applicability of results}
\label{subsec:limitations}

\begin{figure}
\begin{center}
    \includegraphics[width=\columnwidth]{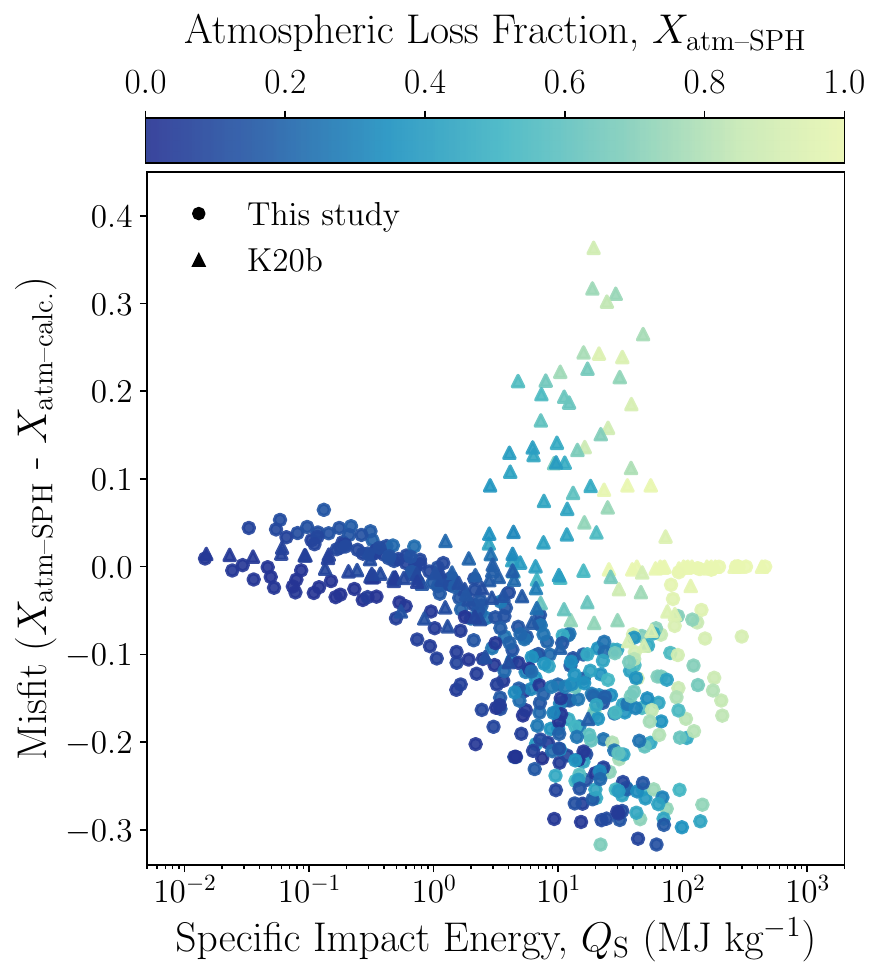}
    \caption{Scaling laws should not be used to predict the outcome of simulations outside of the parameter space from which they were constructed. The difference in combined loss (misfit) simulated directly from SPH simulations in this study (circles) or in \citet{kegerreis2020atmosphericB} (triangles) and the combined loss predicted by the scaling law derived by \citet{kegerreis2020atmosphericB}. This scaling law was constructed based on simulations that involved targets with less massive atmospheres than considered here and was fitted in log space rather than linear space. Misfits are plotted against the specific impact energy (Eqn.~\ref{eq:QS})  markers coloured by their SPH-simulated loss.}
    \label{fig:K20b}
\end{center}
\end{figure}

We now turn to discuss the applicability of the results and scaling law presented in this work. Here we have considered impacts onto planets with 5\% mass fraction H$_2$–He atmospheres, such as could be acquired by terrestrial planets if they grow large enough whilst the proto-planetary nebula is present \citep{ikoma2006constraints}. As we have only considered ocean-less rocky planets with one atmospheric mass fraction, temperature, and composition, our results and scaling law should be used with caution when studying planets with different atmospheric properties and impacts with parameters outside of the ones considered here. For instance, 1D hydrodynamic simulations and impedance match calculations \citep{lock2024atmospheric,genda2005enhanced} have shown that atmospheric loss is highly dependent on pre-impact surface conditions and atmospheric properties (namely, atmospheric temperature, composition, and pressure, and the presence/absence of an ocean). The influence of such parameters on loss have not been systematically explored in 3D simulations and no existing scaling laws derived from 3D SPH simulations account for atmospheric properties or the presence of an ocean. Previous work has developed independent scaling laws for collisions between larger planets with massive, dense primordial atmospheres \citep[i.e., mini-Neptunes,][]{denman2022atmosphere} and for loss from thinner terrestrial atmospheres \citep[1\% mass fraction H$_2$–He atmospheres,][]{kegerreis2020atmosphericB}. Figure \ref{fig:K20b} provides a demonstration of the magnitude of the error that can be introduced by using a scaling law (here, that from \citealt{kegerreis2020atmosphericB}) outside of the parameter space in which they were fitted. The \citet{kegerreis2020atmosphericB} scaling law over-predicts the expected loss for some of our impact scenarios by up to $\sim$30\%, particularly for low-loss impacts. It is important to note that the \citet{kegerreis2020atmosphericB} scaling law also under-predicts loss for some impact scenarios in \citet{kegerreis2020atmosphericB} by nearly 40\%, likely due to their scaling law being fitted in log space rather than linear space and not including sufficient terms to account for the complex interplay of varying near- and far-field loss efficiency (Section~\ref{subsec:mechanisms}). Using the scaling law presented here for other types of atmospheres could similarly significantly underestimate or overestimate the efficiency of loss, and caution is advised.

The susceptibility of even an individual planet to loss could change drastically throughout its evolution. Due to some combination of loss processes \citep[e.g.,][]{lopez2012thermal, lopez2013role, ginzburg2018core, zahnle2017cosmic, owen2013kepler, schlichting2015atmospheric, kegerreis2020atmosphericB} and outgassing from their interiors, terrestrial planets can develop thinner, higher-mean-molecular-weight secondary atmospheres throughout accretion \citep[e.g.,][]{krissansen-totton_erosion_2024, schaefer_predictions_2016, kite2020exoplanet}. \citet{lock2024atmospheric} showed that atmospheres which are hotter, lower pressure, and lower mean molecular weight are much more easily lost due to shock-kick during giant impacts. Moreover, towards the later stages of accretion, the time between giant impacts can become long enough that protoplanets can cool enough to condense surface oceans \citep{abe1988evolution, genda2005enhanced,matsui1986evolution, elkins2011formation}. For ocean-bearing planets, the atmosphere-to-ocean mass ratio ($\mathcal{R}$ = $M_{\rm atm}$/$M_{\rm oc}$) becomes a critical controlling factor of loss, such that $\mathcal{R}$ $\lesssim$ 1 (i.e., an ocean more massive than the atmosphere) can help promote loss of the atmosphere and $\mathcal{R}$ $\gtrsim$ 1 the ocean would instead inhibit loss \citep{lock2024atmospheric}. Significantly, atmospheric loss studies are yet to quantify the efficiency of loss from those planets that are most susceptible to loss (i.e., planets with thin atmospheres and those with surface oceans) and more work is needed to be able to trace the susceptibility of the atmospheres of terrestrial planets to loss through accretion.

Another important consideration when using our results is that the outcome of giant impacts, including the efficiency of atmospheric loss, can be significantly affected by the choice of EoS \citep{lock2024atmospheric,stewart_2019_3478631,meier2021eos}. For our study we used the most advanced EoS available for modelling the mantles and cores of the colliding bodies \citep{stewart_2019_3478631, 10.1063/12.0000946, sarah_t_stewart_2020_3866507}, but modelled our H$_2$–He atmospheres using the EoS from \citet{hubbard1980structure}. The HM80 EoS is a simple analytical EoS initially designed for modelling the interiors of giant planets and one that does not include known behaviours of hydrogen at the conditions expected during giant impacts (e.g., ionisation or decomposition). We made this choice in part to allow easy comparison to previous work \citep{kegerreis2020atmosphericA,kegerreis2020atmosphericB}, but also due to a scarcity of high-quality, publicly-available EoS for H$_2$–He mixtures suitable for modelling dynamical systems. There is currently considerable effort in developing new, publicly-available EoS for H$_2$–He mixtures \citep[e.g.,][]{chabrier2019new, chabrier2021new}, and it is important that the effect of a more accurate EoS on the calculated efficiency of atmospheric loss is quantified in future work.

\subsection{The potential of using coupled 1D–3D calculations to predict loss}
\label{subsec:1D_3D_coupling}

\citet{lock2024atmospheric} showed that atmospheric loss, at least that driven by shock-kick, is highly sensitive to the surface conditions on the colliding bodies (for further detail see Section~\ref{subsec:limitations}). There is therefore a need to quantify atmospheric loss efficiency over a large parameter space of not just impact parameters but also atmospheric compositions, temperatures, and masses, as well as the relative mass of an ocean and atmosphere. Full 3D SPH simulations as presented here, and in other works \citep{kegerreis2020atmosphericA,kegerreis2020atmosphericB,denman2020atmosphere,denman2022atmosphere}, are computationally expensive, particularly at high enough resolutions to model thin atmospheres like those of the solar system terrestrial planets, making such an endeavour incredibly challenging. Alternative methods for approximating atmospheric loss from planets with different surface properties would be very valuable to substitute for 3D simulations in parameter spaces where they are not yet available and to target cases where more computationally expensive simulations are required to better quantify loss. 

We have shown that far-field loss, as it is driven by shock kick, can be well-approximated by coupling 1D impedance match and loss calculations with the pressure field derived from 3D impact simulations (Figures \ref{fig:1D_loss_fractions} \& \ref{fig:FF_vs_1D}). It is straightforward to perform coupled 1D–3D calculations for different atmospheres using ground pressures sampled from existing SPH simulations, including from SPH simulations involving atmosphere-less bodies, providing an easy way to estimate loss. However, we do not currently have a similarly straightforward method to approximate loss in the near field. Further work is still required to understand how to account for changes in the efficiency of near-field loss with atmospheric parameters before the coupled 1D–3D approach can be effectively used.

There are also improvements that need to be made to the 1D–3D coupled approach that we have implemented in this work. Existing 1D shock-kick simulations \citep{lock2024atmospheric,genda2003survival,genda2005enhanced} model purely radial propagation of the impact shock into the atmosphere. In reality, the impact shock wave often arrives at the surface at more oblique angles, as dictated by the geometry of the impact. It is not clear how much this change in geometry affects the efficiency of loss. Our results suggest that it may be possible, to first order, to approximate an angled shock as a radial shock into an atmosphere with a larger scale height. Although the relationship between shock strength and loss is highly sensitive to atmospheric properties, the relationship between ground velocity and loss is not \citep{lock2024atmospheric,genda2003survival}. Hence, the efficiency of loss could be largely dependent on the total shock strength and not on the geometry of the shock. We find good agreement in both absolute far-field loss and spatial distribution of loss between the full 3D and coupled 1D–3D calculations (Figures~\ref{fig:1D_loss_fractions} \& \ref{fig:FF_vs_1D}), implying the validity of this logic. However, the coupled calculations tend to slightly underestimate the loss close to the impact site and overestimate loss further away. The underestimate near the impact site is likely due to an imperfect separation of the influence of  near- and far-field processes (see Section~\ref{subsec:mechanisms}), but the overestimated loss elsewhere on the target is due to other effects, such as a dependence on impact angle. Another factor that could contribute to an overestimate of loss in the coupled calculations is that we have assumed a planar geometry in our impedance-match calculations, and not accounted for the small increase in radius between the near-surface mantle, where we determined the strength of the shock, and the base of the atmosphere where the shock is released. Future work is needed to explore these effects and continue to improve the estimates of loss from coupled 1D–3D simulations. 

\subsection{Implications for the volatile evolution of terrestrial planets}
\label{subsec:implications_volatiles}

\begin{figure}[t]
\begin{center}
\includegraphics[width=\columnwidth]{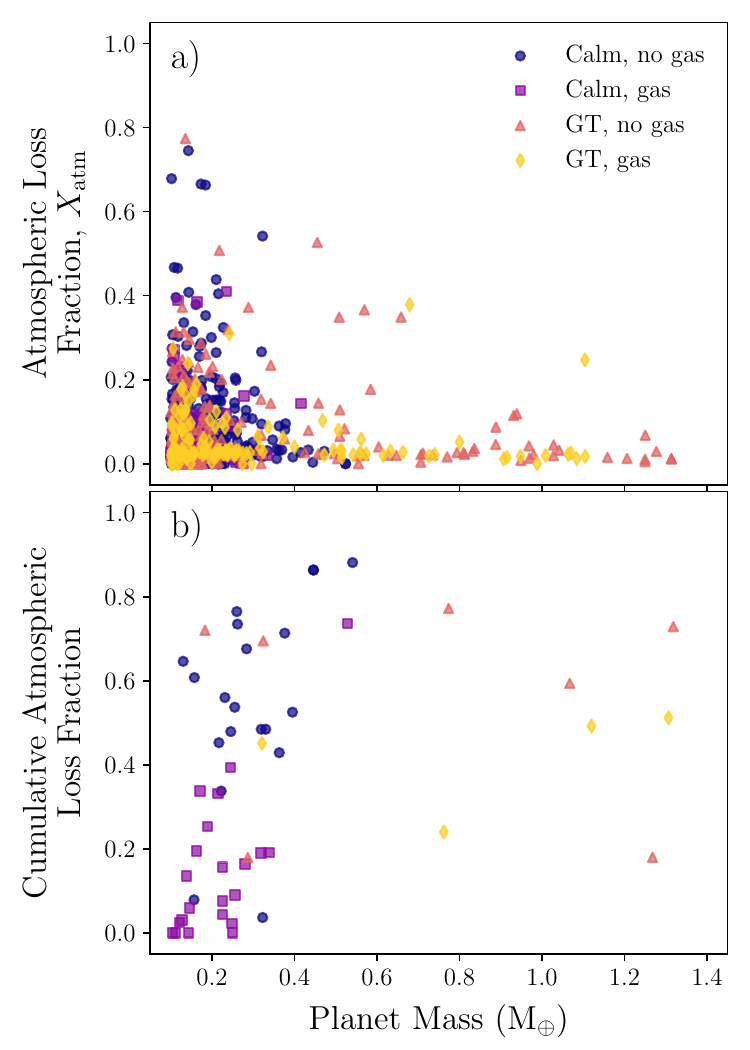}
    \caption{For planets with massive ($>$5\% mass fraction) primordial atmospheres, significant atmospheric loss for individual giant impacts onto planets of $\sim$1 M$_{\oplus}$ rarely cause $>$10\% loss, but the cumulative effect of the many giant impacts during accretion can result in substantial total loss. Atmospheric loss fraction experienced from individual giant impacts (a) and the combined effect of loss through accretion (b) in $N$-body simulations of \citet{carter2015compositional} and \citet{carter2022did} for four different planet formation scenarios (colours/shapes). For the definition of cumulative loss see text.}
    \label{fig:N_body}
\end{center}
\end{figure}

Quantifying the fraction of atmosphere lost in different styles of impacts during accretion is key to gaining a more comprehensive understanding of the volatile evolution of planets. We examine the effects of atmospheric erosion by applying our scaling law to the giant impacts in $N$-body simulations from \citet{carter2015compositional} and \citet{carter2022did}. Our aim is not to accurately trace the volatile budgets of planets through accretion, but rather gain insight into the role that loss during giant impacts could play in volatile evolution.

The 14 $N$-body simulations are separated into two broad scenarios, calm --- in which the influence of the giant planets is ignored --- and Grand Tack (GT) --- in which Jupiter migrates into the inner solar system, truncating and exciting the inner disk, before migrating back outward \citep{walsh2011low}. These two scenarios are likely end-member cases for excitation of the terrestrial planet forming region during accretion. Each of these scenarios is further separated into two simulation types, `gas' and `no gas', which include and ignore aerodynamic drag from the nebular gas, respectively. Note that (proto)planets grew larger in the Grand Tack simulations due to the increased mass delivered to the terrestrial planet region by Jupiter's migration. See \citet{carter2015compositional} for a complete description of the methods and dynamical results. 
The collision data have been nominally corrected for the expanded radii (expansion factor, $f$) used in the simulations to speed up the collisional evolution and thus reduce computation time. The collision times have been multiplied by $f^2$, the body radii have been divided by $f$, and the velocities have been increased to account for infall from $f (R_\mathrm{t}+R_\mathrm{i})$ to a realistic impact point at $R_\mathrm{t}+R_\mathrm{i}$. 
As mentioned in Section~\ref{sec:intro}, we treat as giant impacts all collisions with refractory target masses of at least 0.1 M$_\oplus$ and refractory impactor-to-target mass ratios of at least 0.01 (the mass ratio proposed by \citet{carter2020energy} to define a giant impact), and apply our atmospheric erosion scaling law to each. 
While smaller impacts may have a substantial cumulative effect \citep[e.g.][]{schlichting2015atmospheric}, such impacts lie outside the regime in which our new scaling law is applicable, therefore we neglect loss from smaller projectiles.

To estimate cumulative erosion by giant impacts across the simulations, we compute the ratio of the results of applying the scaling law to all giant impacts to the equivalent case in which no erosion occurs. Since we do not attempt to account for accretion of atmosphere during collisions or outgassing of atmospheres, the mass fraction of atmosphere will decrease even without erosion due to the growth of the `solid' planet. This `relative erosion' ratio therefore captures only the effects of giant impact atmosphere erosion, ranging from 0 (no loss) to 1 (total loss). Bodies that reach a minimum mass of 0.1 M$_\oplus$ within the lifetime of the nebular gas, which is nominally set at 5~Myr uninflated time, acquire a 5\% mass fraction atmosphere (see Sectio~\ref{subsec:limitations} for further discussion). Note that the lifetime of the nebula for atmospheric accretion is not the same as the lifetime of the gas used in the $N$-body simulations and only affects the cumulative erosion calculations. First we assign the compositions of the target and projectile in each collision. If this is their first collision, we apply the initial composition, otherwise compositions are taken from the appropriate remnant of the previous collision each body was involved in. Then, we calculate the effect of each collision on the composition of the remnants, removing atmosphere according to our scaling law. If there is a second large remnant produced from a collision, it is assigned atmosphere from the projectile (partial accretion and hit-and-run collisions) or the remaining portion of the target (erosive and super catastrophic collisions) proportionally to its mass. Note that we do not account for changes in the efficiency of atmospheric loss with changes in the mass of the atmosphere (see Section~\ref{subsec:limitations}) and use the scaling law derived in Section~\ref{subsec:scaling} throughout. Finally, we retrieve the compositions of the final protoplanets after their last collisions.

The results of this calculation are reported in Figure \ref{fig:N_body}. In general, simulations that include the effects of aerodynamic drag from the nebula (gas) have lower velocity collisions, giving a lower degree of erosion for those simulations compared to those without gas. In addition, decreasing the lifetime of the nebular gas results in an increase in the relative cumulative erosion, as the bodies undergo more collisions after their last opportunity to capture gas from the nebula. However, if the lifetime is reduced too far, very few bodies were large enough to capture an atmosphere while the nebular gas remained.

We find that individual loss events that substantially change atmosphere mass fraction (by $>$10\%) are not uncommon for small planet masses ($<$ 0.6 M$_{\oplus}$) but are generally rare for larger planet masses ($>$ 0.8 M$_{\oplus}$) (Figure \ref{fig:N_body}a) -- a result also previously identified in \citet{quintana2016frequency}. Despite this, the cumulative effect of many giant impacts during accretion can lead to much greater atmospheric loss (40–70\%) for larger planet masses with an initial atmosphere mass fraction of 5\% (Figure \ref{fig:N_body}b). The cumulative atmospheric loss driven by giant impacts varies significantly between bodies of a similar final mass, even within simulations with the same giant planet behaviour and effect of gas. For example, the two $\sim$1.3~M$_{\oplus}$ planets from `GT, no gas' simulations experience 20\% and 70\% cumulative loss, respectively (orange upward triangles in Figure~\ref{fig:N_body}b).

However, these results represent a weak lower limit for, and likely a significant underestimate of, the amount of loss driven by giant impacts during solar system accretion. By using our scaling law for massive, 5\% mass fraction atmospheres for all impacts, we are considering a low efficiency loss scenario and have neglected the fact that thinner atmospheres are much more easily eroded \citep{lock2024atmospheric}. If a planet underwent some amount of atmospheric loss, the next impact it experienced would be more effective at removing atmosphere, leading to an increased cumulative loss and more effective single loss events later in accretion -- an effect that would be exacerbated by condensation of oceans. This feedback could also increase the variation between planets within the same or between different systems that experienced similar growth environments but different impact histories. Further, in this work we have not considered the fate of volatiles carried on the impactor. Since more of the smaller body experiences higher shock pressures, we expect that the impactor's atmosphere is less likely to be retained by the final remnant and we will quantify this in future work. Including loss from the impactor in our calculations would amplify the role of giant impacts on the volatile budgets of terrestrial planets.

Our results have significant implications for the survival of the primordial atmospheres accreted by Earth's progenitor protoplanets. If the protoplanets that combined to form the terrestrial planets accreted primordial atmospheres from the nebula that were $>$5\% of their mass, giant impacts alone cannot be responsible for completely removing those atmospheres. 
This is significant in the case of the Earth as \citet{tucker2014evidence} proposed that the elevated $^3$He/$^{22}$Ne ratio of the depleted mantle compared to the plume source provided geochemical evidence that the proto-Earth experienced multiple episodes of atmospheric loss, mantle melting, and mantle degassing. They suggested that giant impacts onto the growing proto-Earth could provide the simplest explanation for this observation, as such events could simultaneously eject the pre-impact atmosphere and lead to extensive melting and degassing of the upper mantle. In addition, \citet{tucker2014evidence} also suggested that atmospheric loss events, potentially in the presence of a surface ocean, associated with giant impacts could also help to explain the Earth's subchondritic C/H, N/H, and Cl/F ratios. However, as a primordial atmosphere would have been the overwhelmingly dominant reservoir of noble gases, near total loss of the primordial atmosphere is required to significantly elevate the $^3$He/$^{22}$Ne ratio. We have shown that if the proto-Earth accreted a $>$5\% primordial atmosphere such total loss in giant impacts is highly unlikely. Therefore, if Earth's building blocks initially held massive nebula-captured atmospheres, potentially as large as 20\% the mass of the embryos as proposed by \citet[][]{young2023earth}, a large cumulative amount of volatile loss from small, sub-giant impacts would likely be required to explain the mantle $^3$He/$^{22}$Ne ratios \citep{schlichting2018atmosphere,PARAI2025513,peron2021deep}. Various other loss mechanisms are likely acting during accretion \citep[hydrodynamic escape, jeans escape, core-powered mass loss etc.,][]{biersteker2019atmospheric, biersteker2021losing, ginzburg2018core, lopez2012thermal, lopez2013role, ginzburg2018core, zahnle2017cosmic, owen2013kepler}, but such mechanisms can significantly fractionate the isotopes of volatile species, and it has been argued that such a large amount of fractionating loss is inconsistent with terrestrial noble gas signatures \citep{pepin1991origin,marty2012origins,schlichting2018atmosphere,PARAI2025513,peron2021deep}. The relative roles of different loss mechanisms that could explain the volatile geochemistry of Earth therefore depends critically on the initial mass of the nebula atmosphere.

Regardless of their effect on primordial atmospheres, giant impacts likely still played a large, or even dominant, role in removing subsequent secondary atmospheres and fractionating key terrestrial element ratios, as originally suggested by \citet{tucker2014evidence}. Even for the massive atmospheres we have considered here, our calculations suggest that atmospheric loss due to giant impacts would play a considerable cumulative role in shaping the volatile budgets of planets, and could contribute significantly to the variation between planets in our solar system (e.g., between Earth and Venus) and in exosystems \citep{bonomo2019giant, liu2015giant}.

\section{Conclusions}
\label{sec:conclusions}
We ran a suite of 3D SPH giant impact simulations to study the mechanisms by which atmosphere is lost in such events. We show that atmospheric loss is controlled principally by two mechanisms --- ejecta plumes and shock-kick --- which drive loss in the near and far field of the planet, respectively, with a small contribution from aur shocks in higher-velocity collisions. The relative contributions of these two mechanisms to loss change substantially as a function of the impact scenario. By parametrising these mechanisms separately as a function of the given impact parameters, we have developed a highly precise scaling law for loss for a wide range of terrestrial planet masses, impact velocities, and angles, concerning a target planet with a 5\% mass fraction H$_2$–He atmosphere. Given the very different dependences of near- and far-field loss, we encourage future works to use this approach of parametrising loss in the near and far field separately when developing new scaling laws. Our scaling law can be easily incorporated into larger-scale models of planet formation, however should be used with caution if being applied to planets with properties (e.g., masses, compositions, spin, atmospheric properties etc.) outside the range of those considered in this study.

We have studied in detail how the properties of the `ground surface' of an SPH planet respond during an impact, and thus how best to couple the results of SPH simulations with existing 1D impedance match and loss calculations \citep{lock2024atmospheric, genda2003survival, genda2005enhanced}. We show that order-of-magnitude estimates of combined loss, and good first-order approximations of far-field loss, can be made using computationally inexpensive 1D impedance match calculations coupled with existing 3D simulations. This result opens the door for the exploration of a wider range of atmospheric parameters without having to run as many expensive SPH simulations, including exploring thin atmospheres unresolvable in the current generation of giant impact simulations. 

Finally, we have placed limits on the role of atmospheric loss in giant impacts in shaping the volatile budget of solar system planets by applying our scaling law to $N$-body simulations of four different solar system formation scenarios. We find that individual giant impacts onto planets that are a substantial fraction of an Earth mass and that host the massive primordial atmospheres considered in this work rarely cause substantial ($>$10\%) atmospheric loss. However, the cumulative effect of multiple giant impacts can lead to significant loss (up to 70\%). These estimates likely represent lower limits on the amount of loss due to giant impacts, and our results therefore suggest that atmospheric loss due to giant impacts plays a significant role in shaping the volatile budgets of planets, and could potentially explain the variation between planets in our solar system.

Future work will look to develop scaling laws that account for variation in atmospheric composition, temperature, and mass fraction, and seek to explore the efficiency of loss from planets with much thinner atmospheres and/or surface oceans.

\vspace{1cm}
\noindent\textbf{Funding:} MJR acknowledges financial support from the STFC, UK (grant number: ST/X508251/1). SJL acknowledges financial support from NERC, UK (grant number: NE/V014129/1). JD acknowledges funding support from the Chinese Scholarship Council (No. 202008610218). JAK acknowledges support from a NASA Postdoctoral Program Fellowship. ZML and PJC acknowledge financial support from the STFC, UK (grant numbers: ST/V000454/1 and
ST/Y002024/1). All simulations were carried out using the computational facilities of the Advanced Computing Research Centre, University of Bristol – \url{http://www.bristol.ac.uk/acrc/}. We thank Sarah Stewart for useful discussions and Lucy Taylor for assistance in calculating the spatial distribution of vaporised material. \textbf{Author contributions:} Both MJR and SJL devised the project. MJR ran and analysed 3D SPH simulations, developed the scaling law, performed 1D–3D coupling, and wrote the text. SJL developed the near-/far-field division and wrote Appendix \ref{sec:NF_sep}, performed 1D impedance match calculations, discussed results, and edited the text. JD assisted in running simulations, discussed results, and edited the text. PJC conducted analysis of $N$-body simulations, discussed results, and edited the text. JAK assisted in running simulations, discussed results, and edited the text. ZML discussed results and edited the text. \textbf{Competing interests:} The authors declare that they have no competing interests. \textbf{Software:} This work made use of \texttt{SWIFT} \citep{schaller2016swift, schaller2023swift, kegerreis2019planetary}, \texttt{WoMa} \citep{ruiz2021effect}, \texttt{SEAGen} \citep{kegerreis2019planetary}, \texttt{numpy} \citep{van2011numpy}, \texttt{matplotlib} \citep{hunter2007matplotlib}, \texttt{scipy} \citep{virtanen2020scipy}, \texttt{pandas} \citep{mckinney2011pandas}, \texttt{jupyter} \citep{kluyver2016jupyter}, \texttt{mayavi} \citep{Ramachandran2011mayavi}, and \texttt{cmcrameri} \citep{crameri_2023_8409685}. \textbf{Supplementary materials:} The simulation data and code required to reproduce the figures and results from this paper are available at \citet{roche_2025_15211594} and \url{https://github.com/mjr230600/Roche_et_al_2025_PSJ_atmospheric_loss_mechanisms.git}. 

\appendix

\section{Near/Far Field Separation}
\label{sec:NF_sep}

In this section we describe how we define a tangent surface to divide the target's atmosphere into a near and far field. The first step is to define a contact curve that traces the first intersection of the path of the impactor with the target's surface (e.g., green line in Figure~\ref{fig:NF_schematic}). This is the area over which impactor material would directly hit the surface of the target unless forces act to deflect impactor material from its initial course. We can consider the path of the impactor as a cylinder of radius $R_{\rm i}^{\rm r}$ parallel to the $z$ axis that intersects the target, a sphere of radius $R_{\rm t}^{\rm r}$ (Figure \ref{fig:NF_schematic}). Following previous work, for non-head on impacts we notate the offset between the centre of the sphere and the centre of the cylinder (always taken as being in the negative $x$ direction for the purposes of this derivation) as $B$, which is related to the impactor parameter, $b$, by
\begin{equation}
    B=(R_{\rm t}^{\rm r}+R_{\rm i}^{\rm r})\,b \; .
\end{equation}
The intersection between a cylinder and a sphere on the upper hemisphere of the target can be defined by the parametric curve:
\begin{align}
\label{eqn:contact_curve}
    \boldsymbol{r}_{\rm cc}(\phi) = & (x_{\rm cc},y_{\rm cc},z_{\rm cc}) \nonumber \\
    = & \bigg(\left.R_{\rm i}^{\rm r}\cos \phi - B ,\right . \nonumber \\
    & \left . R_{\rm i}^{\rm r}\sin \phi , {\sqrt{2B(b_{\rm int}+R_{\rm i}^{\rm r}\cos \phi )}}\right) \; ,
\end{align}

\noindent where $\phi$ is the parametric parameter, and 
\begin{equation}
    b_{\rm int} = \frac {R_{\rm t}^{\rm r ^ 2}-{R_{\rm i}^{\rm r ^ 2}}-B^{2}}{2B} \; .
\end{equation} 
Note that the intersection between the sphere and cylinder on the surface of the lower hemisphere of the target is described by the same expression with $z \rightarrow -z$. If the impactor is grazing (i.e., $ R_{\rm t}^{\rm r} < R_{\rm i}^{\rm r} +B$, Figure~\ref{fig:NF_schematic}d) then the contact curve is open and $\phi$ is bounded such that $ -\phi _{0}<\phi <+\phi _{0}$, where $\cos \phi _{0}=-b_{\rm int}/R_{\rm i}^{\rm r}$. The points given by $\phi = \pm \phi _{0}$ are where the surfaces of the cylinder and sphere intersect in the $z=0$ plane (the plane perpendicular to the path of the projectile that passes through the centre of the target). To close the contact surface, we join these two extrema with a line that follows the great circle path between the two points (dashed green line in Figure~\ref{fig:NF_schematic}d). If the whole cylinder passes through the sphere (i.e., $ R_{\rm t}^{\rm r} > R_{\rm i}^{\rm r} +B$, Figure~\ref{fig:NF_schematic}a–c), then the contact curve is closed and the value of $\phi$ is unrestricted with the contact curve traced out every $2\pi$. We take $\phi_0=\pi$ as an arbitrary bound. In the special case that $R_{\rm t}^{\rm r}=R_{\rm i}^{\rm r}+B$ and $R_{\rm t}^{\rm r}=2 R_{\rm i}^{\rm r}$, the contact curve just touches the $z=0$ plane at one point and is a shape known as Viviani's curve.

In describing a tangent surface we wish to extend the concept of a tangent plane, the plane parallel to the target's surface at the point of contact, which is often used for describing loss from smaller impactors where $ R_{\rm t}^{\rm r} \gg R_{\rm i}^{\rm r}$ \citep[e.g.,][]{schlichting2015atmospheric}. Except in some special cases, we define the tangent surface as the surface described by the set of vectors that are parallel to the target's surface but perpendicular to the contact curve. Any vector that lies in the tangent surface and passes through a given point on the contact curve (e.g., green line in Figure~\ref{fig:NF_schematic}) must be perpendicular to the position vector of the specified point on the contact curve and perpendicular to the tangent of the contact curve at the same point. A parametric equation for the tangent to the contact curve can be found by using the derivative of Equation~\ref{eqn:contact_curve} with respect to $\phi$:
\begin{equation}
    \boldsymbol{r}_{\rm t} = (x_{\rm t},y_{\rm t},z_{\rm t}) = \frac{\mathrm{d}\boldsymbol{r}_{\rm cc}(\phi)}{\mathrm{d}\phi} \,\, t + \boldsymbol{r}_{\rm cc} \; ,
\end{equation}
where we have translated the vector to pass through the specified point on the contact surface and $t$ is a parametric parameter describing the distance from that point. Using Equation~\ref{eqn:contact_curve} this can be expressed in terms of $\phi$ and other parameters
\begin{align}
    \boldsymbol{r}_{\rm t} = \bigg(&\left.-R_{\rm i}^{\rm r} \sin{\phi} \,\, t + x_{\rm cc} , \right. \nonumber \\
       & \left. R_{\rm i}^{\rm r} \cos{\phi} \,\, t +y_{\rm cc} , \right. \nonumber \\
    & - R_{\rm i}^{\rm r} \sin{\phi} \sqrt{\frac{B}{2 \left ( b_{\rm int}+R_{\rm i}^{\rm r} \cos{\phi} \right )}} \,\, t + z_{\rm cc} \bigg)
\end{align}
Unfortunately, the prefactor to $t$ in the expression for $z_{\mathrm{t}}$ is undefined as $\phi \rightarrow \phi_0$ for grazing impacts. One way to avoid this singularity would be to normalize $\boldsymbol{r}_{\mathrm t}$, but this expression is somewhat messy. Here, we therefore take the approach of redefining the parametric parameter
\begin{equation}
    t \rightarrow 2 \sqrt{b_{\rm int}+R_{\rm i}^{\rm r} \cos{\phi}} \;t \; ,
\end{equation}
which removes the singularity. The resulting expression for the tangent is
\begin{align}
\label{eqn:tangent_vector}
    \boldsymbol{r}_{\rm t} = (x_{\rm t},y_{\rm t},z_{\rm t}) = \bigg(  & \left.-2 R_{\rm i}^{\rm r} \sin{\phi} \sqrt{b_{\rm int}+R_{\rm i}^{\rm r} \cos{\phi}} \,\, t + x_{\rm cc} , \right. \nonumber \\
       & \left. 2 R_{\rm i}^{\rm r} \cos{\phi} \sqrt{b_{\rm int}+R_{\rm i}^{\rm r} \cos{\phi}} \,\, t +y_{\rm cc} , \right. \nonumber \\
   & -R_{\rm i}^{\rm r} \sin{\phi} \sqrt{2B} \,\, t + z_{\rm cc} \bigg) \; .
\end{align}

\noindent A vector in the tangent surface can then be found by taking the cross product of the position vector and the tangent vector at a point on the contact curve
\begin{equation}
    \boldsymbol{r}_{\rm tv} = \boldsymbol{r}_{\rm cc} \times \boldsymbol{r}_{\rm t} + \boldsymbol{r}_{\rm cc}  \, ,
\end{equation}
where again we have translated the vector to pass through the specified point on the contact surface. Expanding
\begin{align}
\label{eqn:tangent_surface}
\boldsymbol{r}_{\rm tv} = \left(x_{\rm tv}, y_{\rm tv}, z_{\rm tv}\right)=( & y_{\rm cc} z_{\rm t}-z_{\rm cc} y_{\rm t} + x_{\rm cc} , \nonumber \\
    & z_{\rm cc} x_{\rm t}-x_{\rm cc} z_{\rm t} +y_{\rm cc} , \nonumber \\
    & x_{\rm cc} y_{\rm t}-y_{\rm cc} x_{\rm t} +z_{\rm cc} ) \, .
\end{align}

For non-grazing impacts ($ R_{\rm t}^{\rm r} > R_{\rm i}^{\rm r} +B$), the tangent surface is described by the infinite set of tangent vectors (given by Equation~\ref{eqn:tangent_surface}) with $\phi$ varying between $\pm \pi$. The tangent plane is defined for all directions away from the intersection of the two bodies. However, for grazing impacts ($ R_{\rm t}^{\rm r} < R_{\rm i}^{\rm r} +B$), the contact curve is open on the grazing side and not completely described by Equation~\ref{eqn:contact_curve}. In between the extrema of Equation~\ref{eqn:contact_curve} ($\phi=\pm \phi_0$) the contact curve is defined by the great circle path joining the great circle path and in this region the tangent surface is described by vectors parallel to the cylinder described by the path of the impactor (i.e., the velocity vector of the impactor relative to the target) passing through points on the great circle path (Figure~\ref{fig:NF_schematic}d). Such vectors are given by
\begin{align}
\label{eqn:tangent_surface_grazing}
    \boldsymbol{r}_{\rm tv}^{\rm gc} = &\left(x_{\rm tv}^{\rm gc}, y_{\rm tv}^{\rm gc}, z_{\rm tv}^{\rm gc}\right) = \left(x_{\rm cc}, \sqrt{R_{\rm t}^{\rm r ^ 2}-x_{\rm cc}^2}, t \right) \; ,
\end{align}
where, as above, $t$ is a parametric parameter with $\boldsymbol{r}_{\rm tv}^{\rm gc}=\boldsymbol{r}_{\rm cc}$ at $t=0$, and $-R_{\rm t}^{\rm r}<x_{\rm cc}<-b_{\rm int}$. The tangent surface defined by Equations~\ref{eqn:tangent_surface} and \ref{eqn:tangent_surface_grazing} leaves a gap in the tangent surface in the down-range direction either side of the great-circle portion of the contact curve. We therefore include another set of vectors to describe the tangent plane which have the same direction as the vector describing the tangent surface at $\phi=\phi_0$ (equation~\ref{eqn:tangent_surface}) but transposed in the $z$ direction to join the two previously defined surfaces. This set of vectors is given by
\begin{align}
\label{eqn:tangent_surface_phi0}
   \boldsymbol{r}_{\rm tv}^{\phi_0} (t,h) =  \boldsymbol{r}_{\rm tv}\left(\phi=\phi_0, t\right) - h\hat{\boldsymbol{k}} \; ,
\end{align}
where $\hat{\boldsymbol{k}}$ is the unit vector in the $z$ direction and $h>0$ is a parametric scalar that sweeps out the set of vectors to describe the tangent surface.

Whether a point ($\boldsymbol{r'}=(x', y', z')$) is above or below the tangent surface, and hence either in the near or the far field, can be determined by comparing the position of the point in the $x$–$y$ plane relative to the intersection of the tangent surface with the $z$=$z'$ plane. For head-on impacts this is relatively straightforward. From the geometry of the system, the height of the contact curve is
\begin{equation}
z_{\rm cc}=\sqrt{R_{\rm t}^{\rm r ^ 2}-R_{\rm i}^{\rm r ^ 2}}
\end{equation}
and the distance of the tangent surface from the $z$ axis is
\begin{equation}
r^{\rm ts}_{xy}=R_{\rm i}^{\rm r} + \frac{R_{\rm t}^{\rm r} \left ( z_{\rm cc} -z' \right )}{R_{\rm i}^{\rm r}}
 \sqrt{1-\left(\frac{R_{\rm i}^{\rm r}}{R_{\rm t}^{\rm r}}\right )^2} \; .
\end{equation}
If $z'>z_{\rm cc}$, $r'_{xy}=\sqrt{x'^2+y'^2}>r^{\rm ts}_{xy}$, or the point is in the path of the approaching projectile (i.e., $r'_{xy} \leq R_{\rm i}^{\rm r}$ and $z' \geq 0$) then it is considered to be in the near field. Else, the point is in the far field.

For non-head-on impacts, determining whether a point is in the near or far field is more complex as there are a variety of specific cases that must be considered. We first apply some broad criteria in order that can determine whether a point is in the near or far field with out much computational effort: 
\begin{enumerate}
    \item Any point directly behind the planet from the direction of the impactor's approach ($x'^2+y'^2<R_{\rm t}^2$ and $z'<0$)  is in the far field assuming $R_{\rm t}^{\rm r}\geqq R_{\rm i}^{\rm r}$.
    \item Any point in the cylinder swept out by the impactor upon approach (i.e., where $(x'+B)^2+y'^2)<R_{\rm i}^{\rm r ^ 2}$, $x'+B<R_{\rm i}^{\rm r}$ and $z'\geq0$) is in the near field.
    \item Additionally, in the case of a grazing impact ($R_{\rm i}^{\rm r}+B>R_{\rm t}^{\rm r}$), any point that is within the impactor cylinder but not in the shadow of the planet (i.e., where $x'+B\leq-b_{\rm int}$, $y'^2\leq R_{\rm i}^{\rm r ^ 2}-b_{\rm int}^2$, and $x'^2+y'^2>R_{\rm t}^{\rm r ^ 2}$) is in the near field.
    \item If the contact curve is contained within just one half-hemisphere of the target (i.e., if $B-R_{\rm i}^{\rm r}\geq0$) then any point in the opposite half-hemisphere of the atmosphere (those with $z'<0$ and $x'>0$) are necessarily in the far field.
    \item Many points in the shadow of the planet can be determined to be in the far field based on the maximum slope of the tangent surface as defined by equation~\ref{eqn:tangent_surface}. The maximum angle, $\theta_{\rm ts}^{\rm max}$, between a plane parallel to the $x$–$y$ plane and any vector describing the tangent surface can be found by minimising the function
    \begin{equation}
        f(\phi)=-\theta_{\rm ts}(\phi) \, 
    \end{equation}
    with respect to $\phi$, where
    \begin{equation}
        \begin{aligned}
            \theta_{\rm ts}(\phi) &= \tan^{-1} \bigg\{ \left| z_{\rm tv}(\phi,-2) - z_{\rm cc}(\phi) \right| \\
            & \quad\quad\quad\quad \times \Big[ \left( x_{\rm cc}(\phi) - x_{\rm tv}(\phi,-2) \right)^2 \\
            & \quad\quad\quad\quad + \left( y_{\rm cc}(\phi) - y_{\rm tv}(\phi,-2) \right)^2 \Big]^{-\frac{1}{2}} \bigg\}
        \end{aligned}
    \end{equation}
    and the choice of using $t=-2$ is arbitrary. We perform this minimization using the \texttt{minimize} function from the \texttt{scipy.optimize} package. Points in the lower hemisphere ($z'<0$) of non-grazing impacts can be definitively assigned to the far field if the equivalent angle between a vector connecting the point to the nearest edge of the planet,  
    \begin{equation}
        \theta=\tan^{-1}{\left | \frac{z'}{\sqrt{x'^2+y'^2}-R_{\rm t}} \right | } \, ,
    \end{equation}
    is greater than  $\theta_{\rm ts}^{\rm max}$. 
    \item If the impactor is grazing ($R_{\rm i}^{\rm r}+B>R_{\rm t}^{\rm r}$), it is possible to account for much of the complexity introduced by the non-differentiable regions of the tangent surface (i.e., the transitions to and between those regions not defined by equation~\ref{eqn:tangent_surface}) by assigning points to be in the near field based on their angle to the extrema of equation~\ref{eqn:tangent_surface}. A point is defined as being in the near field if one of the following apply:
    \begin{enumerate}
        \item The point is beyond the edge of the planet in the $x$ direction on the side that the impactor is grazing ($x'<-R_{\rm t}^{\rm r}$).
        \item The point is beyond the extrema of the curve described by equation~\ref{eqn:tangent_surface} in the $x$ direction ($x' \leq-B-b_{\rm int}$) and is either above the midplane ($z>0$) or below the midplane and not in the shadow of the target ($z'\leq0$ and $\sqrt{x'^2+y'^2}>R_{\rm t}^{\rm r}$).
        \item The point lies outside the extrema of the curve described by equation~\ref{eqn:tangent_surface} in the $y$ direction but is positioned so that it lies outside the tangent surface as defined by the set of vectors described by equation~\ref{eqn:tangent_surface_phi0}. As all the said vectors are parallel, we can access based on the angle between the vector $\boldsymbol{r}_{\rm tv}^{\phi_0} (t=0,h)$ and the point in a plane parallel to the $x$-$y$ plane. Therefore if $y'\geq \sqrt{R_{\rm i}^{\rm r ^ 2}-b_{\rm int}^2}$ and $\theta_+<\theta_{\rm t}$, or $y'\leq \sqrt{R_{\rm i}^{\rm r ^ 2}-b_{\rm int}^2}$ and $\theta_-<\theta_{\rm t}$, where 
        \begin{equation}
            \theta_{\pm}=\tan^{-1}{\left | \frac{B-b_{\rm int}-x'}{y' \pm \sqrt{R_{\rm i}^{\rm r ^ 2}-b_{\rm int}^2}} \right | }
        \end{equation}
        and
        \begin{equation}
            \theta_{\rm t}=\tan^{-1}{\left | \frac{x_{\rm tv}\left(\phi_0, -1\right)-(B-b_{\rm int})}{y_{\rm tv}\left(\phi_0, -1\right)-\sqrt{R_{\rm i}^{\rm r ^ 2}-b_{\rm int}^2}} \right | }
        \end{equation}

        the point is in the near field. The choice of using $t=-1$ is arbitrary and the value of $t$ must simply describe a point on the vector in the tangent surface that is not on the contact curve.
    \end{enumerate}

\end{enumerate}

If none of the above criteria apply, we identify whether a parcel of the atmosphere is in the near or far field by comparing its position in the $z=z'$ plane to the intersection of the tangent surface (usually one or more lines) with the same plane, i.e., the point when
\begin{equation}
z_{\rm tv}=x_{\rm cc} y_{\rm t}-y_{\rm cc} x_{\rm t} +z_{\rm cc}=z' \; .
\end{equation}
Subbing in from Equations~\ref{eqn:tangent_vector} and \ref{eqn:contact_curve} we can rewrite this expression as
\begin{align}
z' = & (R_{\rm i}^{\rm r}\cos \phi - B) (2 R_{\rm i}^{\rm r} \cos{\phi} \sqrt{b_{\rm int}+R_{\rm i}^{\rm r} \cos{\phi}} \,\, t_{z'} \nonumber \\ 
 & \hspace{75pt} +R_{\rm i}^{\rm r}\sin \phi )  \nonumber \\ 
    & - R_{\rm i}^{\rm r}\sin \phi (-2 R_{\rm i}^{\rm r} \sin{\phi} \sqrt{b_{\rm int}+R_{\rm i}^{\rm r} \cos{\phi}} \,\, t_{z'}  \nonumber \\
    & \hspace{75pt} + R_{\rm i}^{\rm r}\cos \phi - B) \nonumber \\ 
    & +{\sqrt{2B(b_{\rm int}+R_{\rm i}^{\rm r}\cos \phi )}}  \; .
\end{align}
Rearranging and using the trigonometric identity $\cos^2{\phi}+\sin^2{\phi}=1$ we find
\begin{equation}
\label{eqn:nearfar:tintersect}
t_{z'}=\frac{z'-\sqrt{2B(b_{\rm int}+R_{\rm i}^{\rm r}\cos \phi )}}{2 R_{\rm i}^{\rm r} \sqrt{b_{\rm int}+R_{\rm i}^{\rm r} \cos{\phi}} \,\, \left ( R_{\rm i}^{\rm r} - B \cos{\phi}\right)} \;.
\end{equation}
Note that the sign of $t_{z'}$ changes at $\phi=\phi_0$ and when
\begin{equation}
\label{eqn:nearfar:tsignchange1}
z'=\sqrt{2B(b_{\rm int}+R_{\rm i}^{\rm r}\cos \phi_{t_{z'}=0} )} \;,
\end{equation}
or alternatively
\begin{equation}
\label{eqn:nearfar:tsignchange2}
\phi_{t_{z'}=0} = \cos^{-1}{\left(\frac{z'^2}{2BR_{\rm i}^{\rm r}}-\frac{b_{\rm int}}{R_{\rm i}^{\rm r}}\right )} \;,
\end{equation}
i.e., the angle at which the intersection curve intercepts the $z=z'$ plane. Subbing the expression from Equation~\ref{eqn:nearfar:tintersect} into Equation~\ref{eqn:tangent_vector} we can find the set of solutions $x_t(\phi, z=z')$ and $y_t(\phi, z=z')$ at which the tangent plane crosses the plane $z=z'$. 

The geometry of the solutions on the $z=z'$ plane for different $\phi$ depends on the relative arrangements of the impactor and target as well as the exact value of $z'$. Figure~\ref{fig:near-far-field_examples} shows the possible solutions in each regime and we discuss each of these in the following subsections.

\begin{figure*}
    \centering
    \includegraphics[angle=90,width=0.99\linewidth]{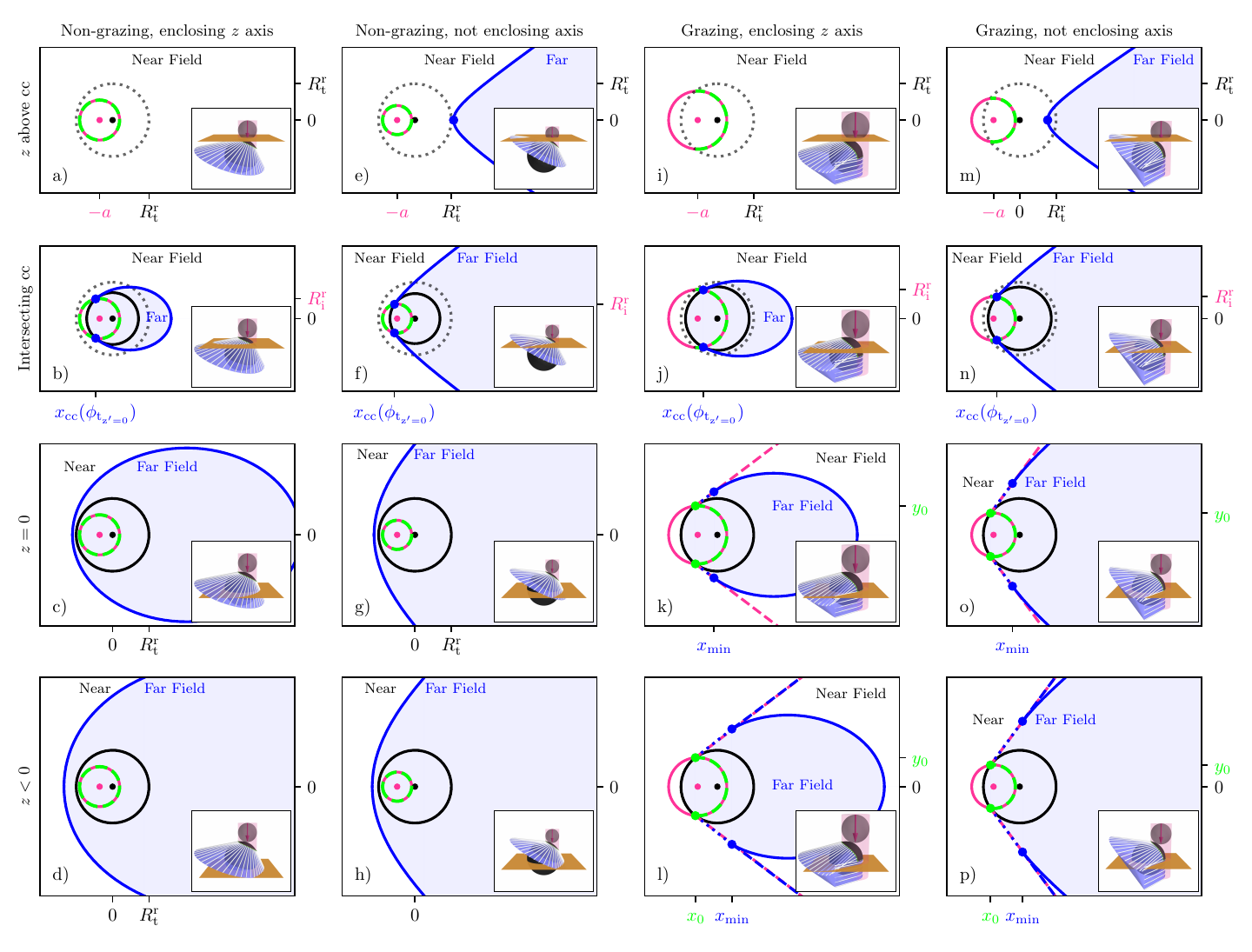}
    \caption{Caption on next page.}
\end{figure*}
\addtocounter{figure}{-1}
\begin{figure*}[t]
    \caption{Determination of whether a point is in the near or far field depends on the relative orientation of the impactor and target, and the $z$ value of the point in question. Each panel shows the division between near and far field in planes at different heights relative to the contact curve and equator (rows) and for different arrangements of the impactor and target (columns). The rows show example planes that (from top to bottom) are above the contact curve, intersect the contact curve, are in the equator, and are below the equator. Columns correspond to each of the cases discussed in Sections~\ref{sec:appendix:ng_enc}–\ref{sec:appendix:g_notenc}. The inset in each panel shows the relative orientation of the impactor, target, and tangent surface as in Figure~\ref{fig:NF_schematic} with the plane being considered in that panel shown in orange. Each panel sows the circumference of the impactor (black dashed line), the intersection of the target's surface and the plane (black solid line, when applicable), the $z$-axis which passes through the centre of the target (black dot), the area traversed by the impactor (pink solid line), the centre of the impactor (pink dot), the contact curve projected into the plane (green dashed line), the intersection of the tangent surface (blue line, when applicable) and select truncation points or extrema of the intersection curve (blue dots, when applicable). The blue shaded region is the region determined to be in the far field and otherwise points are designated as being in the near field. In panels k, l, o, and p, green dots show the intersection of the contact curve and the equator at $\phi=\phi_0$, the dotted blue line is the straight line joining these points with the maxima on the intersection of the tangent surface and the plane, the blue dashed lines are solution branches (referred to as asymptotic `flicks' in the text) that are less restrictive than other solution branches and so not used in our near-far field determination, and the pink dashed lines are tangents to the contact curve in the equator. Key $x$ and $y$ values are noted on the axes and coloured corresponding to the colours of the relevant feature in the panel and are described in text.}
    \label{fig:near-far-field_examples}
\end{figure*}

\subsection{Non-grazing impact with the contact curve enclosing the $z$ axis}
\label{sec:appendix:ng_enc}

In the case that $R_{\rm t} \geq R_{\rm i}^{\rm r} + B$ and $B<R_{\rm i}^{\rm r}$ (Figure~\ref{fig:near-far-field_examples}a–d) the contact curve encloses the pole and is continuous and described exclusively by Equation~\ref{eqn:contact_curve} (i.e., the impact is non-grazing). In this case the tangent surface is always sloped downwards relative to the $z$ axis (i.e., towards smaller or more negative $z$) and any point above the highest point on the contact curve (i.e., $z'>\sqrt{2B\left(b_{\rm int}+R_{\rm i}^{\rm r}\right)}\;$) is in the near field (Figure~\ref{fig:near-far-field_examples}a). If the $z=z'$ curve intersects the contact curve (i.e., $\sqrt{2B\left(b_{\rm int}-R_{\rm i}^{\rm r}\right)} \leq z' \leq \sqrt{2B\left(b_{\rm int}+R_{\rm i}^{\rm r}\right)}$) then the intersection of the tangent surface with the plane is an open arc bounded by $\phi=0$ and $\phi=\phi_{t_{z'}=0}$ (solid blue line in Figure~\ref{fig:near-far-field_examples}b). Otherwise, the intersection is a closed curve marked out between $\phi=\pm\pi$ (Figure~\ref{fig:near-far-field_examples}c, d). In the later two cases, to determine whether a point is in the near or far field we use a bisection method, as implemented in the \texttt{root\_scalar} function from the \texttt{scipy} Python package, to solve $x_{\rm t}(\phi, z=z')-x'=0$ for $\phi$ within the relevant bounds. If $|y'|<|y_{\rm t}(\phi, z=z')|$ then the point is underneath the tangent surface and so in the far field and otherwise is above or on the surface and is in the near field.

\subsection{Non-grazing impact with the contact curve not enclosing the $z$ axis}
\label{sec:appendix:ng_notenc}

In the case that $R_{\rm t}^{\rm r} \geq R_{\rm i}^{\rm r} +a$ and $B<R_{\rm i}^{\rm r}$ the contact curve is continuous and described exclusively by Equation~\ref{eqn:contact_curve} (i.e., the impact is non-grazing) but does not enclose the $z$ axis (Figure~\ref{fig:near-far-field_examples}e–h). This situation is more complicated than when the contact curve encloses the pole (Section~\ref{sec:appendix:ng_enc}) as the tangent surface can be sloped both upwards and downwards relative to the $z$ axis and every possible $z=z'$ plane is therefore intersected by the tangent surface (see inset in Figure~\ref{fig:near-far-field_examples}e). Further, each plane holds solutions for both positive and negative $t_{z'}(\phi)$ (i.e., tangents that leave the contact curve both towards and away from the contact point) and so not all solutions represent the intersection of the tangent surface with the plane. 

The set of desired solutions describe a hyperbola with asymptotes at $\phi=\pm\phi_{\rm lim}=\pm\cos^{-1}{\left(R_{\rm i}^{\rm r}/B\right)}$ and vertices at $\phi=0$ and $\phi=\pm\pi$. If the $z=z'$ plane is above the highest point on the contact curve (i.e., $z'>\sqrt{2B\left(b_{\rm int}+R_{\rm i}^{\rm r}\right)}\;$), then the intersection of the tangent surface with the plane is bounded by $\phi=0$ and $\phi=\phi_{\rm lim}$ (Figure~\ref{fig:near-far-field_examples}e). Any point for which $x'<x_{\rm tv}(\phi=0,t_{z'}(\phi=0))$ is in the near field. If the $z=z'$ curve intersects the contact curve (i.e., $\sqrt{2B\left(b_{\rm int}-R_{\rm i}^{\rm r}\right)} \leq z' \leq \sqrt{2B\left(b_{\rm int}+R_{\rm i}^{\rm r}\right)}$) then the intersection of the tangent surface with the plane is bounded by $\phi=\phi_{t_{z'}=0}$ and $\phi=\phi_{\rm lim}$ (Figure~\ref{fig:near-far-field_examples}f). Any point for which $x'<x_{\rm cc}(\phi=\phi_{t_{z'}=0})$ is in the near field. Otherwise, the point is below the lowest point on the contact curve, the intersection is bounded by $\phi=\pi$ and $\phi=\phi_{\rm lim}$, and any points for which $x'<x_{\rm tv}(\phi=\pi,t_{z'}(\phi=\pi))$ are automatically in the near field (Figure~\ref{fig:near-far-field_examples}g, h).

To determine whether a point is in the near or far field we use a bisection method, as implemented in the \texttt{root\_scalar} function from the \texttt{scipy} Python package, to solve $x_{\rm t}(\phi, z=z')-x'=0$ for $\phi$ within the relevant bounds. If $|y'|<|y_{\rm t}(\phi, z=z')|$ then the point is underneath the tangent surface and so in the far field and otherwise is above or on the surface and is in the near field. 

For numerical feasibility we do not consider solutions up to the limit of $\phi=\phi_{\rm lim}$ but instead $\phi=(1\pm\epsilon)\phi_{\rm lim}$ where $\epsilon=10^{-13}$ is small with the sign depending on whether the limit is being approached from above or below.

\subsection{Grazing impact with the contact curve enclosing the $z$ axis}
\label{sec:appendix:g_enc}

In the case that $R_{\rm t}^{\rm r} \leq R_{\rm i}^{\rm r} + B$ and $B<R_{\rm i}^{\rm r}$ the impact is grazing but the contact curve still encloses the pole (Figure~\ref{fig:near-far-field_examples}i–l). The tangent surface is always pointing downwards relative to the $z$ axis or is perpendicular to the $z$ axis. Therefore, if the $z=z'$ plane is above the highest point on the contact curve (i.e., $z'>\sqrt{2B\left(b_{\rm int}+R_{\rm i}^{\rm r}\right)}\;$) then any point is in the near field (Figure~\ref{fig:near-far-field_examples}i). 

If the $z=z'$ plane intersects the contact curve (i.e., $0 < z' \leq \sqrt{2B\left(b_{\rm int}+R_{\rm i}^{\rm r}\right)}$)  then the intersection of the tangent plane with the curve is a truncated oval with a maximum in $x$ at $\phi=0$ and a minimum at $\phi_{\rm t_{z'=0}}$ (Figure~\ref{fig:near-far-field_examples}j). Any point with $x'<x_{\rm cc}\left ( \phi_{\rm t_{z'=0}} \right)$ is in the near field. Otherwise we again determine whether a point is in the near or far field using a bisection method to solve $x_{\rm t}(\phi, z=z')-x'=0$ for $\phi$ within the relevant bounds and comparing the root to the value of $y'$. 

For $z'\leq0$, the intersection of the tangent surface and the plane is in two parts due to the two-part definition of the tangent surface: one part defined by the set of tangents extended from the edge of the planet (Equation~\ref{eqn:tangent_surface_grazing}), and the other by Equation~\ref{eqn:tangent_surface}. Whether particles are in the near field based on the first of these parts has already been taken care by the general criteria 6 above. For the second set of tangents, the intersection of the tangent surface with the plane is mostly a truncated ellipsoid (Figure~\ref{fig:near-far-field_examples}l) but with backwards facing asymptotic `flicks' if $z'<0$ as $\phi\rightarrow\pm\phi_0$ after the ellipse reaches a minima in $x$ at $\phi_{\rm min}$ with a value $x_{\rm min}>x_{\rm cc}(\phi_0)$ (blue dashed lines in Figure~\ref{fig:near-far-field_examples}l). This complicated intersection arises due to rapid changes in gradient in the tangents to the contact curve as $\phi \rightarrow \phi_0$, but the asymptotic solution branches are less restrictive in terms of the definition of the far field that the main solution branch and so we ignore the asymptotic solution branch here. For points with $x'\geq x_{\rm min}$, whether a point is in the near or far field was determined using a bisection method to solve $x_{\rm t}(\phi, z=z')-x'=0$ for $\phi$ and comparing the root to the value of $y'$. The bounds were set so as not to include the asymptotic regions of the intersection curve, with $\phi_{\rm min}$ found by minimizing $x_{\rm t}(\phi, z=z')$ using the  Broyden–Fletcher–Goldfarb–Shanno (BFGS) algorithm as implemented in the \texttt{minimize} function from the \texttt{scipy} Python package. For points with $x_{\rm cc}(\phi_0)<x'<x_{\rm min}$, a point was determined as being in the near field if it had a $\left |y'\right|$ value above line joining the points $(x_{\rm tv}(\phi_0),y_{\rm tv}(\phi_0),z')$ and $(x_{\rm min},y_{\rm min}, z')$ (blue dotted line in Figure~\ref{fig:near-far-field_examples}k, l). Note that this line is generally very close to the set of tangents described in Equation~\ref{eqn:tangent_surface_grazing} between the two points (pink dashed line in Figure~\ref{fig:near-far-field_examples}k, l), but there can be some small deviations.

\subsection{Grazing impact with the contact curve not enclosing the $z$ axis}
\label{sec:appendix:g_notenc}

In the case that $R_{\rm t}^{\rm r} \leq R_{\rm i}^{\rm r} + B$ and $B<R_{\rm i}^{\rm r}$ the impact is grazing but the contact curve does not enclose the $z$ axis (Figure~\ref{fig:near-far-field_examples}m–p). We treat this case ostensibly the same as the grazing case where the contact curve does enclose the $z$ axis. The principle differences are that the shape of the intersection between the tangent plane and the $z=z'$ frame are truncated hyperboles, not truncated ellipses, and that there are far-field solutions for $z'>\sqrt{2B\left(b_{\rm int}+R_{\rm i}^{\rm r}\right)}$. If the $z=z'$ plane is above the highest point on the contact curve (i.e., $z'>\sqrt{2B\left(b_{\rm int}+R_{\rm i}^{\rm r}\right)}\;$) then the intersection between the tangent surface and the $z=z'$ planet is hyperbole with a maximum in $x$ at $\phi=0$ and asymptotes at $\phi=\pm\phi_{\rm lim}=\pm\cos^{-1}{\left(R_{\rm i}^{\rm r}/B\right)}$ (Figure~\ref{fig:near-far-field_examples}m).

If the $z=z'$ plane intersects the contact curve (i.e., $0 < z' \leq \sqrt{2B\left(b_{\rm int}+R_{\rm i}^{\rm r}\right)}$)  then the intersection of the tangent plane with the curve is a truncated hyperbole defined between $\phi=\phi_{t_{z'=0}}$ and $\phi=\phi_{\rm lim}$ (Figure~\ref{fig:near-far-field_examples}n). Any point with $x'<x_{\rm cc}\left ( \phi_{t_{z'=0}} \right)$ is in the near field. Otherwise we again determine whether a point is in the near or far field using a bisection method to solve $x_{\rm t}(\phi, z=z')-x'=0$ and comparing the root to the value of $y'$. 

For $z'\leq0$, the intersection of the tangent surface and the plane is in two parts due to the two-part definition of the tangent surface: one part defined by the set of tangents extended from the edge of the planet (Equation~\ref{eqn:tangent_surface_grazing}), and the other by Equation~\ref{eqn:tangent_surface} (Figure~\ref{fig:near-far-field_examples}o, p). Whether particles are in the near field based on the first of these parts has already been taken care of in the general condition 6 above. For the second set of tangents, the intersection of the tangent surface with the plane is mostly a truncated hyperbole (Figure~\ref{fig:near-far-field_examples}p) but with backwards facing asymptotic `flicks' if $z'<0$ as $\phi\rightarrow\pm\phi_0$ after the hyperbole originally reaches a minima in $x$ at $\phi_{\rm min}$ with a value $x_{\rm min}>x_{\rm cc}(\phi_0)$ (blue dashed line in Figure~\ref{fig:near-far-field_examples}p). As in the previous case, this complicated intersection arises due to rapid changes in gradient in the tangents to the contact curve as $\phi \rightarrow \phi_0$, but the asymptotic solution branches are less restrictive in terms of the definition of the far field that the main solution branch and so we ignore the asymptotic solution branch here. For points with $x'\geq x_{\rm min}$, whether a point is in the near or far field was determined using a bisection method to solve $x_{\rm t}(\phi, z=z')-x'=0$ for $\phi$ and comparing the root to the value of $y'$. The bounds were set so as not to include the asymptotic regions of the intersection curve, with $\phi_{\rm min}$ found by minimizing $x_{\rm t}(\phi, z=z')$ using the Broyden–Fletcher–Goldfarb–Shanno (BFGS) algorithm as implemented in the \texttt{minimize} function from the \texttt{scipy} Python package. For points with $x_{\rm cc}(\phi_0)<x'<x_{\rm min}$, a point was determined as being in the near field if it had a $\left |y'\right|$ value above line joining the points $(x_{\rm tv}(\phi_0),y_{\rm tv}(\phi_0),z')$ and $(x_{\rm min},y_{\rm min}, z')$ (blue dotted line in Figure~\ref{fig:near-far-field_examples}o, p). Note that this line is generally very close to the set of tangents described in Equation~\ref{eqn:tangent_surface_grazing} between the two points, but there can be some small deviations (pink dashed line in Figure~\ref{fig:near-far-field_examples}o, p). 

As in Section~\ref{sec:appendix:ng_notenc}, for numerical feasibility we do not consider solutions up to the asymptotic limits at $\phi=\phi_{\rm lim}$ but instead $\phi=(1\pm\epsilon)\phi_{\rm lim}$ where $\epsilon=10^{-13}$ is small with the sign depending on whether the limit is being approached from above or below.


\section{Impact Scenarios}
\label{sec:appendix}
\setcounter{table}{0}

The initial conditions for each simulation along with the resultant near-field, far-field, and combined atmospheric losses are listed below in Table \ref{table:initial_conditions}.

\renewcommand{\thetable}{\thesection\arabic{table}}
\begin{longtable*}{ccccccccccccc}
    \caption{Planet masses and radii (including atmosphere), mass ratios, impact velocities, impact parameters, and resultant near-field, far-field, and combined atmospheric losses for each impact scenario performed in this study. Target masses with a superscript $^{*}$ denote the simulations run without the effects of tidal deformation. A machine readable form of this table is available in the Supplement.} \\
    \toprule
    \toprule
    $M_{\rm t}^{\rm tot}$ (M$_{\oplus}$) & $R_{\rm t}^{\rm tot}$ (R$_{\oplus}$) & $M_{\rm i}^{\rm r}$ (M$_{\oplus}$) & $R_{\rm i}^{\rm r}$ (R$_{\oplus}$) & $\gamma$ $\left(M_{\rm i}^{\rm r}/M_{\rm tot}^{\rm r}\right)$ & $v_{\rm c}$ $\left(\text{km s}^{-1}\right)$ & $v_{\rm c}$ ($v_{\rm esc}$) & $b$ & $\theta$ $\left(^\circ\right)$ & $X_{\rm NF}$ & $X_{\rm FF}$ & $X_{\rm 1D}$ & $X_{\rm atm}$ \\
    \midrule
    \endfirsthead

    \toprule
    \toprule
    $M_{\rm t}^{\rm tot}$ (M$_{\oplus}$) & $R_{\rm t}^{\rm tot}$ (R$_{\oplus}$) & $M_{\rm i}^{\rm r}$ (M$_{\oplus}$) & $R_{\rm i}^{\rm r}$ (R$_{\oplus}$) & $\gamma$ $\left(M_{\rm i}^{\rm r}/M_{\rm tot}^{\rm r}\right)$ & $v_{\rm c}$ $\left(\text{km s}^{-1}\right)$ & $v_{\rm c}$ ($v_{\rm esc}$) & $b$ & $\theta$ $\left(^\circ\right)$ & $X_{\rm NF}$ & $X_{\rm FF}$ & $X_{\rm 1D}$ & $X_{\rm atm}$ \\
    \midrule
    \endhead

    \bottomrule
    \endfoot
    
    \begin{filecontents*}{data_X_NF_FF.csv}
M_t_tot,R_t_tot,M_i,R_i,gamma,v_c,v_c_v_esc,b,theta,X_NF,X_FF,X_1D,X_atm
0.367,1.154,0.039,0.371,0.1,6.75,1.0,0.0,0.0,0.018,0.002,0.000,0.020
0.367,1.154,0.039,0.371,0.1,6.75,1.0,0.3,17.5,0.039,0.002,0.000,0.041
0.367,1.154,0.039,0.371,0.1,6.75,1.0,0.5,30.0,0.041,0.001,0.000,0.042
0.367,1.154,0.039,0.371,0.1,6.75,1.0,0.7,44.4,0.037,0.004,0.000,0.041
0.367,1.154,0.039,0.371,0.1,6.75,1.0,0.9,64.2,0.023,0.006,0.000,0.029
0.367,1.154,0.039,0.371,0.1,10.12,1.5,0.0,0.0,0.154,0.069,0.015,0.224
0.367,1.154,0.039,0.371,0.1,10.12,1.5,0.3,17.5,0.130,0.027,0.010,0.157
0.367,1.154,0.039,0.371,0.1,10.12,1.5,0.5,30.0,0.130,0.011,0.003,0.141
0.367,1.154,0.039,0.371,0.1,10.12,1.5,0.7,44.4,0.101,0.019,0.001,0.120
0.367,1.154,0.039,0.371,0.1,10.12,1.5,0.9,64.2,0.060,0.018,0.000,0.078
0.367,1.154,0.039,0.371,0.1,13.50,2.0,0.0,0.0,0.222,0.140,0.129,0.362
0.367,1.154,0.039,0.371,0.1,13.50,2.0,0.3,17.5,0.195,0.111,0.091,0.306
0.367,1.154,0.039,0.371,0.1,13.50,2.0,0.5,30.0,0.194,0.053,0.034,0.248
0.367,1.154,0.039,0.371,0.1,13.50,2.0,0.7,44.4,0.142,0.030,0.006,0.172
0.367,1.154,0.039,0.371,0.1,13.50,2.0,0.9,64.2,0.082,0.021,0.000,0.103
0.367,1.154,0.039,0.371,0.1,20.25,3.0,0.0,0.0,0.231,0.531,0.556,0.762
0.367,1.154,0.039,0.371,0.1,20.25,3.0,0.3,17.5,0.240,0.378,0.395,0.618
0.367,1.154,0.039,0.371,0.1,20.25,3.0,0.5,30.0,0.256,0.185,0.181,0.441
0.367,1.154,0.039,0.371,0.1,20.25,3.0,0.7,44.4,0.189,0.070,0.044,0.259
0.367,1.154,0.039,0.371,0.1,20.25,3.0,0.9,64.2,0.109,0.039,0.001,0.148
0.367,1.154,0.087,0.483,0.2,6.81,1.0,0.0,0.0,0.091,0.008,0.001,0.098
0.367,1.154,0.087,0.483,0.2,6.81,1.0,0.3,17.5,0.079,0.002,0.000,0.081
0.367,1.154,0.087,0.483,0.2,6.81,1.0,0.5,30.0,0.060,0.001,0.000,0.061
0.367,1.154,0.087,0.483,0.2,6.81,1.0,0.7,44.4,0.048,0.006,0.000,0.054
0.367,1.154,0.087,0.483,0.2,6.81,1.0,0.9,64.2,0.023,0.002,0.000,0.025
0.367,1.154,0.087,0.483,0.2,10.21,1.5,0.0,0.0,0.279,0.071,0.036,0.350
0.367,1.154,0.087,0.483,0.2,10.21,1.5,0.3,17.5,0.241,0.024,0.017,0.265
0.367,1.154,0.087,0.483,0.2,10.21,1.5,0.5,30.0,0.191,0.021,0.006,0.212
0.367,1.154,0.087,0.483,0.2,10.21,1.5,0.7,44.4,0.132,0.028,0.001,0.160
0.367,1.154,0.087,0.483,0.2,10.21,1.5,0.9,64.2,0.071,0.022,0.000,0.093
0.367,1.154,0.087,0.483,0.2,13.61,2.0,0.0,0.0,0.298,0.287,0.245,0.585
0.367,1.154,0.087,0.483,0.2,13.61,2.0,0.3,17.5,0.335,0.140,0.129,0.475
0.367,1.154,0.087,0.483,0.2,13.61,2.0,0.5,30.0,0.262,0.092,0.056,0.354
0.367,1.154,0.087,0.483,0.2,13.61,2.0,0.7,44.4,0.177,0.041,0.009,0.218
0.367,1.154,0.087,0.483,0.2,13.61,2.0,0.9,64.2,0.093,0.025,0.000,0.118
0.367,1.154,0.087,0.483,0.2,20.42,3.0,0.0,0.0,0.301,0.622,0.631,0.923
0.367,1.154,0.087,0.483,0.2,20.42,3.0,0.3,17.5,0.360,0.440,0.447,0.799
0.367,1.154,0.087,0.483,0.2,20.42,3.0,0.5,30.0,0.317,0.260,0.263,0.577
0.367,1.154,0.087,0.483,0.2,20.42,3.0,0.7,44.4,0.225,0.101,0.073,0.326
0.367,1.154,0.087,0.483,0.2,20.42,3.0,0.9,64.2,0.120,0.050,0.002,0.170
0.367,1.154,0.15,0.575,0.3,7.00,1.0,0.0,0.0,0.171,0.051,0.001,0.222
0.367,1.154,0.15,0.575,0.3,7.00,1.0,0.3,17.5,0.107,0.005,0.000,0.112
0.367,1.154,0.15,0.575,0.3,7.00,1.0,0.5,30.0,0.077,0.005,0.000,0.081
0.367,1.154,0.15,0.575,0.3,7.00,1.0,0.7,44.4,0.053,0.005,0.000,0.058
0.367,1.154,0.15,0.575,0.3,7.00,1.0,0.9,64.2,0.021,0.002,0.000,0.023
0.367,1.154,0.15,0.575,0.3,10.50,1.5,0.0,0.0,0.343,0.093,0.057,0.436
0.367,1.154,0.15,0.575,0.3,10.50,1.5,0.3,17.5,0.313,0.035,0.027,0.349
0.367,1.154,0.15,0.575,0.3,10.50,1.5,0.5,30.0,0.239,0.055,0.009,0.294
0.367,1.154,0.15,0.575,0.3,10.50,1.5,0.7,44.4,0.156,0.034,0.001,0.191
0.367,1.154,0.15,0.575,0.3,10.50,1.5,0.9,64.2,0.079,0.023,0.000,0.102
0.367,1.154,0.15,0.575,0.3,14.00,2.0,0.0,0.0,0.363,0.321,0.298,0.684
0.367,1.154,0.15,0.575,0.3,14.00,2.0,0.3,17.5,0.408,0.203,0.170,0.611
0.367,1.154,0.15,0.575,0.3,14.00,2.0,0.5,30.0,0.317,0.120,0.077,0.436
0.367,1.154,0.15,0.575,0.3,14.00,2.0,0.7,44.4,0.204,0.052,0.012,0.255
0.367,1.154,0.15,0.575,0.3,14.00,2.0,0.9,64.2,0.103,0.027,0.001,0.130
0.367,1.154,0.15,0.575,0.3,21.00,3.0,0.0,0.0,0.372,0.588,0.563,0.961
0.367,1.154,0.15,0.575,0.3,21.00,3.0,0.3,17.5,0.420,0.483,0.465,0.903
0.367,1.154,0.15,0.575,0.3,21.00,3.0,0.5,30.0,0.358,0.325,0.323,0.683
0.367,1.154,0.15,0.575,0.3,21.00,3.0,0.7,44.4,0.252,0.135,0.103,0.388
0.367,1.154,0.15,0.575,0.3,21.00,3.0,0.9,64.2,0.129,0.057,0.003,0.186
0.367,1.154,0.232,0.66,0.4,7.31,1.0,0.0,0.0,0.221,0.073,0.001,0.294
0.367,1.154,0.232,0.66,0.4,7.31,1.0,0.3,17.5,0.145,0.008,0.001,0.153
0.367,1.154,0.232,0.66,0.4,7.31,1.0,0.5,30.0,0.107,0.017,0.000,0.124
0.367,1.154,0.232,0.66,0.4,7.31,1.0,0.7,44.4,0.058,0.003,0.000,0.061
0.367,1.154,0.232,0.66,0.4,7.31,1.0,0.9,64.2,0.019,0.000,0.000,0.020
0.367,1.154,0.232,0.66,0.4,10.96,1.5,0.0,0.0,0.399,0.095,0.075,0.495
0.367,1.154,0.232,0.66,0.4,10.96,1.5,0.3,17.5,0.359,0.047,0.041,0.407
0.367,1.154,0.232,0.66,0.4,10.96,1.5,0.5,30.0,0.296,0.078,0.014,0.374
0.367,1.154,0.232,0.66,0.4,10.96,1.5,0.7,44.4,0.183,0.045,0.002,0.228
0.367,1.154,0.232,0.66,0.4,10.96,1.5,0.9,64.2,0.090,0.024,0.000,0.113
0.367,1.154,0.232,0.66,0.4,14.62,2.0,0.0,0.0,0.448,0.293,0.313,0.741
0.367,1.154,0.232,0.66,0.4,14.62,2.0,0.3,17.5,0.453,0.257,0.209,0.710
0.367,1.154,0.232,0.66,0.4,14.62,2.0,0.5,30.0,0.364,0.157,0.103,0.522
0.367,1.154,0.232,0.66,0.4,14.62,2.0,0.7,44.4,0.232,0.068,0.017,0.300
0.367,1.154,0.232,0.66,0.4,14.62,2.0,0.9,64.2,0.114,0.030,0.001,0.144
0.367,1.154,0.232,0.66,0.4,21.93,3.0,0.0,0.0,0.455,0.539,0.481,0.994
0.367,1.154,0.232,0.66,0.4,21.93,3.0,0.3,17.5,0.464,0.488,0.456,0.952
0.367,1.154,0.232,0.66,0.4,21.93,3.0,0.5,30.0,0.389,0.392,0.374,0.781
0.367,1.154,0.232,0.66,0.4,21.93,3.0,0.7,44.4,0.276,0.176,0.135,0.453
0.367,1.154,0.232,0.66,0.4,21.93,3.0,0.9,64.2,0.139,0.068,0.005,0.207
0.367,1.154,0.349,0.747,0.5,7.75,1.0,0.0,0.0,0.250,0.058,0.002,0.308
0.367,1.154,0.349,0.747,0.5,7.75,1.0,0.3,17.5,0.196,0.024,0.001,0.220
0.367,1.154,0.349,0.747,0.5,7.75,1.0,0.5,30.0,0.150,0.032,0.001,0.183
0.367,1.154,0.349,0.747,0.5,7.75,1.0,0.7,44.4,0.069,0.008,0.000,0.077
0.367,1.154,0.349,0.747,0.5,7.75,1.0,0.9,64.2,0.020,0.000,0.000,0.020
0.367,1.154,0.349,0.747,0.5,11.63,1.5,0.0,0.0,0.452,0.082,0.084,0.534
0.367,1.154,0.349,0.747,0.5,11.63,1.5,0.3,17.5,0.385,0.080,0.058,0.466
0.367,1.154,0.349,0.747,0.5,11.63,1.5,0.5,30.0,0.352,0.114,0.023,0.466
0.367,1.154,0.349,0.747,0.5,11.63,1.5,0.7,44.4,0.215,0.060,0.003,0.276
0.367,1.154,0.349,0.747,0.5,11.63,1.5,0.9,64.2,0.101,0.027,0.000,0.128
0.367,1.154,0.349,0.747,0.5,15.51,2.0,0.0,0.0,0.564,0.182,0.272,0.746
0.367,1.154,0.349,0.747,0.5,15.51,2.0,0.3,17.5,0.497,0.303,0.240,0.800
0.367,1.154,0.349,0.747,0.5,15.51,2.0,0.5,30.0,0.403,0.220,0.137,0.623
0.367,1.154,0.349,0.747,0.5,15.51,2.0,0.7,44.4,0.264,0.092,0.026,0.356
0.367,1.154,0.349,0.747,0.5,15.51,2.0,0.9,64.2,0.126,0.036,0.001,0.162
0.367,1.154,0.349,0.747,0.5,23.26,3.0,0.0,0.0,0.569,0.429,0.366,0.998
0.367,1.154,0.349,0.747,0.5,23.26,3.0,0.3,17.5,0.506,0.473,0.425,0.979
0.367,1.154,0.349,0.747,0.5,23.26,3.0,0.5,30.0,0.414,0.459,0.419,0.873
0.367,1.154,0.349,0.747,0.5,23.26,3.0,0.7,44.4,0.299,0.234,0.176,0.533
0.367,1.154,0.349,0.747,0.5,23.26,3.0,0.9,64.2,0.149,0.080,0.007,0.229
,,,,,,,,,,,,
1.05,1.375,0.111,0.518,0.1,9.76,1.0,0.0,0.0,0.005,0.000,0.002,0.005
1.05,1.375,0.111,0.518,0.1,9.76,1.0,0.3,17.5,0.020,0.000,0.002,0.020
1.05,1.375,0.111,0.518,0.1,9.76,1.0,0.5,30.0,0.027,0.000,0.001,0.027
1.05,1.375,0.111,0.518,0.1,9.76,1.0,0.7,44.4,0.029,0.002,0.000,0.031
1.05,1.375,0.111,0.518,0.1,9.76,1.0,0.9,64.2,0.018,0.004,0.000,0.022
1.05,1.375,0.111,0.518,0.1,14.63,1.5,0.0,0.0,0.117,0.068,0.062,0.185
1.05,1.375,0.111,0.518,0.1,14.63,1.5,0.3,17.5,0.117,0.034,0.048,0.151
1.05,1.375,0.111,0.518,0.1,14.63,1.5,0.5,30.0,0.117,0.013,0.020,0.130
1.05,1.375,0.111,0.518,0.1,14.63,1.5,0.7,44.4,0.091,0.017,0.004,0.109
1.05,1.375,0.111,0.518,0.1,14.63,1.5,0.9,64.2,0.049,0.019,0.000,0.067
1.05,1.375,0.111,0.518,0.1,19.51,2.0,0.0,0.0,0.209,0.219,0.244,0.428
1.05,1.375,0.111,0.518,0.1,19.51,2.0,0.3,17.5,0.178,0.159,0.177,0.337
1.05,1.375,0.111,0.518,0.1,19.51,2.0,0.5,30.0,0.180,0.070,0.080,0.250
1.05,1.375,0.111,0.518,0.1,19.51,2.0,0.7,44.4,0.124,0.025,0.018,0.149
1.05,1.375,0.111,0.518,0.1,19.51,2.0,0.9,64.2,0.064,0.019,0.001,0.083
1.05,1.375,0.111,0.518,0.1,29.27,3.0,0.0,0.0,0.212,0.609,0.618,0.821
1.05,1.375,0.111,0.518,0.1,29.27,3.0,0.3,17.5,0.224,0.432,0.454,0.656
1.05,1.375,0.111,0.518,0.1,29.27,3.0,0.5,30.0,0.232,0.222,0.239,0.454
1.05,1.375,0.111,0.518,0.1,29.27,3.0,0.7,44.4,0.166,0.069,0.071,0.235
1.05,1.375,0.111,0.518,0.1,29.27,3.0,0.9,64.2,0.086,0.034,0.002,0.120
1.05,1.375,0.249,0.67,0.2,9.85,1.0,0.0,0.0,0.025,0.003,0.006,0.028
1.05,1.375,0.249,0.67,0.2,9.85,1.0,0.3,17.5,0.050,0.004,0.003,0.053
1.05,1.375,0.249,0.67,0.2,9.85,1.0,0.5,30.0,0.037,0.000,0.001,0.037
1.05,1.375,0.249,0.67,0.2,9.85,1.0,0.7,44.4,0.038,0.005,0.001,0.042
1.05,1.375,0.249,0.67,0.2,9.85,1.0,0.9,64.2,0.017,0.001,0.000,0.018
1.05,1.375,0.249,0.67,0.2,14.77,1.5,0.0,0.0,0.262,0.099,0.129,0.361
1.05,1.375,0.249,0.67,0.2,14.77,1.5,0.3,17.5,0.219,0.033,0.068,0.252
1.05,1.375,0.249,0.67,0.2,14.77,1.5,0.5,30.0,0.169,0.016,0.032,0.185
1.05,1.375,0.249,0.67,0.2,14.77,1.5,0.7,44.4,0.116,0.026,0.006,0.141
1.05,1.375,0.249,0.67,0.2,14.77,1.5,0.9,64.2,0.055,0.021,0.000,0.076
1.05,1.375,0.249,0.67,0.2,19.69,2.0,0.0,0.0,0.283,0.385,0.387,0.668
1.05,1.375,0.249,0.67,0.2,19.69,2.0,0.3,17.5,0.316,0.203,0.230,0.520
1.05,1.375,0.249,0.67,0.2,19.69,2.0,0.5,30.0,0.241,0.116,0.121,0.357
1.05,1.375,0.249,0.67,0.2,19.69,2.0,0.7,44.4,0.153,0.037,0.026,0.190
1.05,1.375,0.249,0.67,0.2,19.69,2.0,0.9,64.2,0.074,0.023,0.001,0.096
1.05,1.375,0.249,0.67,0.2,29.54,3.0,0.0,0.0,0.285,0.679,0.660,0.963
1.05,1.375,0.249,0.67,0.2,29.54,3.0,0.3,17.5,0.342,0.495,0.494,0.836
1.05,1.375,0.249,0.67,0.2,29.54,3.0,0.5,30.0,0.290,0.311,0.325,0.601
1.05,1.375,0.249,0.67,0.2,29.54,3.0,0.7,44.4,0.199,0.104,0.111,0.303
1.05,1.375,0.249,0.67,0.2,29.54,3.0,0.9,64.2,0.096,0.040,0.003,0.136
1.05,1.375,0.428,0.79,0.3,10.14,1.0,0.0,0.0,0.115,0.020,0.010,0.136
1.05,1.375,0.428,0.79,0.3,10.14,1.0,0.3,17.5,0.071,0.002,0.005,0.074
1.05,1.375,0.428,0.79,0.3,10.14,1.0,0.5,30.0,0.052,0.002,0.002,0.054
1.05,1.375,0.428,0.79,0.3,10.14,1.0,0.7,44.4,0.041,0.003,0.001,0.044
1.05,1.375,0.428,0.79,0.3,10.14,1.0,0.9,64.2,0.014,0.001,0.000,0.015
1.05,1.375,0.428,0.79,0.3,15.21,1.5,0.0,0.0,0.339,0.105,0.179,0.444
1.05,1.375,0.428,0.79,0.3,15.21,1.5,0.3,17.5,0.283,0.040,0.096,0.324
1.05,1.375,0.428,0.79,0.3,15.21,1.5,0.5,30.0,0.214,0.064,0.043,0.278
1.05,1.375,0.428,0.79,0.3,15.21,1.5,0.7,44.4,0.136,0.033,0.008,0.170
1.05,1.375,0.428,0.79,0.3,15.21,1.5,0.9,64.2,0.061,0.022,0.001,0.083
1.05,1.375,0.428,0.79,0.3,20.28,2.0,0.0,0.0,0.351,0.401,0.425,0.752
1.05,1.375,0.428,0.79,0.3,20.28,2.0,0.3,17.5,0.387,0.275,0.278,0.662
1.05,1.375,0.428,0.79,0.3,20.28,2.0,0.5,30.0,0.288,0.156,0.154,0.445
1.05,1.375,0.428,0.79,0.3,20.28,2.0,0.7,44.4,0.179,0.050,0.035,0.229
1.05,1.375,0.428,0.79,0.3,20.28,2.0,0.9,64.2,0.082,0.025,0.002,0.107
1.05,1.375,0.428,0.79,0.3,30.42,3.0,0.0,0.0,0.357,0.642,0.588,1.000
1.05,1.375,0.428,0.79,0.3,30.42,3.0,0.3,17.5,0.397,0.535,0.504,0.932
1.05,1.375,0.428,0.79,0.3,30.42,3.0,0.5,30.0,0.332,0.380,0.384,0.712
1.05,1.375,0.428,0.79,0.3,30.42,3.0,0.7,44.4,0.224,0.145,0.149,0.369
1.05,1.375,0.428,0.79,0.3,30.42,3.0,0.9,64.2,0.104,0.046,0.005,0.150
1.05,1.375,0.665,0.902,0.4,10.60,1.0,0.0,0.0,0.159,0.050,0.016,0.208
1.05,1.375,0.665,0.902,0.4,10.60,1.0,0.3,17.5,0.104,0.010,0.008,0.114
1.05,1.375,0.665,0.902,0.4,10.60,1.0,0.5,30.0,0.072,0.017,0.003,0.089
1.05,1.375,0.665,0.902,0.4,10.60,1.0,0.7,44.4,0.042,0.002,0.001,0.044
1.05,1.375,0.665,0.902,0.4,10.60,1.0,0.9,64.2,0.014,0.000,0.000,0.014
1.05,1.375,0.665,0.902,0.4,15.90,1.5,0.0,0.0,0.409,0.069,0.202,0.478
1.05,1.375,0.665,0.902,0.4,15.90,1.5,0.3,17.5,0.338,0.056,0.123,0.394
1.05,1.375,0.665,0.902,0.4,15.90,1.5,0.5,30.0,0.263,0.088,0.057,0.352
1.05,1.375,0.665,0.902,0.4,15.90,1.5,0.7,44.4,0.160,0.044,0.010,0.204
1.05,1.375,0.665,0.902,0.4,15.90,1.5,0.9,64.2,0.070,0.024,0.001,0.094
1.05,1.375,0.665,0.902,0.4,21.20,2.0,0.0,0.0,0.433,0.375,0.408,0.808
1.05,1.375,0.665,0.902,0.4,21.20,2.0,0.3,17.5,0.432,0.336,0.310,0.768
1.05,1.375,0.665,0.902,0.4,21.20,2.0,0.5,30.0,0.336,0.202,0.187,0.538
1.05,1.375,0.665,0.902,0.4,21.20,2.0,0.7,44.4,0.204,0.067,0.046,0.271
1.05,1.375,0.665,0.902,0.4,21.20,2.0,0.9,64.2,0.090,0.028,0.002,0.118
1.05,1.375,0.665,0.902,0.4,31.81,3.0,0.0,0.0,0.440,0.560,0.503,1.000
1.05,1.375,0.665,0.902,0.4,31.81,3.0,0.3,17.5,0.439,0.538,0.492,0.977
1.05,1.375,0.665,0.902,0.4,31.81,3.0,0.5,30.0,0.361,0.447,0.429,0.808
1.05,1.375,0.665,0.902,0.4,31.81,3.0,0.7,44.4,0.248,0.190,0.185,0.437
1.05,1.375,0.665,0.902,0.4,31.81,3.0,0.9,64.2,0.113,0.052,0.007,0.165
1.05,1.375,0.997,1.013,0.5,11.26,1.0,0.0,0.0,0.199,0.009,0.022,0.208
1.05,1.375,0.997,1.013,0.5,11.26,1.0,0.3,17.5,0.146,0.020,0.013,0.165
1.05,1.375,0.997,1.013,0.5,11.26,1.0,0.5,30.0,0.096,0.018,0.005,0.114
1.05,1.375,0.997,1.013,0.5,11.26,1.0,0.7,44.4,0.049,0.003,0.001,0.052
1.05,1.375,0.997,1.013,0.5,11.26,1.0,0.9,64.2,0.014,0.000,0.000,0.015
1.05,1.375,0.997,1.013,0.5,16.89,1.5,0.0,0.0,0.451,0.033,0.187,0.484
1.05,1.375,0.997,1.013,0.5,16.89,1.5,0.3,17.5,0.368,0.081,0.152,0.449
1.05,1.375,0.997,1.013,0.5,16.89,1.5,0.5,30.0,0.317,0.121,0.077,0.437
1.05,1.375,0.997,1.013,0.5,16.89,1.5,0.7,44.4,0.189,0.060,0.014,0.249
1.05,1.375,0.997,1.013,0.5,16.89,1.5,0.9,64.2,0.079,0.026,0.001,0.105
1.05,1.375,0.997,1.013,0.5,22.52,2.0,0.0,0.0,0.549,0.264,0.337,0.812
1.05,1.375,0.997,1.013,0.5,22.52,2.0,0.3,17.5,0.472,0.386,0.330,0.857
1.05,1.375,0.997,1.013,0.5,22.52,2.0,0.5,30.0,0.375,0.274,0.227,0.649
1.05,1.375,0.997,1.013,0.5,22.52,2.0,0.7,44.4,0.233,0.092,0.061,0.325
1.05,1.375,0.997,1.013,0.5,22.52,2.0,0.9,64.2,0.101,0.032,0.003,0.133
1.05,1.375,0.997,1.013,0.5,33.77,3.0,0.0,0.0,0.558,0.442,0.385,1.000
1.05,1.375,0.997,1.013,0.5,33.77,3.0,0.3,17.5,0.478,0.519,0.461,0.997
1.05,1.375,0.997,1.013,0.5,33.77,3.0,0.5,30.0,0.386,0.513,0.463,0.899
1.05,1.375,0.997,1.013,0.5,33.77,3.0,0.7,44.4,0.270,0.253,0.229,0.523
1.05,1.375,0.997,1.013,0.5,33.77,3.0,0.9,64.2,0.122,0.061,0.010,0.183
,,,,,,,,,,,,
1.050$^{*}$,1.375,0.111,0.518,0.1,9.76,1.0,0.0,0.0,--,--,--,0.021
1.050$^{*}$,1.375,0.111,0.518,0.1,9.76,1.0,0.3,17.5,--,--,--,0.039
1.050$^{*}$,1.375,0.111,0.518,0.1,9.76,1.0,0.5,30.0,--,--,--,0.048
1.050$^{*}$,1.375,0.111,0.518,0.1,14.63,1.5,0.0,0.0,--,--,--,0.201
1.050$^{*}$,1.375,0.111,0.518,0.1,14.63,1.5,0.3,17.5,--,--,--,0.163
1.050$^{*}$,1.375,0.111,0.518,0.1,14.63,1.5,0.5,30.0,--,--,--,0.138
1.050$^{*}$,1.375,0.111,0.518,0.1,14.63,1.5,0.7,44.4,--,--,--,0.120
1.050$^{*}$,1.375,0.111,0.518,0.1,14.63,1.5,0.9,64.2,--,--,--,0.087
1.050$^{*}$,1.375,0.111,0.518,0.1,19.51,2.0,0.0,0.0,--,--,--,0.436
1.050$^{*}$,1.375,0.111,0.518,0.1,19.51,2.0,0.3,17.5,--,--,--,0.341
1.050$^{*}$,1.375,0.111,0.518,0.1,19.51,2.0,0.5,30.0,--,--,--,0.252
1.050$^{*}$,1.375,0.111,0.518,0.1,19.51,2.0,0.7,44.4,--,--,--,0.156
1.050$^{*}$,1.375,0.111,0.518,0.1,19.51,2.0,0.9,64.2,--,--,--,0.090
1.050$^{*}$,1.375,0.111,0.518,0.1,29.27,3.0,0.0,0.0,--,--,--,0.817
1.050$^{*}$,1.375,0.111,0.518,0.1,29.27,3.0,0.3,17.5,--,--,--,0.659
1.050$^{*}$,1.375,0.111,0.518,0.1,29.27,3.0,0.5,30.0,--,--,--,0.455
1.050$^{*}$,1.375,0.111,0.518,0.1,29.27,3.0,0.7,44.4,--,--,--,0.238
1.050$^{*}$,1.375,0.111,0.518,0.1,29.27,3.0,0.9,64.2,--,--,--,0.123
1.050$^{*}$,1.375,0.249,0.67,0.2,9.85,1.0,0.0,0.0,--,--,--,0.111
1.050$^{*}$,1.375,0.249,0.67,0.2,9.85,1.0,0.3,17.5,--,--,--,0.078
1.050$^{*}$,1.375,0.249,0.67,0.2,9.85,1.0,0.5,30.0,--,--,--,0.070
1.050$^{*}$,1.375,0.249,0.67,0.2,14.77,1.5,0.0,0.0,--,--,--,0.384
1.050$^{*}$,1.375,0.249,0.67,0.2,14.77,1.5,0.3,17.5,--,--,--,0.269
1.050$^{*}$,1.375,0.249,0.67,0.2,14.77,1.5,0.5,30.0,--,--,--,0.205
1.050$^{*}$,1.375,0.249,0.67,0.2,14.77,1.5,0.7,44.4,--,--,--,0.151
1.050$^{*}$,1.375,0.249,0.67,0.2,14.77,1.5,0.9,64.2,--,--,--,0.094
1.050$^{*}$,1.375,0.249,0.67,0.2,19.69,2.0,0.0,0.0,--,--,--,0.677
1.050$^{*}$,1.375,0.249,0.67,0.2,19.69,2.0,0.3,17.5,--,--,--,0.527
1.050$^{*}$,1.375,0.249,0.67,0.2,19.69,2.0,0.5,30.0,--,--,--,0.366
1.050$^{*}$,1.375,0.249,0.67,0.2,19.69,2.0,0.7,44.4,--,--,--,0.199
1.050$^{*}$,1.375,0.249,0.67,0.2,19.69,2.0,0.9,64.2,--,--,--,0.102
1.050$^{*}$,1.375,0.249,0.67,0.2,29.54,3.0,0.0,0.0,--,--,--,0.964
1.050$^{*}$,1.375,0.249,0.67,0.2,29.54,3.0,0.3,17.5,--,--,--,0.839
1.050$^{*}$,1.375,0.249,0.67,0.2,29.54,3.0,0.5,30.0,--,--,--,0.605
1.050$^{*}$,1.375,0.249,0.67,0.2,29.54,3.0,0.7,44.4,--,--,--,0.309
1.050$^{*}$,1.375,0.249,0.67,0.2,29.54,3.0,0.9,64.2,--,--,--,0.141
1.050$^{*}$,1.375,0.428,0.79,0.3,10.14,1.0,0.0,0.0,--,--,--,0.190
1.050$^{*}$,1.375,0.428,0.79,0.3,10.14,1.0,0.3,17.5,--,--,--,0.105
1.050$^{*}$,1.375,0.428,0.79,0.3,10.14,1.0,0.5,30.0,--,--,--,0.082
1.050$^{*}$,1.375,0.428,0.79,0.3,10.14,1.0,0.7,44.4,--,--,--,0.101
1.050$^{*}$,1.375,0.428,0.79,0.3,15.21,1.5,0.0,0.0,--,--,--,0.473
1.050$^{*}$,1.375,0.428,0.79,0.3,15.21,1.5,0.3,17.5,--,--,--,0.356
1.050$^{*}$,1.375,0.428,0.79,0.3,15.21,1.5,0.5,30.0,--,--,--,0.285
1.050$^{*}$,1.375,0.428,0.79,0.3,15.21,1.5,0.7,44.4,--,--,--,0.183
1.050$^{*}$,1.375,0.428,0.79,0.3,15.21,1.5,0.9,64.2,--,--,--,0.097
1.050$^{*}$,1.375,0.428,0.79,0.3,20.28,2.0,0.0,0.0,--,--,--,0.760
1.050$^{*}$,1.375,0.428,0.79,0.3,20.28,2.0,0.3,17.5,--,--,--,0.672
1.050$^{*}$,1.375,0.428,0.79,0.3,20.28,2.0,0.5,30.0,--,--,--,0.455
1.050$^{*}$,1.375,0.428,0.79,0.3,20.28,2.0,0.7,44.4,--,--,--,0.239
1.050$^{*}$,1.375,0.428,0.79,0.3,20.28,2.0,0.9,64.2,--,--,--,0.115
1.050$^{*}$,1.375,0.428,0.79,0.3,30.42,3.0,0.0,0.0,--,--,--,1.000
1.050$^{*}$,1.375,0.428,0.79,0.3,30.42,3.0,0.3,17.5,--,--,--,0.934
1.050$^{*}$,1.375,0.428,0.79,0.3,30.42,3.0,0.5,30.0,--,--,--,0.715
1.050$^{*}$,1.375,0.428,0.79,0.3,30.42,3.0,0.7,44.4,--,--,--,0.375
1.050$^{*}$,1.375,0.428,0.79,0.3,30.42,3.0,0.9,64.2,--,--,--,0.156
1.050$^{*}$,1.375,0.665,0.902,0.4,10.60,1.0,0.0,0.0,--,--,--,0.238
1.050$^{*}$,1.375,0.665,0.902,0.4,10.60,1.0,0.3,17.5,--,--,--,0.146
1.050$^{*}$,1.375,0.665,0.902,0.4,10.60,1.0,0.5,30.0,--,--,--,0.126
1.050$^{*}$,1.375,0.665,0.902,0.4,10.60,1.0,0.7,44.4,--,--,--,0.097
1.050$^{*}$,1.375,0.665,0.902,0.4,15.90,1.5,0.0,0.0,--,--,--,0.511
1.050$^{*}$,1.375,0.665,0.902,0.4,15.90,1.5,0.3,17.5,--,--,--,0.414
1.050$^{*}$,1.375,0.665,0.902,0.4,15.90,1.5,0.5,30.0,--,--,--,0.376
1.050$^{*}$,1.375,0.665,0.902,0.4,15.90,1.5,0.7,44.4,--,--,--,0.215
1.050$^{*}$,1.375,0.665,0.902,0.4,15.90,1.5,0.9,64.2,--,--,--,0.102
1.050$^{*}$,1.375,0.665,0.902,0.4,21.20,2.0,0.0,0.0,--,--,--,0.815
1.050$^{*}$,1.375,0.665,0.902,0.4,21.20,2.0,0.3,17.5,--,--,--,0.779
1.050$^{*}$,1.375,0.665,0.902,0.4,21.20,2.0,0.5,30.0,--,--,--,0.551
1.050$^{*}$,1.375,0.665,0.902,0.4,21.20,2.0,0.7,44.4,--,--,--,0.285
1.050$^{*}$,1.375,0.665,0.902,0.4,21.20,2.0,0.9,64.2,--,--,--,0.125
1.050$^{*}$,1.375,0.665,0.902,0.4,31.81,3.0,0.0,0.0,--,--,--,1.000
1.050$^{*}$,1.375,0.665,0.902,0.4,31.81,3.0,0.3,17.5,--,--,--,0.979
1.050$^{*}$,1.375,0.665,0.902,0.4,31.81,3.0,0.5,30.0,--,--,--,0.811
1.050$^{*}$,1.375,0.665,0.902,0.4,31.81,3.0,0.7,44.4,--,--,--,0.446
1.050$^{*}$,1.375,0.665,0.902,0.4,31.81,3.0,0.9,64.2,--,--,--,0.171
1.050$^{*}$,1.375,0.997,1.013,0.5,11.26,1.0,0.0,0.0,--,--,--,0.237
1.050$^{*}$,1.375,0.997,1.013,0.5,11.26,1.0,0.3,17.5,--,--,--,0.199
1.050$^{*}$,1.375,0.997,1.013,0.5,11.26,1.0,0.5,30.0,--,--,--,0.157
1.050$^{*}$,1.375,0.997,1.013,0.5,11.26,1.0,0.7,44.4,--,--,--,0.087
1.050$^{*}$,1.375,0.997,1.013,0.5,16.89,1.5,0.0,0.0,--,--,--,0.514
1.050$^{*}$,1.375,0.997,1.013,0.5,16.89,1.5,0.3,17.5,--,--,--,0.473
1.050$^{*}$,1.375,0.997,1.013,0.5,16.89,1.5,0.5,30.0,--,--,--,0.468
1.050$^{*}$,1.375,0.997,1.013,0.5,16.89,1.5,0.7,44.4,--,--,--,0.263
1.050$^{*}$,1.375,0.997,1.013,0.5,16.89,1.5,0.9,64.2,--,--,--,0.113
1.050$^{*}$,1.375,0.997,1.013,0.5,22.52,2.0,0.0,0.0,--,--,--,0.833
1.050$^{*}$,1.375,0.997,1.013,0.5,22.52,2.0,0.3,17.5,--,--,--,0.867
1.050$^{*}$,1.375,0.997,1.013,0.5,22.52,2.0,0.5,30.0,--,--,--,0.666
1.050$^{*}$,1.375,0.997,1.013,0.5,22.52,2.0,0.7,44.4,--,--,--,0.343
1.050$^{*}$,1.375,0.997,1.013,0.5,22.52,2.0,0.9,64.2,--,--,--,0.141
1.050$^{*}$,1.375,0.997,1.013,0.5,33.77,3.0,0.0,0.0,--,--,--,1.000
1.050$^{*}$,1.375,0.997,1.013,0.5,33.77,3.0,0.3,17.5,--,--,--,0.997
1.050$^{*}$,1.375,0.997,1.013,0.5,33.77,3.0,0.5,30.0,--,--,--,0.901
1.050$^{*}$,1.375,0.997,1.013,0.5,33.77,3.0,0.7,44.4,--,--,--,0.534
1.050$^{*}$,1.375,0.997,1.013,0.5,33.77,3.0,0.9,64.2,--,--,--,0.191
,,,,,,,,,,,,
2.098,1.571,0.221,0.642,0.1,12.50,1.0,0.0,0.0,0.003,0.000,0.001,0.003
2.098,1.571,0.221,0.642,0.1,12.50,1.0,0.3,17.5,0.011,0.000,0.005,0.011
2.098,1.571,0.221,0.642,0.1,12.50,1.0,0.5,30.0,0.021,0.000,0.002,0.021
2.098,1.571,0.221,0.642,0.1,12.50,1.0,0.7,44.4,0.023,0.002,0.001,0.025
2.098,1.571,0.221,0.642,0.1,12.50,1.0,0.9,64.2,0.017,0.002,0.000,0.019
2.098,1.571,0.221,0.642,0.1,18.74,1.5,0.0,0.0,0.084,0.049,0.067,0.133
2.098,1.571,0.221,0.642,0.1,18.74,1.5,0.3,17.5,0.116,0.038,0.071,0.154
2.098,1.571,0.221,0.642,0.1,18.74,1.5,0.5,30.0,0.111,0.015,0.032,0.125
2.098,1.571,0.221,0.642,0.1,18.74,1.5,0.7,44.4,0.086,0.016,0.007,0.102
2.098,1.571,0.221,0.642,0.1,18.74,1.5,0.9,64.2,0.044,0.020,0.000,0.064
2.098,1.571,0.221,0.642,0.1,24.99,2.0,0.0,0.0,0.201,0.251,0.260,0.452
2.098,1.571,0.221,0.642,0.1,24.99,2.0,0.3,17.5,0.174,0.179,0.201,0.352
2.098,1.571,0.221,0.642,0.1,24.99,2.0,0.5,30.0,0.173,0.078,0.097,0.251
2.098,1.571,0.221,0.642,0.1,24.99,2.0,0.7,44.4,0.117,0.024,0.022,0.140
2.098,1.571,0.221,0.642,0.1,24.99,2.0,0.9,64.2,0.056,0.021,0.001,0.077
2.098,1.571,0.221,0.642,0.1,37.49,3.0,0.0,0.0,0.206,0.643,0.614,0.849
2.098,1.571,0.221,0.642,0.1,37.49,3.0,0.3,17.5,0.221,0.455,0.453,0.676
2.098,1.571,0.221,0.642,0.1,37.49,3.0,0.5,30.0,0.223,0.237,0.249,0.460
2.098,1.571,0.221,0.642,0.1,37.49,3.0,0.7,44.4,0.154,0.064,0.073,0.217
2.098,1.571,0.221,0.642,0.1,37.49,3.0,0.9,64.2,0.073,0.034,0.003,0.107
2.098,1.571,0.498,0.823,0.2,12.62,1.0,0.0,0.0,0.019,0.006,0.005,0.024
2.098,1.571,0.498,0.823,0.2,12.62,1.0,0.3,17.5,0.038,0.002,0.007,0.040
2.098,1.571,0.498,0.823,0.2,12.62,1.0,0.5,30.0,0.031,0.000,0.003,0.032
2.098,1.571,0.498,0.823,0.2,12.62,1.0,0.7,44.4,0.033,0.007,0.001,0.039
2.098,1.571,0.498,0.823,0.2,12.62,1.0,0.9,64.2,0.014,0.000,0.000,0.015
2.098,1.571,0.498,0.823,0.2,18.93,1.5,0.0,0.0,0.253,0.115,0.170,0.368
2.098,1.571,0.498,0.823,0.2,18.93,1.5,0.3,17.5,0.211,0.038,0.099,0.250
2.098,1.571,0.498,0.823,0.2,18.93,1.5,0.5,30.0,0.159,0.019,0.049,0.178
2.098,1.571,0.498,0.823,0.2,18.93,1.5,0.7,44.4,0.109,0.025,0.010,0.134
2.098,1.571,0.498,0.823,0.2,18.93,1.5,0.9,64.2,0.049,0.023,0.001,0.072
2.098,1.571,0.498,0.823,0.2,25.24,2.0,0.0,0.0,0.276,0.430,0.410,0.706
2.098,1.571,0.498,0.823,0.2,25.24,2.0,0.3,17.5,0.307,0.229,0.256,0.536
2.098,1.571,0.498,0.823,0.2,25.24,2.0,0.5,30.0,0.231,0.127,0.142,0.358
2.098,1.571,0.498,0.823,0.2,25.24,2.0,0.7,44.4,0.142,0.033,0.033,0.176
2.098,1.571,0.498,0.823,0.2,25.24,2.0,0.9,64.2,0.063,0.024,0.002,0.087
2.098,1.571,0.498,0.823,0.2,37.86,3.0,0.0,0.0,0.282,0.716,0.665,0.997
2.098,1.571,0.498,0.823,0.2,37.86,3.0,0.3,17.5,0.339,0.522,0.495,0.862
2.098,1.571,0.498,0.823,0.2,37.86,3.0,0.5,30.0,0.278,0.331,0.334,0.609
2.098,1.571,0.498,0.823,0.2,37.86,3.0,0.7,44.4,0.184,0.103,0.116,0.288
2.098,1.571,0.498,0.823,0.2,37.86,3.0,0.9,64.2,0.083,0.039,0.004,0.122
2.098,1.571,0.855,0.966,0.3,13.01,1.0,0.0,0.0,0.074,0.027,0.014,0.101
2.098,1.571,0.855,0.966,0.3,13.01,1.0,0.3,17.5,0.057,0.002,0.013,0.059
2.098,1.571,0.855,0.966,0.3,13.01,1.0,0.5,30.0,0.043,0.001,0.006,0.044
2.098,1.571,0.855,0.966,0.3,13.01,1.0,0.7,44.4,0.034,0.002,0.002,0.036
2.098,1.571,0.855,0.966,0.3,13.01,1.0,0.9,64.2,0.013,0.001,0.000,0.014
2.098,1.571,0.855,0.966,0.3,19.51,1.5,0.0,0.0,0.337,0.120,0.222,0.457
2.098,1.571,0.855,0.966,0.3,19.51,1.5,0.3,17.5,0.270,0.035,0.132,0.305
2.098,1.571,0.855,0.966,0.3,19.51,1.5,0.5,30.0,0.203,0.066,0.064,0.270
2.098,1.571,0.855,0.966,0.3,19.51,1.5,0.7,44.4,0.127,0.033,0.012,0.160
2.098,1.571,0.855,0.966,0.3,19.51,1.5,0.9,64.2,0.054,0.024,0.001,0.078
2.098,1.571,0.855,0.966,0.3,26.02,2.0,0.0,0.0,0.346,0.444,0.450,0.790
2.098,1.571,0.855,0.966,0.3,26.02,2.0,0.3,17.5,0.377,0.308,0.304,0.684
2.098,1.571,0.855,0.966,0.3,26.02,2.0,0.5,30.0,0.276,0.169,0.177,0.445
2.098,1.571,0.855,0.966,0.3,26.02,2.0,0.7,44.4,0.166,0.048,0.047,0.214
2.098,1.571,0.855,0.966,0.3,26.02,2.0,0.9,64.2,0.072,0.027,0.002,0.099
2.098,1.571,0.855,0.966,0.3,39.03,3.0,0.0,0.0,0.354,0.646,0.594,1.000
2.098,1.571,0.855,0.966,0.3,39.03,3.0,0.3,17.5,0.390,0.560,0.510,0.951
2.098,1.571,0.855,0.966,0.3,39.03,3.0,0.5,30.0,0.320,0.404,0.392,0.724
2.098,1.571,0.855,0.966,0.3,39.03,3.0,0.7,44.4,0.209,0.145,0.154,0.354
2.098,1.571,0.855,0.966,0.3,39.03,3.0,0.9,64.2,0.091,0.043,0.005,0.134
2.098,1.571,1.327,1.095,0.4,13.61,1.0,0.0,0.0,0.134,0.062,0.034,0.196
2.098,1.571,1.327,1.095,0.4,13.61,1.0,0.3,17.5,0.085,0.010,0.021,0.094
2.098,1.571,1.327,1.095,0.4,13.61,1.0,0.5,30.0,0.059,0.011,0.009,0.071
2.098,1.571,1.327,1.095,0.4,13.61,1.0,0.7,44.4,0.035,0.001,0.002,0.036
2.098,1.571,1.327,1.095,0.4,13.61,1.0,0.9,64.2,0.011,0.001,0.000,0.012
2.098,1.571,1.327,1.095,0.4,20.42,1.5,0.0,0.0,0.411,0.081,0.244,0.492
2.098,1.571,1.327,1.095,0.4,20.42,1.5,0.3,17.5,0.327,0.059,0.164,0.385
2.098,1.571,1.327,1.095,0.4,20.42,1.5,0.5,30.0,0.249,0.089,0.082,0.338
2.098,1.571,1.327,1.095,0.4,20.42,1.5,0.7,44.4,0.150,0.045,0.016,0.195
2.098,1.571,1.327,1.095,0.4,20.42,1.5,0.9,64.2,0.062,0.026,0.001,0.088
2.098,1.571,1.327,1.095,0.4,27.23,2.0,0.0,0.0,0.427,0.414,0.432,0.840
2.098,1.571,1.327,1.095,0.4,27.23,2.0,0.3,17.5,0.421,0.374,0.337,0.795
2.098,1.571,1.327,1.095,0.4,27.23,2.0,0.5,30.0,0.320,0.224,0.213,0.543
2.098,1.571,1.327,1.095,0.4,27.23,2.0,0.7,44.4,0.192,0.065,0.059,0.257
2.098,1.571,1.327,1.095,0.4,27.23,2.0,0.9,64.2,0.081,0.029,0.003,0.110
2.098,1.571,1.327,1.095,0.4,40.84,3.0,0.0,0.0,0.435,0.565,0.511,1.000
2.098,1.571,1.327,1.095,0.4,40.84,3.0,0.3,17.5,0.430,0.558,0.503,0.988
2.098,1.571,1.327,1.095,0.4,40.84,3.0,0.5,30.0,0.350,0.477,0.438,0.827
2.098,1.571,1.327,1.095,0.4,40.84,3.0,0.7,44.4,0.233,0.197,0.192,0.430
2.098,1.571,1.327,1.095,0.4,40.84,3.0,0.9,64.2,0.100,0.048,0.006,0.147
2.098,1.571,1.996,1.226,0.5,14.48,1.0,0.0,0.0,0.179,0.003,0.045,0.183
2.098,1.571,1.996,1.226,0.5,14.48,1.0,0.3,17.5,0.126,0.016,0.033,0.142
2.098,1.571,1.996,1.226,0.5,14.48,1.0,0.5,30.0,0.083,0.025,0.014,0.108
2.098,1.571,1.996,1.226,0.5,14.48,1.0,0.7,44.4,0.042,0.002,0.003,0.044
2.098,1.571,1.996,1.226,0.5,14.48,1.0,0.9,64.2,0.011,0.000,0.000,0.011
2.098,1.571,1.996,1.226,0.5,21.72,1.5,0.0,0.0,0.453,0.028,0.220,0.480
2.098,1.571,1.996,1.226,0.5,21.72,1.5,0.3,17.5,0.363,0.079,0.192,0.443
2.098,1.571,1.996,1.226,0.5,21.72,1.5,0.5,30.0,0.299,0.127,0.105,0.426
2.098,1.571,1.996,1.226,0.5,21.72,1.5,0.7,44.4,0.179,0.058,0.022,0.237
2.098,1.571,1.996,1.226,0.5,21.72,1.5,0.9,64.2,0.070,0.026,0.002,0.096
2.098,1.571,1.996,1.226,0.5,28.96,2.0,0.0,0.0,0.545,0.303,0.352,0.847
2.098,1.571,1.996,1.226,0.5,28.96,2.0,0.3,17.5,0.459,0.424,0.358,0.883
2.098,1.571,1.996,1.226,0.5,28.96,2.0,0.5,30.0,0.363,0.304,0.252,0.666
2.098,1.571,1.996,1.226,0.5,28.96,2.0,0.7,44.4,0.219,0.093,0.073,0.312
2.098,1.571,1.996,1.226,0.5,28.96,2.0,0.9,64.2,0.088,0.032,0.004,0.121
2.098,1.571,1.996,1.226,0.5,43.44,3.0,0.0,0.0,0.555,0.445,0.390,1.000
2.098,1.571,1.996,1.226,0.5,43.44,3.0,0.3,17.5,0.466,0.533,0.476,1.000
2.098,1.571,1.996,1.226,0.5,43.44,3.0,0.5,30.0,0.374,0.544,0.473,0.918
2.098,1.571,1.996,1.226,0.5,43.44,3.0,0.7,44.4,0.256,0.268,0.238,0.523
2.098,1.571,1.996,1.226,0.5,43.44,3.0,0.9,64.2,0.108,0.053,0.009,0.162
,,,,,,,,,,,,
5.245,1.989,0.554,0.845,0.1,17.45,1.0,0.0,0.0,0.002,0.000,0.008,0.002
5.245,1.989,0.554,0.845,0.1,17.45,1.0,0.3,17.5,0.004,0.000,0.014,0.004
5.245,1.989,0.554,0.845,0.1,17.45,1.0,0.5,30.0,0.017,0.000,0.006,0.017
5.245,1.989,0.554,0.845,0.1,17.45,1.0,0.7,44.4,0.017,0.002,0.001,0.019
5.245,1.989,0.554,0.845,0.1,17.45,1.0,0.9,64.2,0.014,0.001,0.000,0.016
5.245,1.989,0.554,0.845,0.1,26.18,1.5,0.0,0.0,0.056,0.041,0.107,0.098
5.245,1.989,0.554,0.845,0.1,26.18,1.5,0.3,17.5,0.110,0.042,0.085,0.152
5.245,1.989,0.554,0.845,0.1,26.18,1.5,0.5,30.0,0.103,0.018,0.041,0.121
5.245,1.989,0.554,0.845,0.1,26.18,1.5,0.7,44.4,0.081,0.019,0.008,0.100
5.245,1.989,0.554,0.845,0.1,26.18,1.5,0.9,64.2,0.038,0.022,0.001,0.059
5.245,1.989,0.554,0.845,0.1,34.91,2.0,0.0,0.0,0.198,0.273,0.268,0.471
5.245,1.989,0.554,0.845,0.1,34.91,2.0,0.3,17.5,0.173,0.192,0.206,0.365
5.245,1.989,0.554,0.845,0.1,34.91,2.0,0.5,30.0,0.162,0.080,0.107,0.242
5.245,1.989,0.554,0.845,0.1,34.91,2.0,0.7,44.4,0.108,0.025,0.021,0.133
5.245,1.989,0.554,0.845,0.1,34.91,2.0,0.9,64.2,0.046,0.022,0.001,0.068
5.245,1.989,0.554,0.845,0.1,52.36,3.0,0.0,0.0,0.208,0.655,0.574,0.863
5.245,1.989,0.554,0.845,0.1,52.36,3.0,0.3,17.5,0.225,0.474,0.500,0.699
5.245,1.989,0.554,0.845,0.1,52.36,3.0,0.5,30.0,0.212,0.253,0.255,0.465
5.245,1.989,0.554,0.845,0.1,52.36,3.0,0.7,44.4,0.137,0.054,0.062,0.190
5.245,1.989,0.554,0.845,0.1,52.36,3.0,0.9,64.2,0.061,0.031,0.002,0.091
5.245,1.989,1.246,1.07,0.2,17.65,1.0,0.0,0.0,0.013,0.023,0.029,0.036
5.245,1.989,1.246,1.07,0.2,17.65,1.0,0.3,17.5,0.027,0.001,0.015,0.028
5.245,1.989,1.246,1.07,0.2,17.65,1.0,0.5,30.0,0.026,0.001,0.012,0.027
5.245,1.989,1.246,1.07,0.2,17.65,1.0,0.7,44.4,0.024,0.001,0.002,0.025
5.245,1.989,1.246,1.07,0.2,17.65,1.0,0.9,64.2,0.013,0.000,0.001,0.013
5.245,1.989,1.246,1.07,0.2,26.48,1.5,0.0,0.0,0.251,0.106,0.210,0.358
5.245,1.989,1.246,1.07,0.2,26.48,1.5,0.3,17.5,0.201,0.047,0.126,0.248
5.245,1.989,1.246,1.07,0.2,26.48,1.5,0.5,30.0,0.149,0.033,0.064,0.182
5.245,1.989,1.246,1.07,0.2,26.48,1.5,0.7,44.4,0.099,0.027,0.012,0.126
5.245,1.989,1.246,1.07,0.2,26.48,1.5,0.9,64.2,0.042,0.023,0.001,0.066
5.245,1.989,1.246,1.07,0.2,35.30,2.0,0.0,0.0,0.277,0.439,0.394,0.716
5.245,1.989,1.246,1.07,0.2,35.30,2.0,0.3,17.5,0.296,0.255,0.262,0.550
5.245,1.989,1.246,1.07,0.2,35.30,2.0,0.5,30.0,0.218,0.136,0.156,0.354
5.245,1.989,1.246,1.07,0.2,35.30,2.0,0.7,44.4,0.130,0.034,0.034,0.164
5.245,1.989,1.246,1.07,0.2,35.30,2.0,0.9,64.2,0.053,0.024,0.002,0.077
5.245,1.989,1.246,1.07,0.2,52.95,3.0,0.3,17.5,0.336,0.537,0.468,0.873
5.245,1.989,1.246,1.07,0.2,52.95,3.0,0.5,30.0,0.267,0.352,0.388,0.620
5.245,1.989,1.246,1.07,0.2,52.95,3.0,0.7,44.4,0.166,0.094,0.109,0.260
5.245,1.989,1.246,1.07,0.2,52.95,3.0,0.9,64.2,0.069,0.035,0.004,0.104
5.245,1.989,2.133,1.243,0.3,18.22,1.0,0.0,0.0,0.036,0.105,0.050,0.142
5.245,1.989,2.133,1.243,0.3,18.22,1.0,0.3,17.5,0.045,0.002,0.028,0.047
5.245,1.989,2.133,1.243,0.3,18.22,1.0,0.5,30.0,0.033,0.000,0.012,0.034
5.245,1.989,2.133,1.243,0.3,18.22,1.0,0.7,44.4,0.027,0.005,0.005,0.032
5.245,1.989,2.133,1.243,0.3,18.22,1.0,0.9,64.2,0.011,0.001,0.001,0.012
5.245,1.989,2.133,1.243,0.3,27.33,1.5,0.0,0.0,0.340,0.123,0.252,0.463
5.245,1.989,2.133,1.243,0.3,27.33,1.5,0.3,17.5,0.259,0.035,0.163,0.293
5.245,1.989,2.133,1.243,0.3,27.33,1.5,0.5,30.0,0.190,0.068,0.085,0.258
5.245,1.989,2.133,1.243,0.3,27.33,1.5,0.7,44.4,0.118,0.035,0.017,0.152
5.245,1.989,2.133,1.243,0.3,27.33,1.5,0.9,64.2,0.046,0.026,0.002,0.072
5.245,1.989,2.133,1.243,0.3,36.44,2.0,0.3,17.5,0.368,0.328,0.306,0.696
5.245,1.989,2.133,1.243,0.3,36.44,2.0,0.5,30.0,0.263,0.184,0.252,0.447
5.245,1.989,2.133,1.243,0.3,36.44,2.0,0.7,44.4,0.153,0.048,0.047,0.200
5.245,1.989,2.133,1.243,0.3,36.44,2.0,0.9,64.2,0.060,0.027,0.002,0.087
5.245,1.989,2.133,1.243,0.3,54.66,3.0,0.5,30.0,0.310,0.418,0.381,0.729
5.245,1.989,2.133,1.243,0.3,54.66,3.0,0.7,44.4,0.190,0.144,0.151,0.334
5.245,1.989,2.133,1.243,0.3,54.66,3.0,0.9,64.2,0.076,0.038,0.003,0.113
5.245,1.989,3.323,1.401,0.4,19.10,1.0,0.0,0.0,0.121,0.035,0.076,0.156
5.245,1.989,3.323,1.401,0.4,19.10,1.0,0.3,17.5,0.069,0.004,0.044,0.074
5.245,1.989,3.323,1.401,0.4,19.10,1.0,0.5,30.0,0.044,0.013,0.018,0.057
5.245,1.989,3.323,1.401,0.4,19.10,1.0,0.7,44.4,0.030,0.001,0.003,0.030
5.245,1.989,3.323,1.401,0.4,19.10,1.0,0.9,64.2,0.010,0.000,0.000,0.010
5.245,1.989,3.323,1.401,0.4,28.65,1.5,0.3,17.5,0.316,0.037,0.188,0.353
5.245,1.989,3.323,1.401,0.4,28.65,1.5,0.5,30.0,0.234,0.101,0.106,0.335
5.245,1.989,3.323,1.401,0.4,28.65,1.5,0.7,44.4,0.136,0.044,0.022,0.181
5.245,1.989,3.323,1.401,0.4,28.65,1.5,0.9,64.2,0.051,0.026,0.002,0.077
5.245,1.989,3.323,1.401,0.4,38.20,2.0,0.3,17.5,0.413,0.390,0.333,0.803
5.245,1.989,3.323,1.401,0.4,38.20,2.0,0.5,30.0,0.307,0.234,0.224,0.541
5.245,1.989,3.323,1.401,0.4,38.20,2.0,0.7,44.4,0.175,0.065,0.065,0.241
5.245,1.989,3.323,1.401,0.4,38.20,2.0,0.9,64.2,0.066,0.030,0.004,0.096
5.245,1.989,3.323,1.401,0.4,57.31,3.0,0.5,30.0,0.343,0.487,0.417,0.830
5.245,1.989,3.323,1.401,0.4,57.31,3.0,0.7,44.4,0.213,0.207,0.195,0.421
5.245,1.989,3.323,1.401,0.4,57.31,3.0,0.9,64.2,0.084,0.042,0.004,0.125
5.245,1.989,4.991,1.555,0.5,20.35,1.0,0.3,17.5,0.104,0.010,0.062,0.114
5.245,1.989,4.991,1.555,0.5,20.35,1.0,0.5,30.0,0.060,0.016,0.028,0.076
5.245,1.989,4.991,1.555,0.5,20.35,1.0,0.7,44.4,0.036,0.001,0.005,0.036
5.245,1.989,4.991,1.555,0.5,20.35,1.0,0.9,64.2,0.009,0.000,0.001,0.009
5.245,1.989,4.991,1.555,0.5,30.53,1.5,0.3,17.5,0.354,0.085,0.201,0.440
5.245,1.989,4.991,1.555,0.5,30.53,1.5,0.5,30.0,0.284,0.132,0.129,0.416
5.245,1.989,4.991,1.555,0.5,30.53,1.5,0.7,44.4,0.163,0.055,0.030,0.219
5.245,1.989,4.991,1.555,0.5,30.53,1.5,0.9,64.2,0.060,0.028,0.002,0.088
5.245,1.989,4.991,1.555,0.5,40.70,2.0,0.5,30.0,0.352,0.304,0.249,0.657
5.245,1.989,4.991,1.555,0.5,40.70,2.0,0.7,44.4,0.202,0.095,0.083,0.298
5.245,1.989,4.991,1.555,0.5,40.70,2.0,0.9,64.2,0.075,0.033,0.003,0.108
5.245,1.989,4.991,1.555,0.5,61.05,3.0,0.5,30.0,0.367,0.554,0.447,0.920
5.245,1.989,4.991,1.555,0.5,61.05,3.0,0.7,44.4,0.237,0.285,0.308,0.523
5.245,1.989,4.991,1.555,0.5,61.05,3.0,0.9,64.2,0.091,0.046,0.005,0.137
\end{filecontents*}
\csvreader[
        head to column names, 
        late after line=\\, 
        late after last line=\\ 
    ]{data_X_NF_FF.csv}{}
    {%
        \csvcoli & \csvcolii & \csvcoliii & \csvcoliv & \csvcolv & \csvcolvi & \csvcolvii & \csvcolviii & \csvcolix & \csvcolx & \csvcolxi & \csvcolxii & \csvcolxiii
    }
    \label{table:initial_conditions}
\end{longtable*}


\section{Fitting parameters}
\label{sec:fit_params}
\setcounter{table}{0}

The fitting parameters used in the scaling law for near-field (Eqns. \ref{eq:NF}, \ref{eq:NF_params}) and far-field (Eqns. \ref{eq:FF}, \ref{eq:FF_params}) loss are listed below in Tables \ref{table:NF_params} and \ref{table:FF_params}, respectively.

\renewcommand{\thetable}{\thesection\arabic{table}}
\begin{table}
    \caption{Fitting parameters, $k_{ij}$, used in Eqn.~\ref{eq:NF_params} to predict the near-field loss.}
    \label{table:NF_params}
    \centering
    \begin{tabular}{c|cccc}
    \toprule
    \toprule
         & Parameter & Value & Parameter & Value \\
         \midrule
         \multirow{3}{*}{$\xi_1$} & $k_{11}$ & -643 & $k_{14}$ & 643 \\
                             & $k_{12}$ & 0.928 & $k_{15}$ & 0.000183 \\
                             & $k_{13}$ & -0.241 & $k_{16}$ & -0.00837 \\
                                     
         \midrule
         \multirow{3}{*}{$\xi_2$} & $k_{21}$ & -817 & $k_{24}$ & 817 \\
                             & $k_{22}$ & 0.122 & $k_{25}$ & 0.000566 \\
                             & $k_{23}$ & -0.106 & $k_{26}$ & 0.00\\
                                     
         \midrule
         \multirow{3}{*}{$\xi_3$} & $k_{31}$ & -0.468 & $k_{34}$ & 2.47 \\
                             & $k_{32}$ & 1.25 & $k_{35}$ & -4.64 \\
                             & $k_{33}$ & -0.516 & $k_{36}$ & 0.00 \\
                                     
    \bottomrule
    \end{tabular}
\end{table}

\begin{table}
    \caption{Fitting parameters, $s_{ij}$, used in Eqn.~\ref{eq:FF_params} to predict the far-field loss. Note, the $^*$ on $\psi_2^*$ indicates that this parameter is given in units of kg~MJ$^{-1}$ to ensure non-dimensionality of the expression.}
    \label{table:FF_params}
    \centering
    \begin{tabular}{c|cccc}
    \toprule
    \toprule
         & Parameter & Value & Parameter & Value \\
         \midrule
         \multirow{4}{*}{$\psi_1$} & $s_{11}$ & -8150 & $s_{14}$ & 256 \\
                             & $s_{12}$ & 7894 & $s_{15}$ & -0.0000468 \\
                             & $s_{13}$ & 0.000167 & -- & -- \\
                                     
         \midrule
         \multirow{4}{*}{$\psi_2^*$} & $s_{21}$ & 1.79 & $s_{24}$ & 0.00709 \\
                             & $s_{22}$ & -1.79 & $s_{25}$ & -0.484 \\
                             & $s_{23}$ & 0.000132 & -- & --\\
                                     
         \midrule
         \multirow{4}{*}{$\psi_3$} & $s_{31}$ & 928 & $s_{34}$ & 0.00 \\
                             & $s_{32}$ & -928 & $s_{35}$ & 0.00 \\
                             & $s_{33}$ & 0.00144 & -- & -- \\
        \midrule
         \multirow{4}{*}{$\psi_4$} & $s_{41}$ & -75.0 & $s_{44}$ & 75.9 \\
                             & $s_{42}$ & -4.25 & $s_{45}$ & 0.00000563 \\
                             & $s_{43}$ & 1.69 & -- & -- \\
                                     
    \bottomrule
    \end{tabular}
\end{table}

\bibliography{bibliography.bib}{}
\bibliographystyle{aasjournal}

\end{document}